\documentclass[a4paper, fleqn, usenatbib]{mnras}

\usepackage{newtxtext,newtxmath}

\usepackage[T1]{fontenc}
\usepackage{ae,aecompl}
\usepackage{booktabs}  

\usepackage{graphicx}   
\usepackage{amsmath}    
\usepackage{gensymb}
\usepackage[flushleft]{threeparttable}

\usepackage{tabularx}
\newcolumntype{Y}{>{\centering\arraybackslash}X}

\usepackage{pdflscape}
\usepackage{CJK}

\graphicspath{{figs/}}

\newcommand{\angstrom}{\textup{\AA}}

\def\lsim{\mathrel{\rlap{\lower4pt\hbox{\hskip1pt$\sim$}}
    \raise1pt\hbox{$<$}}}                
\def\gsim{\mathrel{\rlap{\lower4pt\hbox{\hskip1pt$\sim$}}
    \raise1pt\hbox{$>$}}}                

\title[X-ray spectral and variability properties of RLQs]{The X-ray
spectral and variability properties of typical radio-loud quasars}

\author[S. F. Zhu et al.]{S. F. Zhu,$^{1,2}$\thanks{E-mail: SFZAstro@gmail.com (PSU)}
John D. Timlin III,$^{1,2}$
and W. N. Brandt$^{1,2,3}$
\\
$^1$Department of Astronomy \& Astrophysics, The Pennsylvania State University, University Park, PA 16802, USA\\
$^2$Institute for Gravitation and the Cosmos, The Pennsylvania State University, University Park, PA 16802, USA\\
$^3$Department of Physics, 104 Davey Lab, The Pennsylvania State University, University Park, PA 16802, USA}

\date{Accepted XXX. Received YYY; in original form ZZZ}

\pubyear{2020}

\begin{document}
\label{firstpage}
\pagerange{\pageref{firstpage}--\pageref{lastpage}}
\maketitle
\begin{abstract}
We present X-ray spectral and long-term variability analyses of an unbiased sample of
361 optically selected radio-loud quasars (RLQs) utilizing sensitive
    serendipitous X-ray data from the {\it Chandra} and {\it XMM-Newton} archives.
The spectral and temporal properties of RLQs are compared with those of 
radio-quiet quasars (RQQs) matched in $L_\mathrm{2500\angstrom}$ and $z$.
The median power-law photon index ($\Gamma$) of RLQs is $1.84_{-0.01}^{+0.01}$,
    which is close to that of matched RQQs ($1.90_{-0.01}^{+0.02}$).
No significant correlations between $\Gamma$ and radio-loudness, 
$L_\mathrm{x}/L_\mathrm{x,rqq}$ (the X-ray luminosity over that expected from the $L_\mathrm{x}$-$L_\mathrm{uv}$ relation for RQQs),
redshift, or Eddington ratio are found for our RLQs.
The stacked X-ray spectra of our RLQs show strong iron-line emission 
and a possible Compton-reflection hump. 
The intrinsic X-ray variability amplitude is $\approx40$\% for RLQs on 
timescales of months-to-years in the rest frame,
which is somewhat smaller than for the matched
    RQQs ($\approx60$\%) on similar timescales,
perhaps due to the larger black-hole masses and lower Eddington ratios in our RLQ sample.
The X-ray spectral and variability results for our RLQs
generally support the idea that the X-ray emission
of typical RLQs is dominated by the disk/corona, as is also indicated by a recent luminosity correlation study.
\end{abstract}




\section{Introduction}
\label{sec:intro}
Radio-loud quasars (RLQs) have powerful relativistic jets that are absent in radio-quiet quasars (RQQs; e.g. \citealt{padovani2017}).
These two types of quasars are observationally distinguished by the radio-loudness parameter, $R\equiv L_\mathrm{5GHz}/L_\mathrm{4400\angstrom}$,
where $L_\mathrm{5GHz}$ and $L_\mathrm{4400\angstrom}$ are monochromatic luminosities at rest-frame 5~GHz and 4400~\angstrom, respectively (\citealt{kellermann1989}).
Only 10--20\% of quasars are RLQs with $R\ge10$, while the rest are RQQs (e.g. \citealt{ivezic2002}).
Typical RLQs have similar near-infrared-to-UV spectral energy distributions (SEDs) to those of RQQs,
showing the so-called big blue bump with strong emission lines superimposed (e.g. \citealt{elvis1994,shang2011}).
However, RLQs are generally brighter X-ray emitters than RQQs and thus have a flatter optical/UV-to-X-ray spectral slope,
$\alpha_\mathrm{ox}$ (e.g. \citealt{worrall1987, miller2011}),
where $\alpha_\mathrm{ox}\equiv\log(L_\mathrm{2keV}/L_\mathrm{2500\angstrom})/\log(\nu_\mathrm{2keV}/\nu_\mathrm{2500\angstrom})$ assuming a power-law spectral shape between rest-frame 2~keV and 2500~\angstrom~(\citealt{tananbaum1979}).
Furthermore, the excess X-ray emission of RLQs above that of RQQs correlates with both $R$ and $L_\mathrm{5GHz}$,
which has been taken as evidence that the nuclear X-ray emission of RLQs 
contains both a standard accretion-disk corona component as well as a distinct X-ray 
component associated with the base of the powerful radio jets (e.g. \citealt{worrall1987, miller2011}).
This explanation seems consistent with some previous studies that
have found that RLQs have systematically flatter X-ray ($<10$ keV) spectra
than those of RQQs (e.g. \citealt{wilkes1987, reeves1997, page2005}),
which could arise due to the mixture of the coronal X-ray emission 
and a generally harder \mbox{X-ray} spectrum from the jet-linked emission
(e.g. \citealt{grandi2004}).

This original two-component model of the nuclear X-ray emission from typical RLQs
has recently been challenged by \citet{zhu2020}, who studied the correlations of the
continuum emission from the corona, disk, and jets of RLQs.
\citet{zhu2020} investigated the correlations between X-ray, optical/UV, 
and radio luminosities\footnote{We use $L_\mathrm{2keV}$, $L_\mathrm{2500\angstrom}$, and $L_\mathrm{5GHz}$ interchangeably with $L_\mathrm{x}$, $L_\mathrm{uv}$, and $L_\mathrm{radio}$ to denote the X-ray, optical/UV, and radio luminosities in this paper.}
using a large and well-characterized
sample consisting of more than 700 optically-selected RLQs. 
The steep-spectrum radio quasars (SSRQs), 
which have a radio spectral slope $\alpha_\mathrm{r}\le-0.5$,
showed a $L_\mathrm{x}$-$L_\mathrm{uv}$ relation that is quantitatively similar to that of RQQs (i.e. $L_\mathrm{x}\propto L_\mathrm{uv}^\gamma$, $\gamma\approx$ 0.6; e.g. \citealt{just2007, risaliti2019}),
despite the fact that these SSRQs are typically a factor of $\approx2$--3 times more X-ray luminous than RQQs at a given $L_\mathrm{uv}$.
The quantitatively similar $L_\mathrm{x}$-$L_\mathrm{uv}$ correlation for SSRQs 
supports the idea that the nuclear X-ray emission of these SSRQs is
still dominated by the corona
that is coupled with the accretion disk (e.g. \citealt{arcodia2019}),
and no jet component is required.
\citet{zhu2020} also showed that the jet-linked component is only important
for a small fraction ($<10\%$) of flat-spectrum radio quasars (FSRQs; $\alpha_\mathrm{r}>-0.5$);
thus, the corona-linked component also dominates the X-ray emission of most typical FSRQs.
The relation between $\alpha_\mathrm{ox}$ and 
the equivalent width (EW) of He {\sc ii} in RLQs is consistent with the result of \citet{zhu2020}
that RLQ X-ray emission is mainly related to the disk/corona instead of the jets (\citealt{timlin2021}).

In light of these new results, it is important to investigate further the X-ray spectral
properties of typical RLQs to assess whether they are indeed consistent with being mainly corona-linked.
The primary X-ray continuum radiated from the hot corona is
approximately described by a power law with a typical photon index of \mbox{$\Gamma=1.7$--2.3}
and an exponential cut-off at $\sim100$ keV (e.g. \citealt{mushotzky1993, kamraj2018, molina2019}),
while the spectrum from the jets in the same band 
is a flatter power law ($\Gamma\sim1.5$; e.g. \citealt{paliya2020}) 
that might extend to $\gamma$-rays (e.g. \citealt{hartman1992}).
\citet{kang2020} recently analyzed the {\it NuSTAR} spectra of 28 radio galaxies and detected the
hard X-ray cut-off for 13 objects with sufficient net counts,
which thus supports the idea that the hard X-rays of these
radio galaxies are dominated by their coronae.
Reprocessed X-ray emission can 
also be used to constrain the origin of the primary X-ray continuum.
The iron emission lines between rest-frame \mbox{6.4--7.0}~keV are of particular interest for this work, 
given their ubiquity in radio-quiet Seyfert galaxies (e.g. \citealt{nandra2007}).
If a beamed jet-linked continuum dominates the X-ray emission of RLQs, the iron-line emission 
is expected to be weak to undetectable owing to flux dilution (e.g. \citealt{reeves1997}); 
otherwise, the corona-dominated interpretation for the X-ray emission of RLQs is favored.
The Compton-reflection hump that broadly peaks at rest-frame 20--30 keV is also consistently expected if strong iron lines are detected (e.g. \citealt{george1991}).

Previous \mbox{X-ray} spectral studies of RLQs have often used small,
heterogeneously-selected samples that might be biased.
For example, previous RLQs selected in the radio band usually
contain a significant portion of extreme FSRQs\footnote{The radio slopes of these most radio luminous objects are flat because their core emission is beamed toward the Earth,
which is generally not true for FSRQs in optically selected samples.}
that have prominent 
jet-linked \mbox{X-ray} emission (see \S~\ref{sec:sampleSelection} for further discussion).
In this paper, we focus on the X-ray spectral properties of a large and optically selected RLQ sample (e.g. \citealt{zhu2020}).
Furthermore, we use only serendipitously obtained X-ray data to reduce potential selection effects 
caused by targeted X-ray observations (see \S~\ref{sec:newRLQs}).

Further insights into the nature of the X-ray emission of RLQs can also be obtained by
studying their X-ray variability properties, in comparison with those of RQQs.
Generally, the X-ray variability properties of RLQs have only been investigated for 
either individual objects (e.g. \citealt{leighly1997, hayashida2015}) or 
small samples (e.g. \citealt{zamorani1984b, sambruna1997, gibson2012})
rather than for large, homogeneous, statistically meaningful samples.
In this work, we thus also investigate the \mbox{X-ray}
variability properties of the typical RLQs selected by \citet{zhu2020}.
See \S~\ref{sec:newRLQs} for the improvements of our sample relative to past work.

The sample construction and X-ray data analysis are discussed in \S~\ref{sec:sampleSelection}.
The method used to fit the X-ray spectra and the fitting results are reported in \S~\ref{sec:spec},
and the long-term X-ray variability of RLQs is discussed in \S~\ref{sec:var}.
We adopt a $\Lambda$CDM cosmology with $H_0=70$~km~s$^{-1}$~Mpc$^{-1}$ and $(\Omega_\Lambda, \Omega_\mathrm{M})=(0.7, 0.3)$.




\section{Quasar Samples and X-ray Data Analysis}
\label{sec:sampleSelection}

\subsection{RLQ sample}
\label{sec:newRLQs}

\citet{zhu2020} constructed a sample of 729 RLQs, utilizing the Sloan Digital Sky Survey (SDSS; \citealt{york2000}),
the Faint Images of the Radio Sky at Twenty-Centimeters (FIRST; \citealt{becker1995}), and the NRAO VLA Sky Survey (NVSS; \citealt{condon1998}).
These 729 quasars were initially selected using SDSS colors and are therefore not biased by their X-ray or radio properties.
Archival {\it Chandra}, {\it XMM-Newton}, and {\it ROSAT} observations are used to calculate the \mbox{X-ray} luminosities of these optically-selected quasars.
Only a small fraction ($\lesssim15\%$) of the RLQ sample is targeted by these X-ray observations;
therefore, this sample is not significantly biased by the target selection of these X-ray observations (e.g. \citealt{gibson2012}).
This RLQ sample also has large fractions of X-ray detections and spectroscopic confirmations by SDSS (see Table~1 of \citealt{zhu2020}).
Furthermore, almost all of the RLQs are classified as either FSRQs or SSRQs, utilizing 
radio surveys at frequencies other than 1.4~GHz, in particular, the Very Large Array Sky Survey (VLASS; \citealt{vlass2020}) at $\approx3$ GHz.
Broad absorption line quasars, quasars that suffer from strong dust extinction,
and quasars with prominent jet emission in the infrared-through-UV bands were excluded from this sample.

In order to investigate the X-ray temporal properties of RLQs using multiple-epoch observations,
we match the RLQ list of \citet{zhu2020} with the observation catalogs
of {\it Chandra} and {\it XMM-Newton}.
We only retain serendipitous observations (i.e. where the quasar position is $>1$ arcmin off-axis),
which further minimizes the impacts of a small portion of RLQs in \citet{zhu2020} 
that might be exceptional objects.
The matching results in 326 {\it Chandra}/ACIS observations of 202 quasars and
474 {\it XMM-Newton}/EPIC observations of 293 quasars.
The X-ray data reduction and quality cuts are described in the following subsections.
Together, 333 RLQs will be used for X-ray spectral studies
and 105 RLQs with more than one {\it Chandra}/{\it XMM-Newton} 
observation will be used to investigate the X-ray variability properties of RLQs.
We list our final RLQ sample in Table~\ref{tab:sample} and X-ray observation sample in Table~\ref{tab:obs}, respectively.
The resulting RLQ sample in the luminosity-redshift plane is shown in Fig.~\ref{fig:sample}.

We compare our spectral RLQ sample with those from previous investigations in Table~\ref{tab:spec},
where the size of our RLQ sample is larger than those of previous samples by about one order of magnitude.
Importantly, our RLQs are all optically selected quasars,
while those previous samples are generally heterogeneous with 
a large portion that are radio selected, 
and the rest being optically or X-ray selected.
Furthermore, all of our utilized X-ray observations were serendipitously observed, rendering 
our RLQ sample an unbiased subset of the optically selected SDSS quasars.
In contrast, the previous samples usually utilized targeted X-ray
observations, where the selection effects are hard to assess.
Similarly, for our X-ray variability investigations,
the sizes of the RLQ sample and X-ray observation sample are also significantly larger (by factors of $\approx3$--100)
than those in the literature, as reported in Table~\ref{tab:var}.

\begin{table*}
\centering
\caption{The RLQ sample used in this paper. The quasar properties are taken from \citet{zhu2020}.}
\label{tab:sample}
\begin{threeparttable}[b]
\begin{tabularx}{\linewidth}{@{}Y@{}}
\begin{tabular}{ccccccccc}
\hline
\hline
    Name &  $z$  & $m_i$& $\log L_\mathrm{2500\angstrom}$ & $\log L_\mathrm{5GHz}$ & $\log L_\mathrm{2keV}$ & $\log R$ & $\alpha_\mathrm{r}$ & $\log (L_\mathrm{x}/L_\mathrm{x,rqq})$\tnote{a}\\
\hline
    000442.18$+$000023.3 & 1.008 & 18.91& 30.22& 32.08& 26.91& 1.74& - &0.56\\
    000622.60$-$000424.4 & 1.038 & 19.51& 30.04& 34.94& 27.31& 4.76&$-$0.54& 1.08\\
    001646.54$-$005151.7 & 2.243 & 20.94& 30.20& 32.66& 26.35& 2.33& 0.03& 0.02\\
    001910.95$+$034844.6 & 2.022 & 20.26& 30.35& 32.91& 26.53& 2.43&$-$0.75& 0.10\\
    003054.63$+$045908.4 & 2.201 & 20.92& 30.18& 33.81& 26.59& 3.49& $-$0.57& 0.26\\
\hline
\end{tabular}
\end{tabularx}
\begin{tablenotes}
\item[a] The X-ray luminosity divided by that predicted by the $L_\mathrm{x}-L_\mathrm{uv}$ relation for RQQs (\citealt{zhu2020}).
\end{tablenotes}
\end{threeparttable}
\end{table*}

\begin{table*}
\centering
\caption{The sample of RLQ X-ray observations.}
\label{tab:obs}
\begin{threeparttable}[b]
\begin{tabularx}{\linewidth}{@{}Y@{}}
\begin{tabular}{ccccccccccc}
\hline
\hline
    Name & ObsID & MJD\tnote{a} & Inst.\tnote{b} & $\log f_\mathrm{det}$\tnote{c} & net\tnote{d} & SNR\tnote{e} & $\log f_\mathrm{x}$\tnote{f} & xdet\tnote{g} & $\Gamma$\tnote{h} & goodness-of-fit\tnote{i} \\
\hline
    000442.18$+$000023.3&  0305751001&  53714.8& MOS2& $-13.92$ &263.1&  15.5& $-$12.59&  1 &  2.00$_{-0.19}^{+0.19}$ &  132.8/152.7/14.7\\
    000622.60$-$000424.4&        4096&  52853.3& ACIS& $-13.74$ &220.8&  14.4& $-$12.35&  1 &  1.76$_{-0.21}^{+0.22}$ &  214.3/188.3/16.1\\
    000622.60$-$000424.4&        5617&  53579.5& ACIS& $-14.35$ &806.8&  28.3& $-$12.21&  1 &  1.69$_{-0.10}^{+0.11}$ &  241.9/235.4/19.8\\
    001646.54$-$005151.7&  0403760101&  54076.0&  pn & $-13.98$ &12.9&  2.6&   $-$13.82&  1 & 1.23$_{-1.11}^{+1.48}$ &  80.1/83.8/10.1\\
    001646.54$-$005151.7&  0403760701&  54295.3&  pn & $-14.38$ &24.3&  4.0&   $-$13.91&  1 & 2.37$_{-0.75}^{+0.93}$ &  78.3/88.7/10.7\\
\hline
\end{tabular}
\end{tabularx}
\begin{tablenotes}
\item[a] Observation start time.
\item[b] The instrument used for the observation: ACIS for {\it Chandra} observations and pn/MOS1/MOS2 for {\it XMM-Newton} observations.
\item[c] The detection flux limit in the 0.5--7 keV band at the position of the quasar on the detector. See~\S~\ref{sec:eflux}.
\item[d] The net source counts after subtracting expected background counts, $\mathrm{net}=s-b/k$, where $k$ is the background-to-source area ratio.
\item[e] The signal-to-noise ratio, $\mathrm{SNR}=\mathrm{net}/\sqrt{s+b/k^2}$. \label{note:snr}
\item[f] The energy flux in the 0.5--7 keV band if the the quasar is detected. Otherwise, the upper bound of the 90\% confidence interval is given.
\item[g] If the quasar is detected in the X-ray observation, $\mathrm{xdet}=1$. Otherwise, $\mathrm{xdet}=0$.
\item[h] The power-law photon index derived from our spectral fitting. $\Gamma=-99$ is used for cases of non-detection.
\item[i] The goodness-of-fit of the spectral fitting, $cstat/\mathrm{E}_{cstat}/\mathrm{S}_{cstat}$, 
    where $cstat$ is the statistic of the fit, and $\mathrm{E}_{cstat}$ and $\mathrm{S}_{cstat}$ are the expectation and standard deviation of $cstat$, respectively (\citealt{kaastra2017}).
    For example, if $(cstat-\mathrm{E}_{cstat})/\mathrm{S}_{cstat}>4$, the fitting result is disfavored at a $4\sigma$ significance level.
\end{tablenotes}
\end{threeparttable}
\end{table*}

\begin{table*}
\centering
\caption{Previous sample-based investigations of the X-ray spectral properties of RLQs.}
\label{tab:spec}
\begin{threeparttable}[b]
\begin{tabularx}{\linewidth}{@{}Y@{}}
\begin{tabular}{lcccc}
\hline
\hline
    Sample & Telescope &No. of RLQs\tnote{a} & Radio/Opt\tnote{b} & FSRQ/SSRQ\tnote{c}  \\
\hline
    This paper ($\mathrm{SNR}>3$) & {\it XMM-Newton}/{\it Chandra} & 333 & 0/333 & 184/134\tnote{d} \\
    \cite{zhou2020}& {\it Chandra} & 43 & 43/0 & 14/29 \\
    \cite{grandi2006} & {\it BeppoSAX} & 22 & 17/5 & 16/6 \\
    \cite{page2005} & {\it XMM-Newton} & 16 & 10/4\tnote{e} & - \\
    \cite{reeves2000} & {\it ASCA} & 35 & 33/2 & - \\
    \cite{sambruna1999} & {\it ASCA} & 5 & 5/0 & - \\
    \cite{lawson1997} & {\it Ginga} & 18 & 17/1 & 10/5\tnote{f}\\
    \cite{reeves1997} & {\it ASCA} & 15 & 13/2 & 8/0\tnote{g}\\
    \cite{lawson1992} & {\it EXOSAT} & 18 & 12/6 & 12/6\\
    \cite{wilkes1987} & {\it Einstein} & 17 & 14/3 & - \\
\hline
\end{tabular}
\end{tabularx}
\vspace{1mm}
{\it Notes:} We use only serendipitous X-ray observations to ensure that 
    our RLQs form a representative subset of the parent sample (i.e. optically selected SDSS RLQs).
In contrast, all previous investigations listed here used targeted X-ray observations,
    which may be subject to complex selection effects (except for \citealt{zhou2020}).
\begin{tablenotes}[flushleft]
\item[a] We provide the number of luminous quasars in this column,
and other types of radio-loud AGNs (i.e. broad- and narrow-line radio galaxies) are excluded.
\item[b] The number of quasars that are selected in the radio/optical band. 
\item[c] We list the number of FSRQs/SSRQs if the relevant paper provides radio-spectral 
    information. \citet{zhou2020} divide their RLQs into core-dominated and lobe-dominated classes, which we associate with FSRQs and SSRQs, respectively.
\item[d] The remaining 15 RLQs do not have radio slope measurements.
\item[e] There are two X-ray selected objects in \citet{page2005}.
\item[f] \citet{lawson1997} separate the three optically violent variables (OVVs) from other FSRQs.
\item[g] The remaining seven RLQs of \citet{reeves1997} are either OVVs or Gigahertz-peaked spectrum (GPS) radio sources.
\end{tablenotes}
\end{threeparttable}
\end{table*}

\begin{table*}
\centering
\caption{Previous investigations of the long-term X-ray variability of RLQs.}
\label{tab:var}
\begin{threeparttable}[b]
\begin{tabularx}{\linewidth}{@{}Y@{}}
\begin{tabular}{lcccc}
\hline
\hline
    Sample & No. of RLQs/Object Name& No. of Observations & No. of pairs & Timescale\tnote{a} \\
\hline
\multicolumn{5}{c}{Studies of ensemble X-ray variability}\\
\hline
    This paper (Down-sampled) & 105 & 297 & 314 & 1.7 years \\
    \citet{gibson2012} (High-quality sample) &  8 & 20 & 15 & 1 year \\
    \citet{zamorani1984b} & 7 & 14 & 7 & 1 year \\
\hline
\multicolumn{5}{c}{Studies of individual objects}\\
\hline
    \citet{marscher2018} & 3C~120 & $\approx110$ & - & 8.3 months\\
    \citet{chatterjee2011} & 3C~111 & 822 & - & 5.9 years \\
    \citet{soldi2008}\tnote{b} & 3C~273 & 1036 & - & 31 years \\
    \citet{marscher2006} & 3C~120 & $\approx550$& - & 3 years \\
    \citet{marscher2002} & 3C~120 & $\approx190$ & - & 3 years \\
    \citet{mchardy1999} & 3C~273 & 84 & - & 1.4 months \\
    \citet{leighly1997} &  3C~390.3 & 90 & - & 4.6 months \\
    \citet{sambruna1997}\tnote{c}
                    & 0923+392 & 7 & - & 2 years \\
                    & 3C~345 & 6 & - & 3 years \\
\hline
\end{tabular}
\end{tabularx}
\vspace{1mm}
    {\it Notes:} In this table we do not include the X-ray monitoring studies of 
    blazars and related highly jet-dominated objects (e.g. \citealt{chatterjee2008}).
\begin{tablenotes}
\item[a] The median rest-frame timescale is reported for the ensemble studies, 
    while the maximum rest-frame timescale is reported for individual studies.
\item[b] We consider the 5~keV light-curve of 3C~273 in \citet{soldi2008}.
\item[c] We do not include 3C~279 and 0208$-$512 in \citet{sambruna1997}, of which the former is a blazar and the data of the latter span only 12 days.
\end{tablenotes}
\end{threeparttable}
\end{table*}

\subsection{Data reduction}
\label{sec:dataReduction}
The {\it Chandra} and {\it XMM-Newton}
observations were reduced using the {\sc ciao} (v4.12) and {\sc sas} (v18.0.0) packages, respectively.
The {\it Chandra} data were reprocessed using {\tt chandra\_repro} and
the most recent {\sc caldb} (v4.9.3).
Background flares were then removed using {\tt deflare}.
From the reprocessed, flare-filtered event files, we create clean images in the 0.5--7 keV band.
Source detection is then performed using {\tt wavdetect} with a threshold of $10^{-6}$ 
to obtain a list of X-ray sources in the field of view (FOV).
If a RLQ is detected (i.e. an \mbox{X-ray} source is found within 2 arcsec of the quasar position),
the X-ray position will be used for the following analysis;
otherwise, the optically-determined quasar position is used.
Note that cases where the quasar position is near the detector edge (i.e. the quasar lies within 40 pixels from the edge),
falls in a detector gap, or is near an X-ray luminous cluster are excluded from further analysis.
An elliptical region that encloses 90 per cent of the counts at the position of the quasar
is created using the {\sc marx} (v5.5.0) package (see \S~3.1 of \citealt{timlin2020}).
The background region is defined as an annulus that is centered at the quasar position,
with inner and outer radii of 15 and 50 arcsec, respectively.
X-ray sources detected in the background region are excluded
from the background region or, in a few cases where the RLQ is in a crowded field, 
a nearby circular background region is used.
The source spectrum and associated response files and background spectrum are
extracted using the {\tt specextract} tool.

The EPIC-pn and EPIC-MOS data from {\it XMM-Newton} observations were reprocessed and cleaned using {\tt epproc} and {\tt emproc}, respectively.
We filtered background-flaring periods using {\tt espfilt}.
Clean images are produced using {\tt evselect}, and source detection is then performed using {\tt edetect\_chain}.
Similar to the {\it Chandra} observations,
cases where the quasar position is near the detector edge (i.e. the source region defined below overlaps with the detector boundary),
falls in a detector gap, or is near an X-ray luminous cluster are excluded.
Furthermore, if the background flaring is consistently strong throughout the exposure, 
we exclude the observation entirely.
The source region is defined by a circle centered at the quasar position (or
the \mbox{X-ray} position if an X-ray source is found within 5 arcsec of the quasar position).
The radius of the circle is determined by {\tt eregionanalyse} to optimize the signal-to-noise ratio.
The background region is chosen to be a source-free circular region on the same chip, with a radius of $\gtrsim40$ arcsec.
The source and background spectra are then extracted using {\tt evselect}.
The corresponding RMFs and ARFs are created using {\tt rmfgen} and {\tt arfgen}, respectively.

In total, we obtain 300 useful {\it Chandra}/ACIS spectra of 195 RLQs 
and 401 useful {\it XMM-Newton}/EPIC spectra\footnote{The number of epochs here reflects the number of unique {\it XMM-Newton} ObsIDs 
rather than the number of spectra (i.e. from pn, MOS1, and MOS2). 
See the last paragraph of \S~\ref{sec:eflux} for details regarding 
how we select the most sensitive spectrum.} of 258 RLQs 
All of the following analyses are uniformly performed
utilizing the spectral and associated response files,
regardless of the observatories and instruments.
Note that the RLQs being analyzed here are generally consistent with being X-ray point sources. 
The extended X-ray jet emission from RLQs, if it is spatially resolved,
typically contributes only 1--3\% to the total X-ray flux being analyzed (e.g. \citealt{marshall2018, schwartz2020}),
and thus we do not expect such emission to affect our X-ray spectral and variability analyses below materially.
Pileup is generally not a concern for serendipitous X-ray 
data with large off-axis angles ($>1$ arcmin).\footnote{We checked the brightest X-ray sources with the highest count rates and found no strong pileup effects.}

\subsection{Calculating energy flux and detection flux limit}
\label{sec:eflux}
Using the instrumental response files of each observation, an energy conversion factor (ECF)
that converts the number of net counts
to 0.5--7 keV energy flux is derived using {\sc sherpa} (v4.13, \citealt{db2020}).
The spectral shape of the photons that enter the X-ray telescopes is assumed to be 
a Galactic absorption-modified power law with $\Gamma=1.8$.\footnote{The adopted photon index is consistent with the typical value from our spectral fitting results in \S~\ref{sec:spec}.}
This method is similar to that adopted in the {\it XMM-Newton}
serendipitous source catalog (e.g. \citealt{rosen2016})
and the {\it Chandra} source catalog (e.g. \citealt{evans2019}).

The background-region counts, $b$, and 
the source-region counts, $s$, are read from the source and background spectra.
The source detection is performed using the binomial no-source probability
\begin{equation}
    P_\mathrm{B}(X\ge s)=\sum_{X=s}^{b+s}\frac{(b+s)!}{b!s!}p^X(1-p)^{b+s-X},
\end{equation}
which is the probability of observing a source-to-background counts ratio as extreme as $s/b$,
assuming that the $b+s$ counts are produced
by a uniformly distributed X-ray source, i.e., the background (e.g. \citealt{broos2007, weisskopf2007, xue2011}).
Here, $p=1/(1+k)$, and $k$ is the ratio of the background-to-source region area.
Since the test is performed at the pre-specified positions of optically bright objects,
the quasar is considered detected if $P_\mathrm{B}\le 0.002$ (2.88~$\sigma$),\footnote{Visual 
inspection of the images of the sources with $P_\mathrm{B}\le 0.002$ confirms 
the existence of a X-ray source.}
and the 1$\sigma$ uncertainties (i.e. 68\% confidence interval) of their net counts
are calculated using the {\sc ciao} {\tt aprates} tool (e.g. \citealt{primini2014}).
If the quasar is not detected, we instead calculate 
the 90\% confidence interval using {\tt aprates}
and consider the upper bound as the upper limit of the net counts.
The net counts (upper limit) is then converted to 
energy flux (upper limit of the energy flux) in the 0.5--7 keV band using the ECF.

A flux limit of each {\it XMM-Newton}/{\it Chandra} observation at the quasar position can be calculated
by reversing the problem of source detection.
We find the limiting value of $s$ (denoted $s_\mathrm{det}$) that satisfies the inequality
\begin{equation}
\sum_{X=s}^{b+s}\frac{(b+s)!}{b!s!}p^X(1-p)^{b+s-X}\le0.002,
\end{equation}
searching through integers from $s=1$.
Note that $s_\mathrm{det}$ depends only on $b$ and $k$.
A detection flux limit (denoted $f_\mathrm{det}$) can then be
derived from $s_\mathrm{det}$ and the ECF calculated above.
We apply a uniform, unbiased cut of $f_\mathrm{det}\le3\times10^{-14}$ erg cm$^{-2}$ s$^{-1}$
for all X-ray observations to ensure a high detection fraction. 
The resulting \mbox{X-ray} observations are listed in Table~\ref{tab:obs}.
Note that if multiple cameras (i.e. pn, MOS1, and MOS2) are operating during an {\it XMM-Newton} observation, 
we keep the data from the camera with the smallest $f_\mathrm{det}$ and ignore the 
rest.\footnote{If we utilize data from all working cameras and perform simultaneous source 
detection, the number of {\it XMM-Newton} detections increases only by one,
which is probably caused by the fact that the {\it XMM-Newton} observations are 
generally background-limited (i.e. sensitivity improves with the square root of the exposure time) 
as opposed to photon-limited observations (i.e. sensitivity improves linearly with exposure time) 
for {\it Chandra}.
We thereby elect not to analyze jointly the data from different cameras onboard {\it XMM-Newton} in this paper,
which also maximizes the uniformity between the flows of the data analyses for {\it XMM-Newton} and {\it Chandra}.}

In total, 641 X-ray observations of 361 RLQs passed the flux-limit cut, 
of which 12 RLQs (3.3\%) are not detected in any epoch and thus are not considered in the following analysis.
Our X-ray spectral (\S~\ref{sec:spec}) and variability (\S~\ref{sec:var}) investigations 
each utilize a subset of the remaining 349 RLQs,
which are detected at least once in 628 \mbox{X-ray} observations.
The vast majority of these observations (98.7\%) result in significant \mbox{X-ray} detections, 
and thus our results are not substantively affected by X-ray upper limits.

\begin{figure*}
\centering
\includegraphics[width=0.8\textwidth, clip]{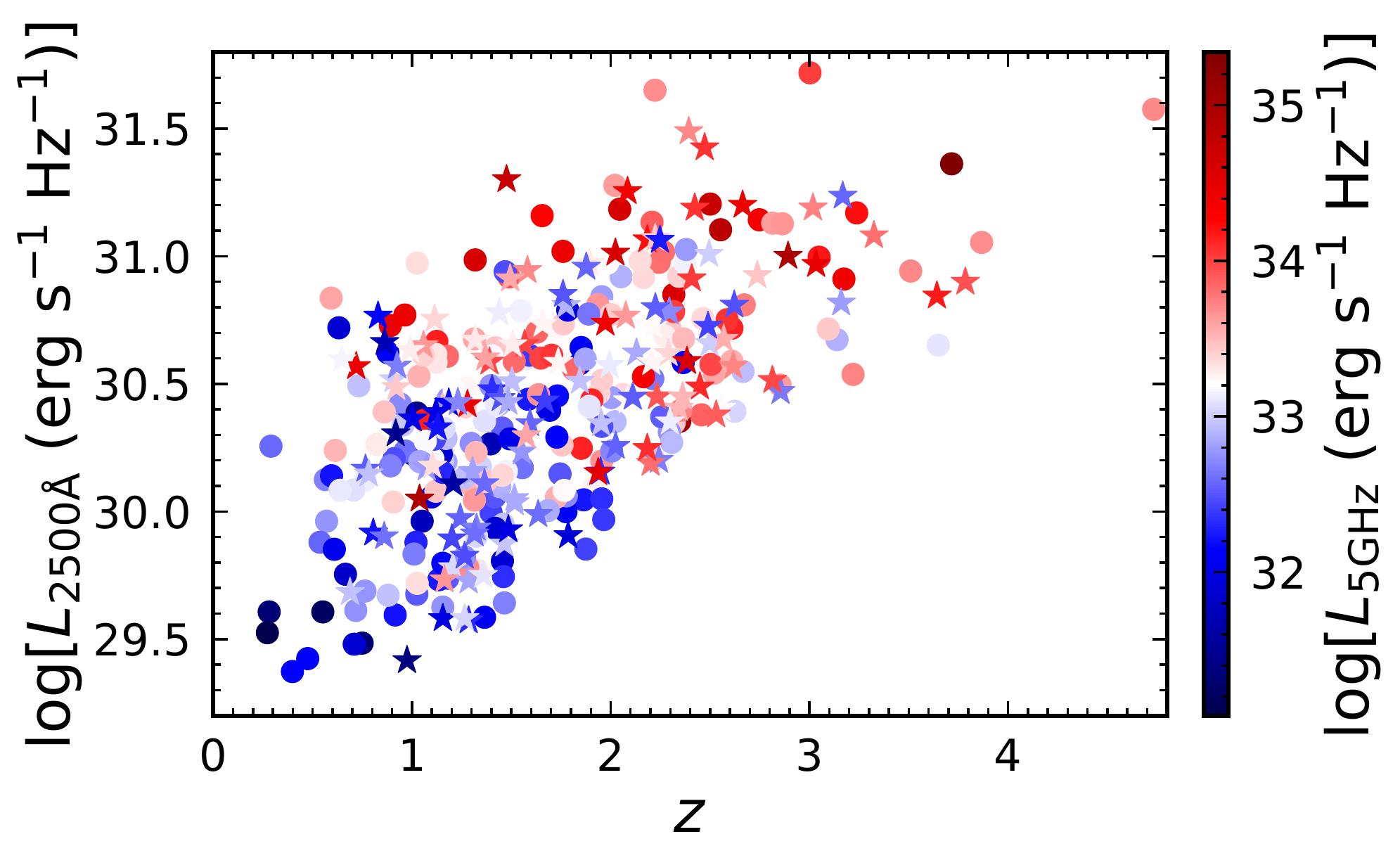}
\caption{The distribution of RLQs from \citet{zhu2020} with serendipitous {\it Chandra}/{\it XMM-Newton} observations 
    in the $L_\mathrm{2500\angstrom}-z$ plane. 
The stars and circles represent quasars having multiple X-ray observations or a single observation, respectively.
    The colors of the data points represent the radio luminosities as per the scale bar.
    Since the X-ray observations are serendipitous, these quasars span wide ranges in $L_\mathrm{2500\angstrom}$, $L_\mathrm{5GHz}$, and $z$.}
\label{fig:sample}
\end{figure*}

\begin{figure*}
\centering
\includegraphics[width=0.90\textwidth, clip]{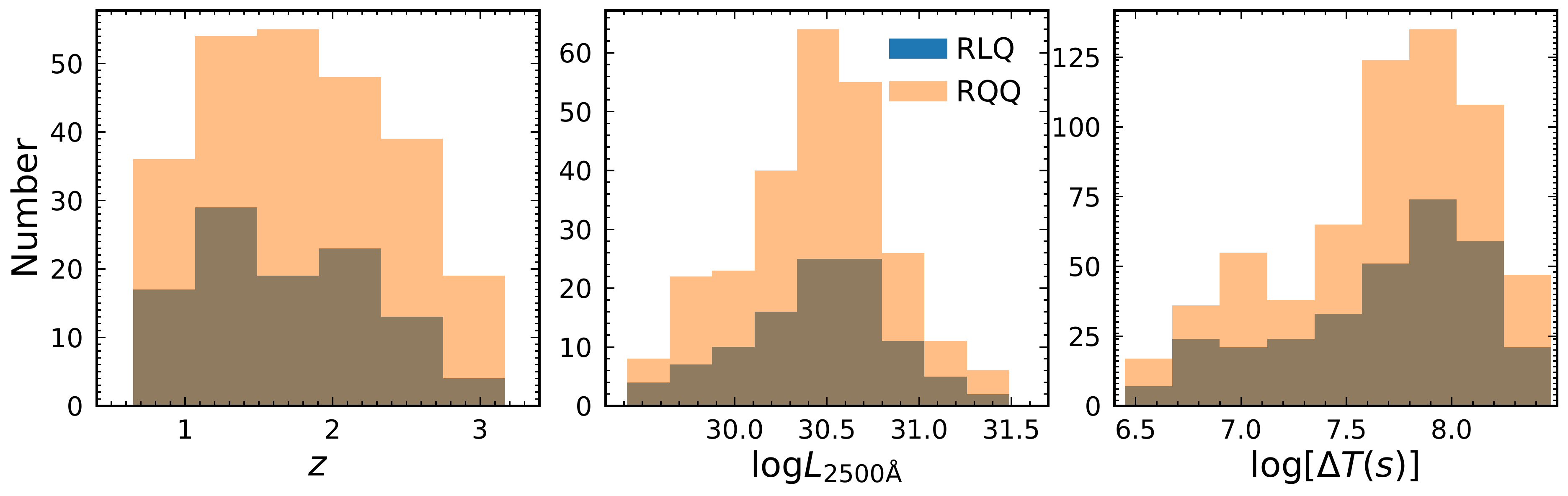}
\caption{Comparisons of the matched RQQ and RLQ distributions of redshift, UV luminosity, and timescale (left, center, and right panels).
    The X-ray light-curves of the quasars in both samples have been down-sampled using the method in \S~\ref{sec:downSample}.
    The RLQ and RQQ distributions in each parameter space are statistically consistent, according to
    AD tests of similarity for the three panels that all result in $p$-values $>0.16$.}
\label{fig:mRQQs}
\end{figure*}

\subsection{Comparison RQQ sample}
\label{sec:rqqSample}

As part of this investigation, we will analyze the X-ray variability properties of our RLQ sample 
and require a suitable matched sample of RQQs with which we can compare our results.
We select RQQs mainly from the sample of \citet{timlin2020},
who analyzed $\approx$~1600 serendipitous {\it Chandra} X-ray observations of
462 typical, non-BAL RQQs.
These spectroscopically-confirmed RQQs span a similar redshift range ($z=0.1$--4) to that of our RLQs (Fig.~\ref{fig:sample});
however, \citet{timlin2020} selected quasars with $m_i\le20.2$,
which is about one magnitude brighter than the RLQs in our sample ($m_i\lesssim21$).
We therefore extended the analysis of \citet{timlin2020} to fainter magnitude
($m_i=20.2$--20.8) quasars in the SDSS DR14Q (\citealt{paris2018})
following the selection method outlined in \S~2 of \citet{timlin2020}.
Note that all observations (including those from \citealt{timlin2020} and newly selected) are consistently analyzed in 
the same way as described above in \S~\ref{sec:dataReduction} and \S~\ref{sec:eflux}.
In total, we obtain 2341 serendipitous {\it Chandra} observations of 606 RQQs.
We list the RQQ sample and corresponding X-ray observations in the tables presented in Appendix~\ref{sec:appendix1}.

Furthermore, to compare appropriately the long-term X-ray variability properties of RLQs with those of RQQs,
we generated a sub-sample of RQQs that is matched in
redshift, optical/UV luminosity, and timescale space to the RLQ sample.
We describe the selection method in more detail in \S~\ref{sec:matchedRQQ}.
The redshift, optical/UV luminosity, and timescale distributions of the matched RQQs are compared with those of RLQs in Fig.~\ref{fig:mRQQs}.
Note that two-sample Anderson-Darling (AD; \citealt{ad1952}) tests are performed for all three panels of Fig.~\ref{fig:mRQQs}, 
and the distributions of RLQs and RQQs are always statistically consistent.
We also selected a RQQ sample that matches with the RLQs used in \S~\ref{sec:spec}
as a reference for RQQ coronal \mbox{X-ray} spectral properties.
To generate this sample, we selected the nearest RQQ in the $L_\mathrm{2500\angstrom}$- $z$ space 
(without replacement) to each RLQ.
An AD test indicates that the $L_\mathrm{2500\angstrom}$ and $z$ 
distributions of this RQQ sample are statistically consistent with those of the RLQs used in \S~\ref{sec:spec}.

\begin{figure}
\centering
\includegraphics[width=0.48\textwidth, clip]{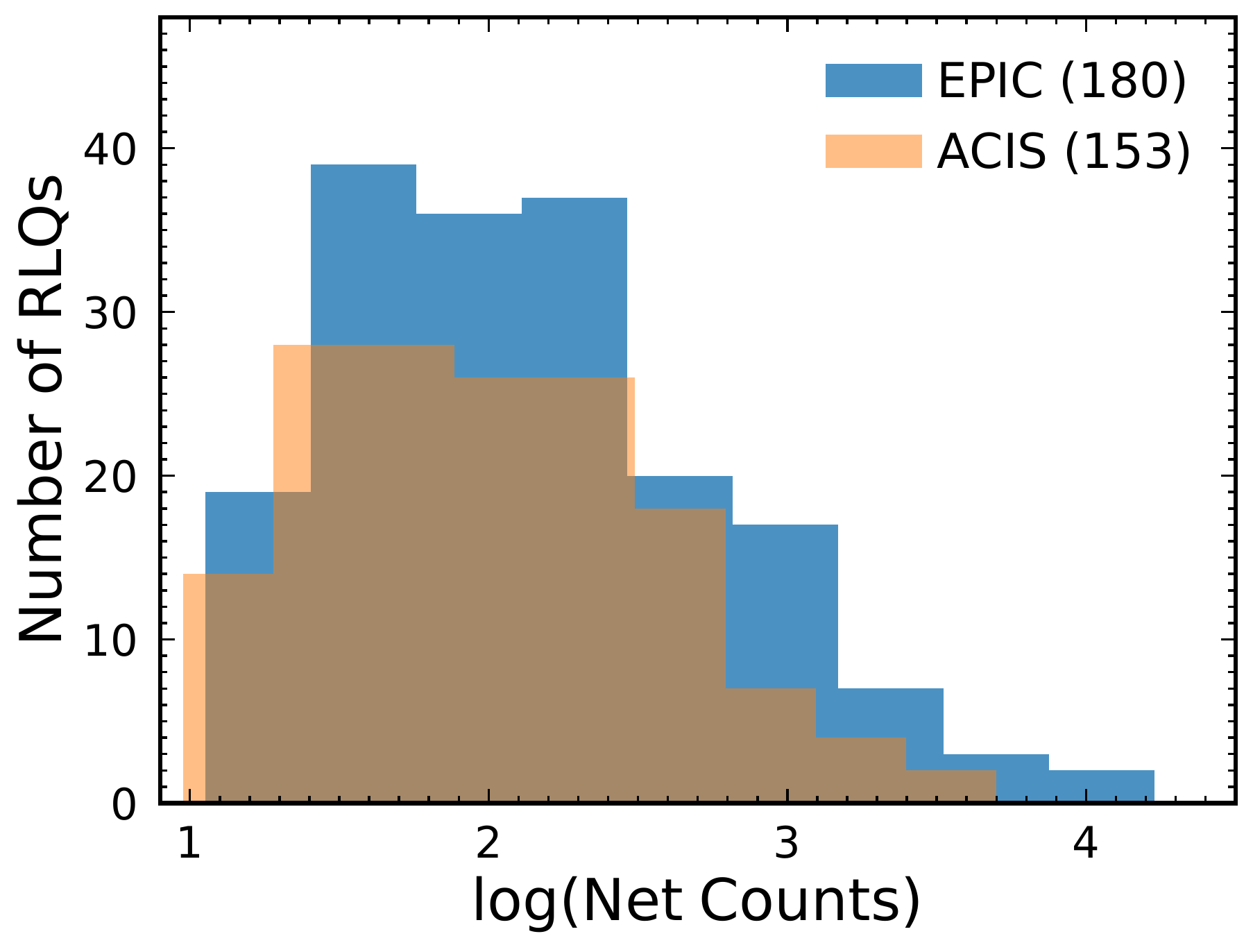}
\includegraphics[width=0.48\textwidth, clip]{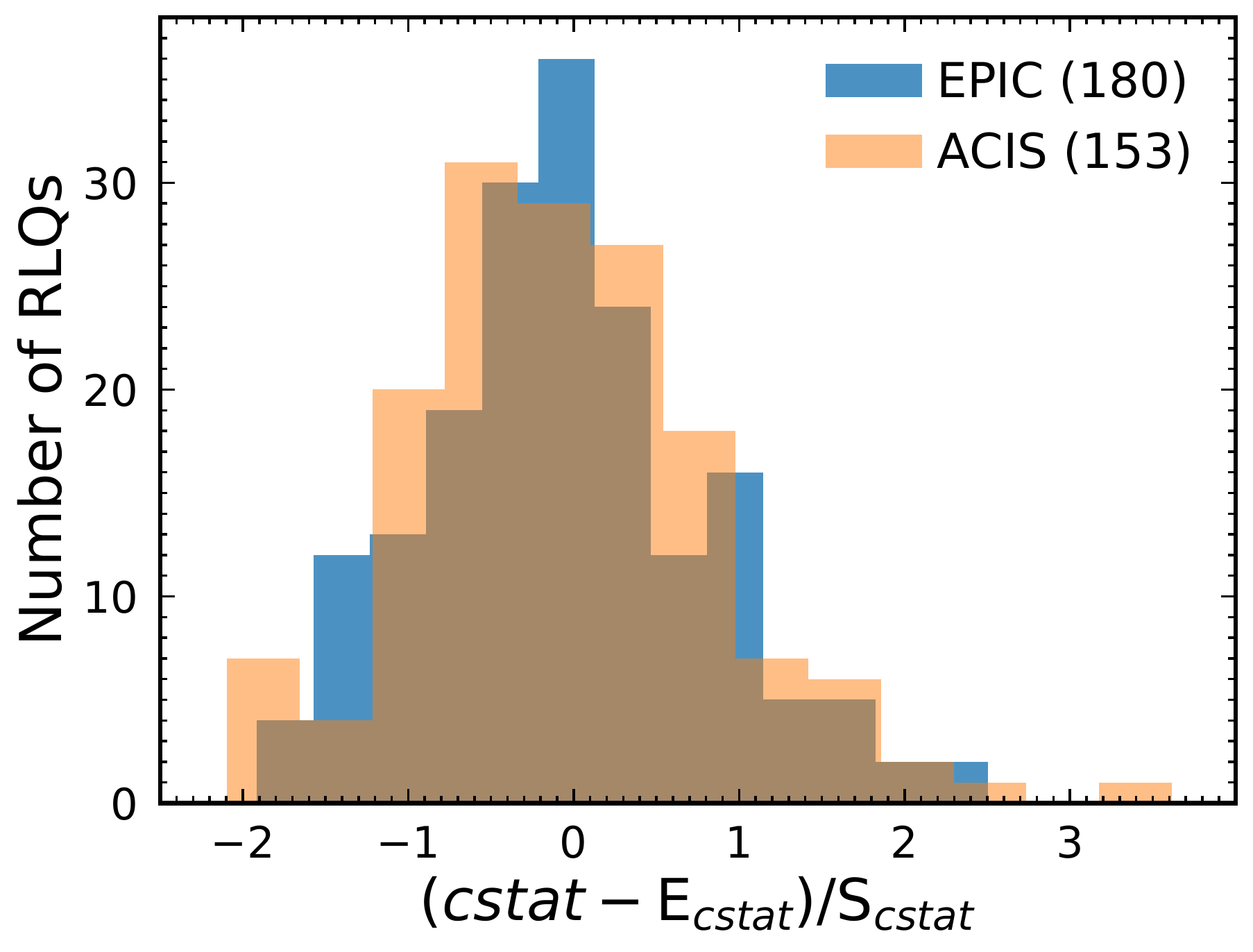}
\includegraphics[width=0.48\textwidth, clip]{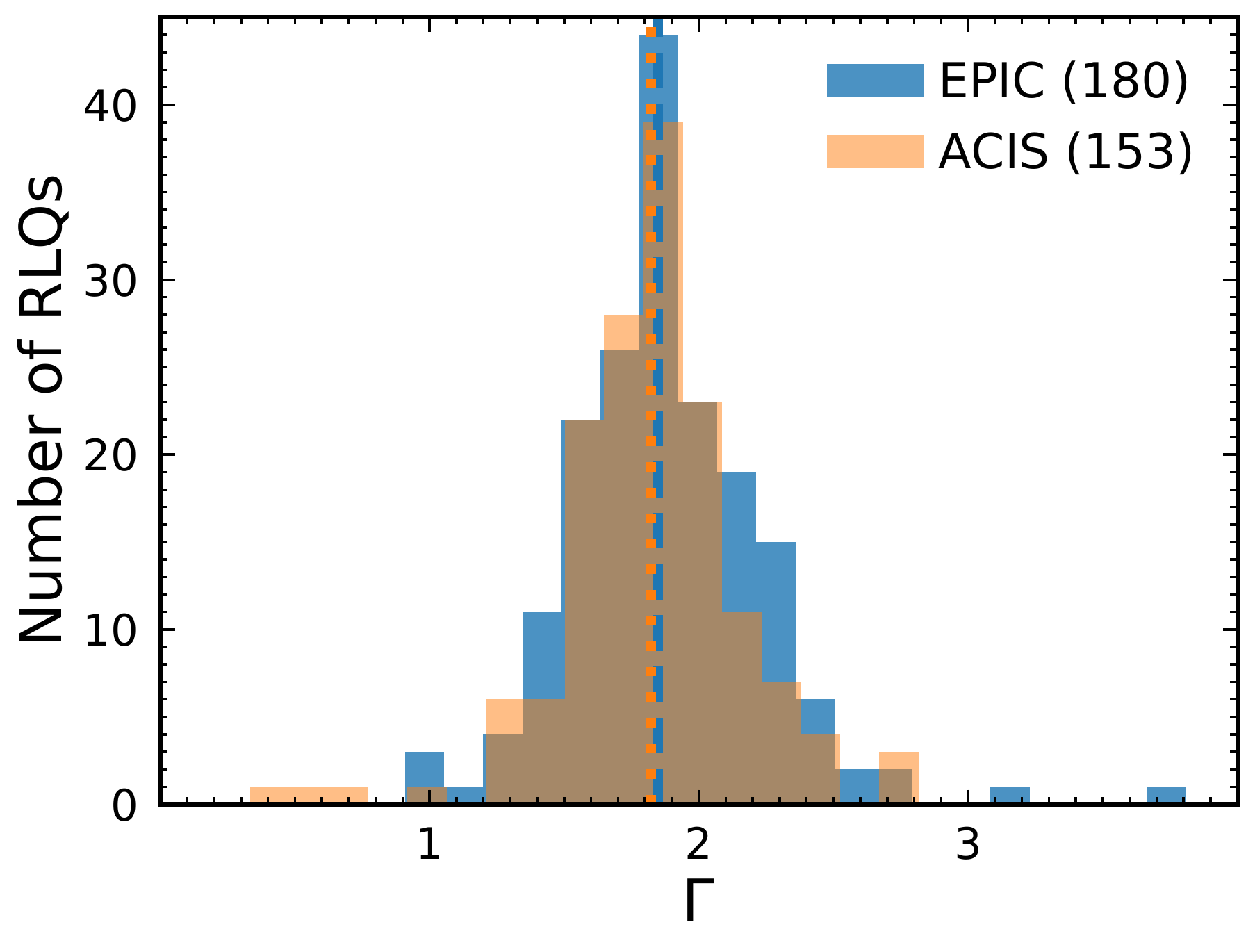}
\caption{Top: The distribution of net counts (0.5--7 keV) of for the 333 RLQ X-ray spectra with SNR $\ge3$, 
    where (and in panels below) we distinguish the 153 {\it Chandra}/ACIS spectra from 
    the remaining 180 {\it XMM-Newton}/EPIC spectra.
    Middle: The distribution of the quality of the spectral fits, $(cstat-\mathrm{E}_{cstat})/\mathrm{S}_{cstat}$. 
    See the main text in \S~\ref{sec:spec}.
    Bottom: The distribution of $\Gamma$, where the median values are $1.82_{-0.01}^{+0.03}$ (dotted line)
    and $1.85_{-0.01}^{+0.02}$ (dashed line) for the ACIS and EPIC spectra, respectively.}
\label{fig:nethist}
\end{figure}

\section{X-ray Spectral Properties}
\label{sec:spec}
We investigate the shape of the primary X-ray continuum in \S~\ref{sec:specFit} 
by fitting the X-ray spectra of individual quasars.
The properties of X-ray reflection features, namely,
the fluorescent iron line and the Compton hump, 
are analyzed in \S~\ref{sec:specStack}, utilizing \mbox{X-ray} spectral stacking and joint spectral fitting.
\subsection{Fitting the X-ray continuum using a power-law model}
\label{sec:specFit}
\subsubsection{Fitting methods}
\label{sec:fit_method}
We perform X-ray spectral fitting using {\sc sherpa},
focusing on the observed-frame 0.5--7 keV band (rest-frame $0.55$--7.7~keV for quasars at $z\sim0.1$ and 2.8--40~keV for quasars at $z\sim4.5$).
We use a simple power-law model to describe the shape of the quasar X-ray spectra,
which is then multiplied by a $z=0$ absorption component with a fixed $N_\mathrm{H}$ value
estimated using the {\sc colden} tool (\citealt{dickey1990}).
The median number of counts between 0.5--7~keV of our X-ray spectra is 111, and the interquartile (25th to 75th percentile) range is (42, 289).
Many X-ray spectra, therefore, do not have a sufficient number of counts to allow for further grouping, 
The c-stat statistic (denoted as $cstat$) is thus utilized,
which provides maximum-likelihood estimates of model parameters
without requiring a minimum number of counts in each bin/channel (\citealt{cash1979, baker1984, humphrey2009}).

When employing the c-stat statistic, the background contribution cannot be directly 
subtracted from the source spectrum;
therefore, we fit the background spectra explicitly using empirical
models.\footnote{The w-stat (a variant of c-stat; \citealt{wachter1979})
does not require the background to be explicitly modeled;
however, to use w-stat, the background spectrum needs to be grouped such that 
each bin contains at least about 5 counts to avoid bias caused by zero/low-counts channels
(e.g. see Appendix~A of \citealt{willis2005}).
Our method is not subject to this binning restriction.}
The background models are derived from principal component analysis using a larger number of
background spectra of each instrument (i.e. ACIS, EPIC-pn, and EPIC-MOS; see Appendix~A of \citealt{simmonds2018}).
After fitting the background spectra, the shape and amplitude of the background
model is fixed,
and a background component is added to the source model with a scaling factor (i.e. $1/k$; see \S~\ref{sec:eflux}).
The source spectra are then fitted with the full model that includes a Galactic absorption-modified
power-law component (with two free parameters) and a fixed background component.

We assess the fit quality using the expectation and variance of the c-stat statistic 
of the best-fit model (\citealt{kaastra2017}).
We denote the expectation and standard deviation of c-stat with $\mathrm{E}_{cstat}$ and $\mathrm{S}_{cstat}$, respectively.
Ideally, the distribution of $(cstat-\mathrm{E}_{cstat})/\mathrm{S}_{cstat}$ is close to a standard normal distribution;
in cases where $(cstat-\mathrm{E}_{cstat})/\mathrm{S}_{cstat}\gg1$, the model is strongly disfavored.
The X-ray spectra of RQQs are analyzed using the same methodology.

\begin{figure}
\centering
\includegraphics[width=0.48\textwidth, clip]{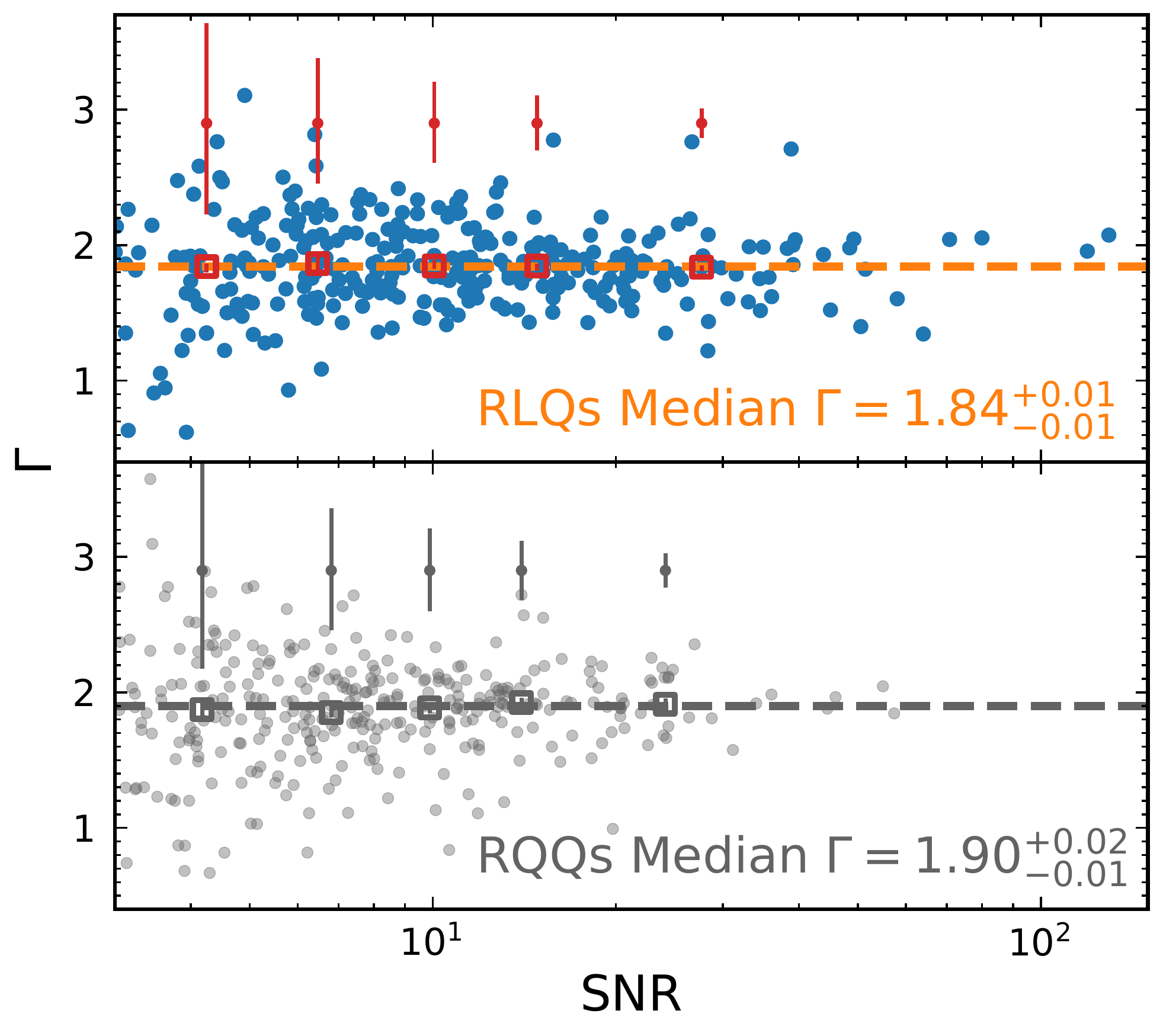}
\caption{The best-fit power-law photon index ($\Gamma$) as a function of the SNR (see Table~\ref{tab:obs}) of the spectrum 
for RLQs (top) and comparison RQQs (bottom). 
Both samples contain 333 quasars, and they are matched in $L_\mathrm{2500\angstrom}$ and $z$.
This figure demonstrates that the photon index does not depend on the quality of the data.
The median photon index of all quasars in each panel is provided at the lower right corner.
The RLQs are grouped into 5 SNR bins of nearly equal size (65--67 quasars per bin),
and the median $\Gamma$ in each bin is calculated and shown as open squares.
The uncertainties of the median $\Gamma$ of each bin are estimated using the bootstrapping method, 
    and these uncertainties are relatively small compared with the size of the square symbols.
Therefore, we additionally show the median uncertainties for individual sources in each bin above the squares.
The binning scheme for RQQs follows that for RLQs.
The X-ray spectra of RLQs are only flatter than those of RQQs by a small amount according to our serendipitous data.}
\label{fig:gamma_snr}
\end{figure}

\begin{figure*}
\centering
\includegraphics[width=0.95\textwidth, clip]{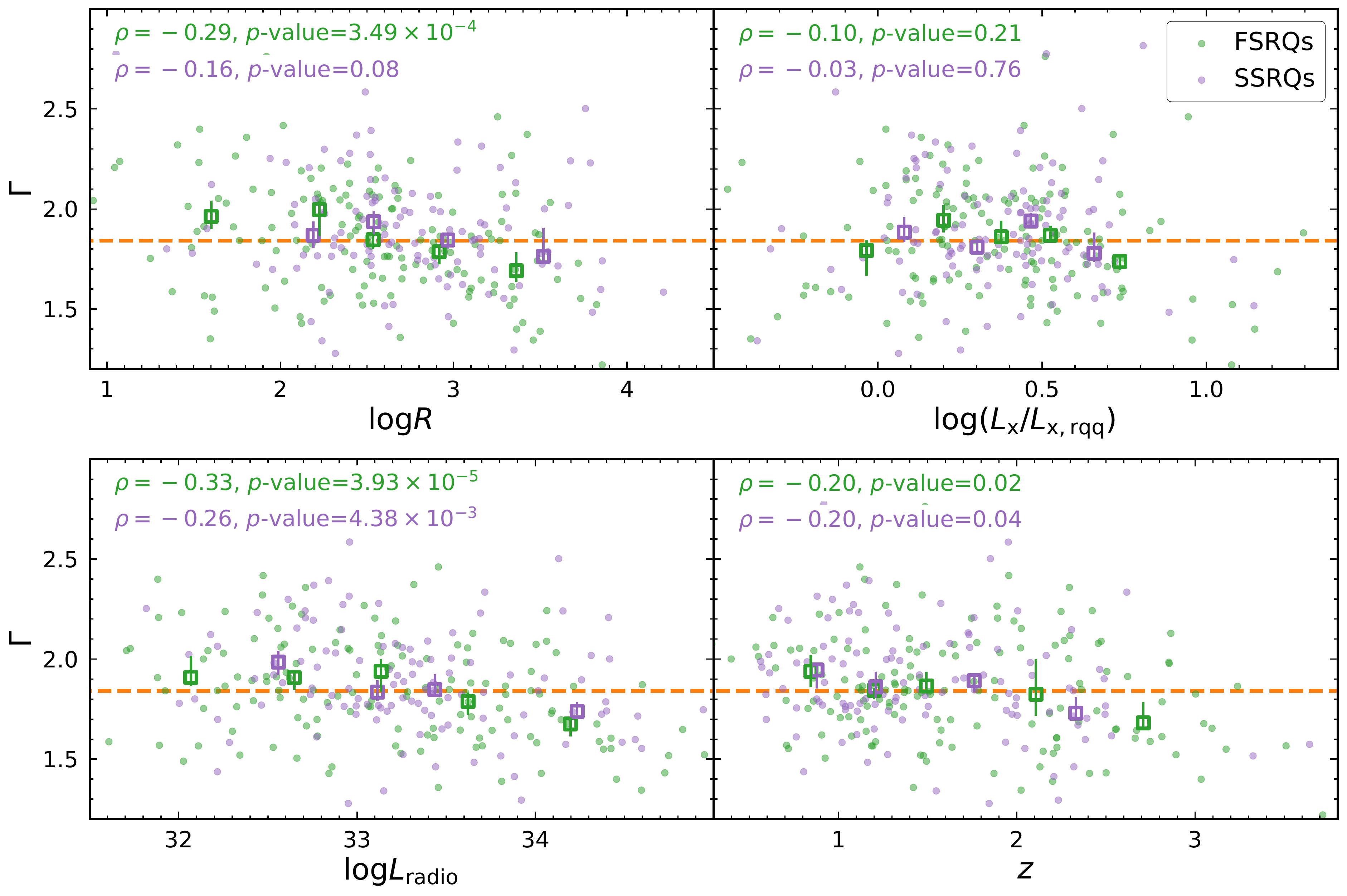}
\caption{
The correlations of photon index ($\Gamma$) with $\log R$, $L_\mathrm{x}/L_\mathrm{x,rqq}$, $L_\mathrm{radio}$, and $z$ for FSRQs (green symbols, 150 quasars) and SSRQs (purple symbols, 114 quasars), separately.
Spectra with $\mathrm{SNR}\le5$ are omitted.
The Spearman rank correlation statistic $\rho$ and the corresponding $p$-value are given at the upper-left corner of each panel.  
The median $\Gamma=1.84$ of the RLQ sample is shown in all four panels using orange dashed lines.
In each panel, FSRQs and SSRQs are grouped into 5 and 4 bins of comparable size, respectively, according to their $x$-axis values.
The median $\Gamma$ of each bin is calculated (open squares), the error bar of which is estimated using bootstrapping.
There are no strong correlations ($p$-value $<10^{-3}$) between $\Gamma$ and $R$, $L_\mathrm{x}/L_\mathrm{x,rqq}$, $L_\mathrm{radio}$, and $z$ for FSRQs and SSRQs, 
except for the anti-correlation between $\Gamma$ and $L_\mathrm{radio}$ for FSRQs (lower left).
A less significant anti-correlation between $\Gamma$ and $R$ for FSRQs (upper left)
is probably driven by the $\Gamma$-$L_\mathrm{radio}$ anti-correlation for these FSRQs.
However, if the 14 FSRQs with $\log L_\mathrm{radio}>34.3$ are removed, both anti-correlations disappear ($p$-value~$>0.05$),
    suggesting that these FSRQs are the rare population of RLQs with a strong jet X-ray component and a flat photon index \mbox{$\Gamma\sim$ 1.5--1.6}.
The lack of a correlation between $\Gamma$ and $\log (L_\mathrm{x}/L_\mathrm{x,rqq})$ 
argues against an important role of jet X-ray emission for general RLQs (e.g. \citealt{zhu2020}).
The photon index of the coronal X-ray component does not depend on 
the quasar properties under consideration here.}
\label{fig:gamma_relation}
\end{figure*}

\begin{figure}
\centering
\includegraphics[width=0.43\textwidth, clip]{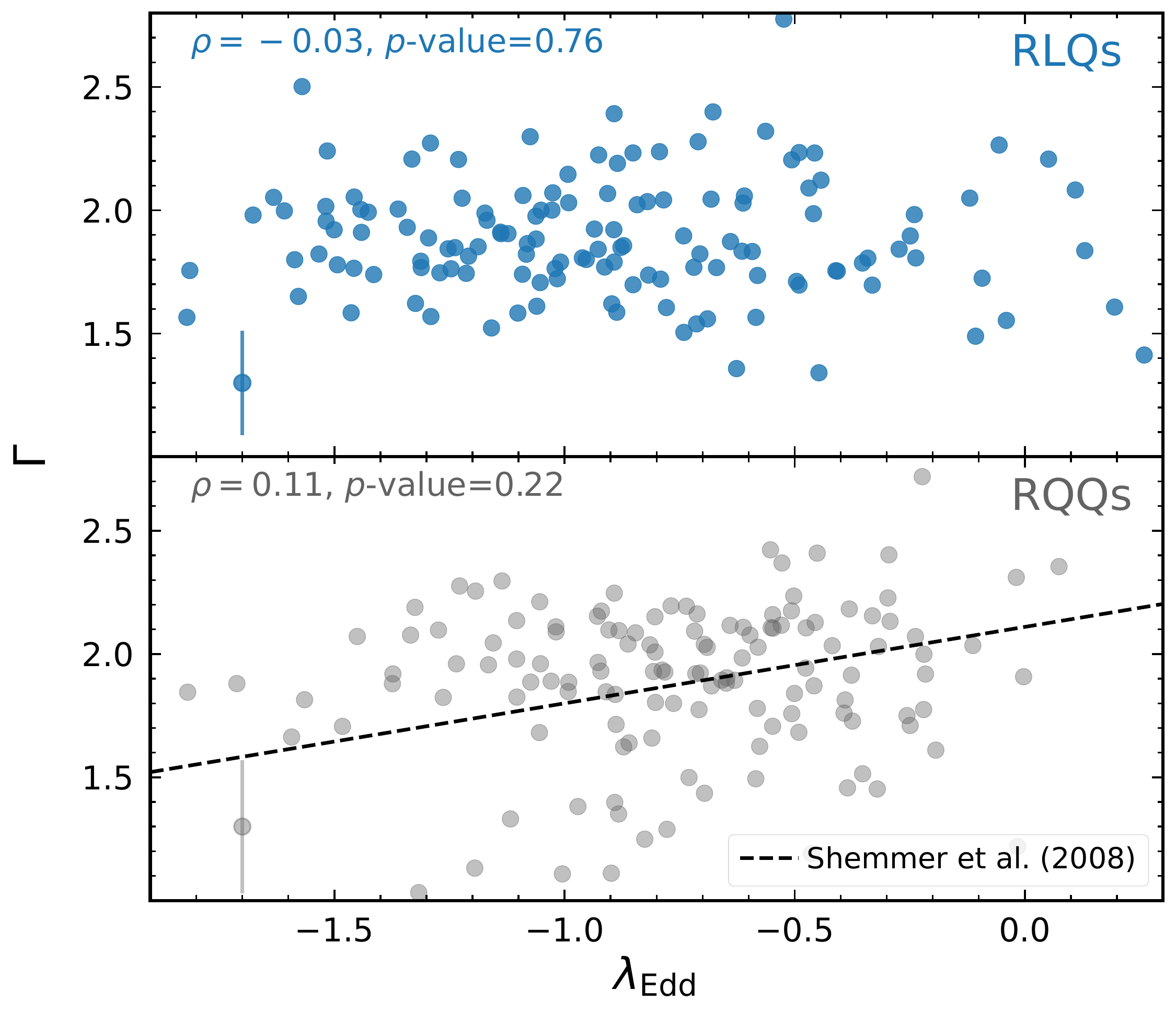}
\caption{Top: The correlation between X-ray power-law photon index and Eddington ratio of RLQs.
The median error bar of $\Gamma$ is shown in the lower-left corner.
The Spearman rank correlation coefficient and associated $p$-value are shown in the upper-left corner.
    Bottom: Same as the top panel for comparison RQQs. 
    The $\Gamma$-$\lambda_\mathrm{Edd}$ relation from \citet{shemmer2008} is also shown for comparison.
No strong correlation between $\Gamma$ and $\lambda_{\mathrm{Edd}}$ is found for either RLQs or RQQs.}
\label{fig:Edd_gamma}
\end{figure}

\begin{figure}
\centering
\includegraphics[width=0.45\textwidth, clip]{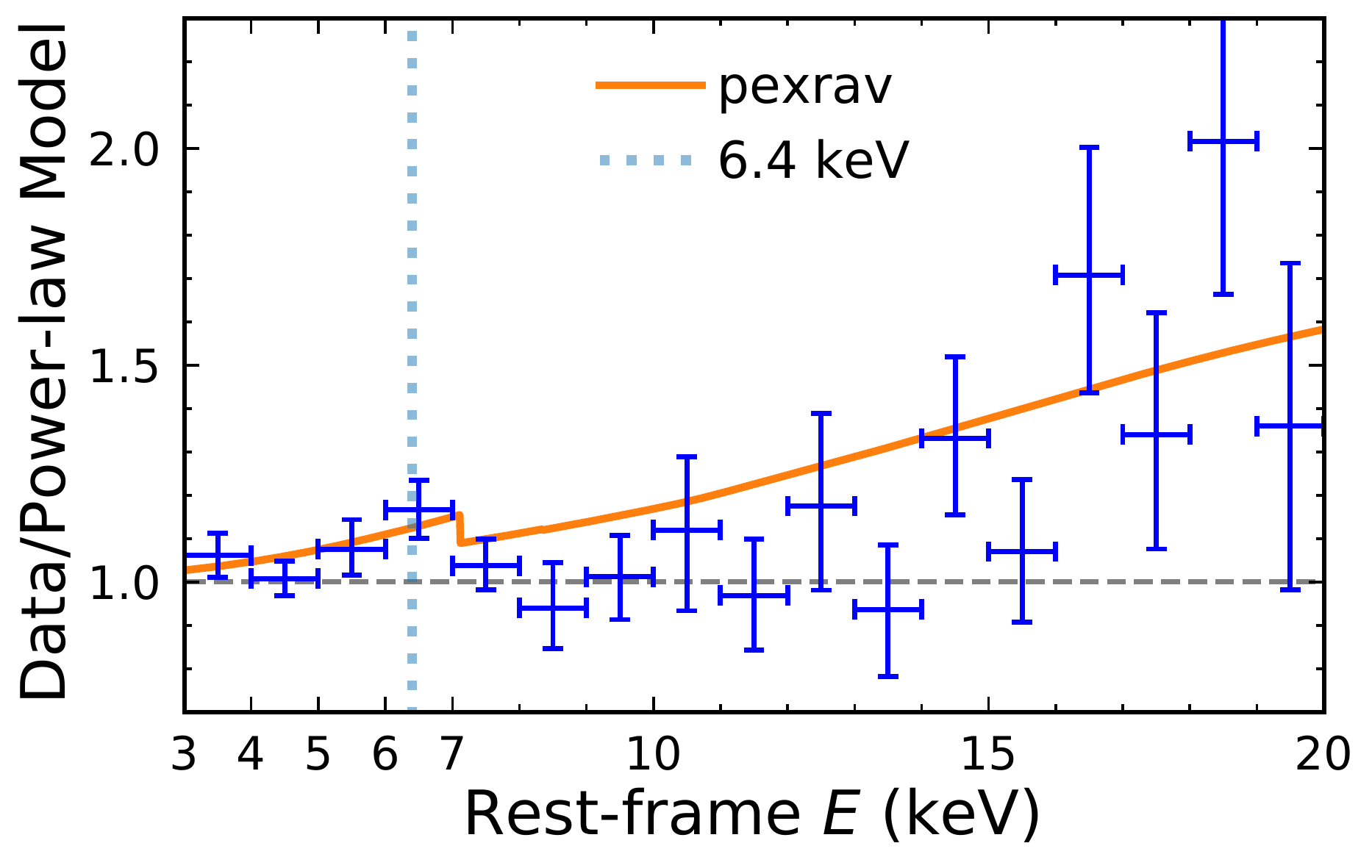}
\includegraphics[width=0.45\textwidth, clip]{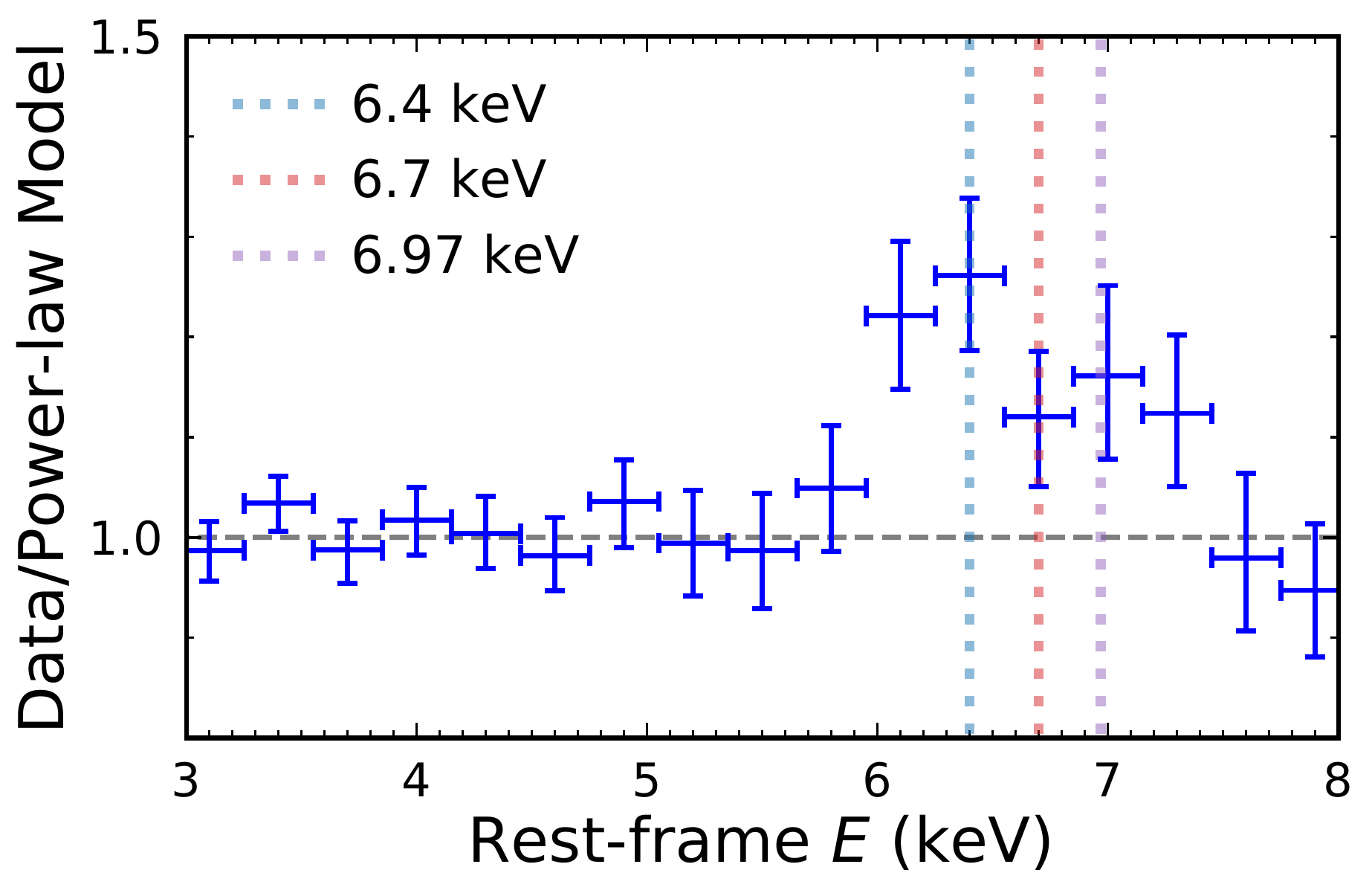}
\includegraphics[width=0.48\textwidth, clip]{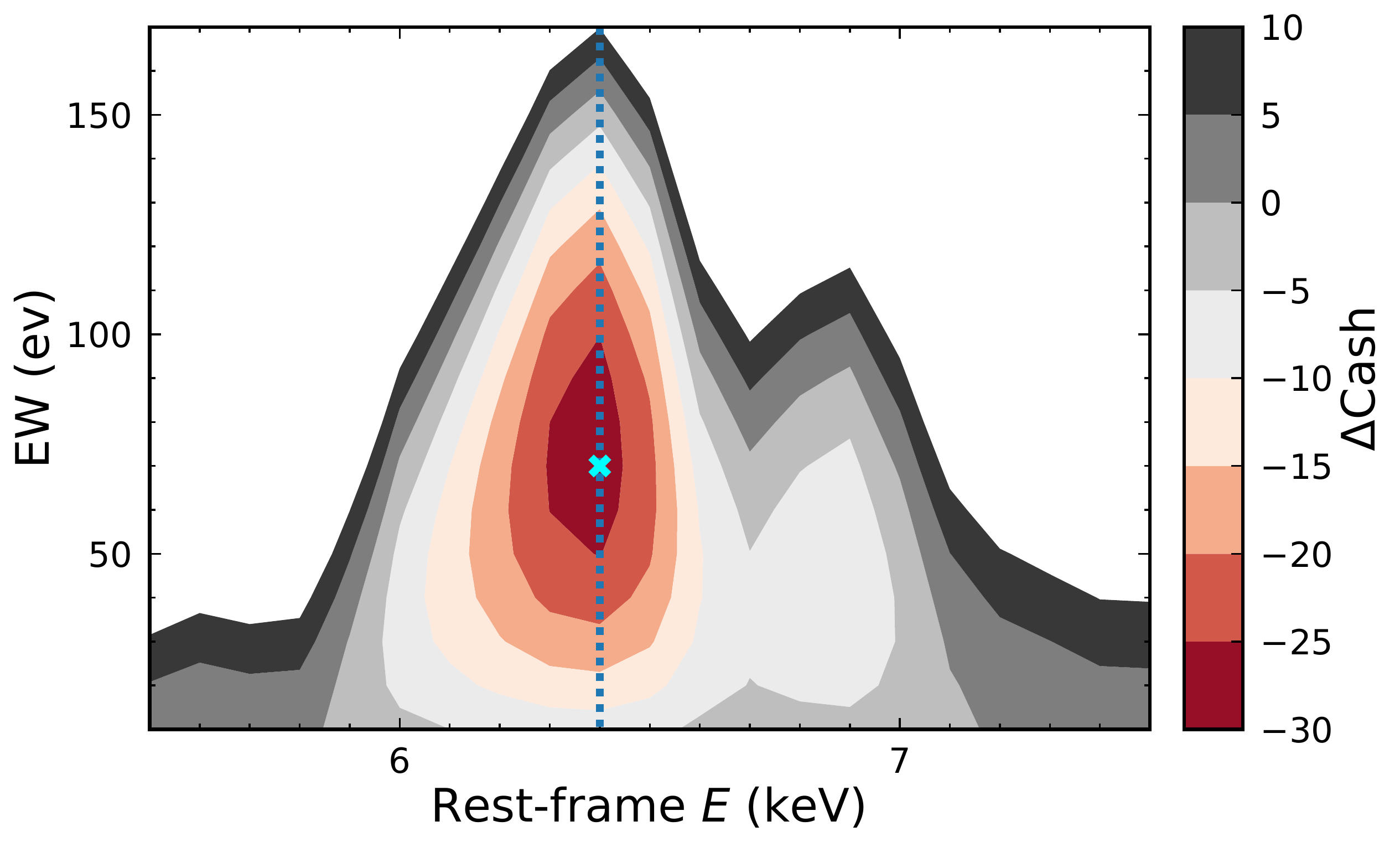}
\caption{Top: The stacked ratio of the background-subtracted X-ray spectrum
    to the best-fit power-law model (with fixed $\Gamma=1.84$).
We stacked the rest-frame 3--20 keV band using a bin size of 1 keV 
    for the 58 RLQs with $z>2$ in our sample.
The orange curve represents the prediction of Compton reflection 
    from neutral material (i.e. pexrav, \citealt{magdziarz1995}).
Middle: The stacked data-to-model ratio in the rest-frame 3--8 keV band for 216 
    RLQs ($0.27<z<3.9$).
Iron-line emission is clearly present in this residual spectrum.
    Bottom: The contour of the improvement of the fitting statistic ($\Delta$Cash) 
    when a narrow emission line (a Gaussian profile with rest-frame FWHM $=20$ eV)
    is added to the power-law model (that is used in the top and middle panels). 
    The cyan cross marks the minimum of the fitting statistic, where $\Delta$Cash$=-27.9$ indicates
a 5.3$\sigma$ detection for a narrow iron line at 6.4~keV.
The EW of the narrow iron line is $70_{-20}^{+30}$ eV.}
\label{fig:reflection}
\end{figure}

\subsubsection{The X-ray power-law photon indices of RLQs}
\label{sec:specResults}

In this section, we present fitting results for the X-ray spectra of 333 RLQs
with SNR (i.e. signal-to-noise ratio) $\ge3$ (16/349 RLQs fail this cut), 
where SNR $=3$ is approximately equivalent to a minimum number of 9 net counts between 
0.5--7 keV.\footnote{\citet{zhou2021} selected 664 RLQs that are in the 3XMM footprint. 
However, only the 160 RLQs with $>200$ X-ray counts were utilized to create their composite X-ray spectrum, which results in an X-ray completeness of only 24\%.}
If a quasar has multiple observations, we utilize the X-ray spectrum from the observation with the smallest $f_\mathrm{det}$ and ignore the rest.
The distribution of net counts for our RLQs 
is presented in the top panel of Fig.~\ref{fig:nethist}.
In the middle panel of Fig.~\ref{fig:nethist}, 
we show the distribution of $(cstat-\mathrm{E}_{cstat})/\mathrm{S}_{cstat}$.
There are 7 objects with $(cstat-\mathrm{E}_{cstat})/\mathrm{S}_{cstat}\ge2$, 
which is consistent with the number expected from a standard normal distribution at $\ge2\sigma$, 
i.e., $2.28\%\times333\approx7.6$.
Therefore, our spectral-fitting results are generally acceptable,
and the power-law model is a good phenomenological description 
of the rest-frame $\approx0.5$--40~keV spectra of our RLQs.
We show the distributions of $\Gamma$ of RLQs in Fig.~\ref{fig:nethist} (bottom).
The median $\Gamma$ for {\it Chandra}/ACIS and {\it XMM-Newton}/EPIC spectra
are $1.82_{-0.01}^{+0.03}$ and $1.85_{-0.01}^{+0.02}$, respectively,
where the uncertainties are estimated using bootstrapping.
Therefore, any cross-calibration uncertainties between the instruments do not affect our results below.

We show the dependence of the resulting photon index $\Gamma$ on the SNR in Fig.~\ref{fig:gamma_snr} (top),
where the orange dashed line represents the median $\Gamma$.
The data points are also grouped into five equally-sized SNR bins, each of which contains $\approx66$ spectra.
The median $\Gamma$ values (as well as uncertainties derived using bootstrapping) of each bin are represented by red open squares
with error bars in Fig.~\ref{fig:gamma_snr} (top).
Since the error bars of each median $\Gamma$ are small compared to the symbol size,
we also depict the median $\Gamma$ uncertainty for individual RLQs in each bin 
as the red error bars above the red squares.
Fig.~\ref{fig:gamma_snr} shows no clear correlation between $\Gamma$ and SNR; 
thus, our methodology of \mbox{X-ray} spectral fitting is not sensitive to the data quality of the spectra.
We repeat the procedure above for RQQs and show the results in Fig.~\ref{fig:gamma_snr} (bottom).

The median photon indexes of RLQs and matched RQQs\footnote{The (16, 50, 84)th percentiles of redshift are (1.01, 1.51, 2.35) and (1.02, 1.62, 2.39) for RLQs and RQQs, respectively.
Therefore, the spectra of both samples mainly probe the rest-frame $\approx1$--24 keV band.}
are $1.84_{-0.01}^{+0.01}$ 
and $1.90_{-0.01}^{+0.02}$, respectively,
the difference between which is small
compared with the intrinsic spread of their $\Gamma$ distributions (0.19 and 0.25).\footnote{The method of \citet{macccaro1988} was used to remove the amount of scatter caused by measurement uncertainties of the $\Gamma$ values.}
The flat-spectrum and steep-spectrum RLQ subsets have median photon indexes of $1.84_{-0.02}^{+0.02}$ and $1.85_{-0.02}^{+0.04}$, respectively, the difference between which is small as well.
However, in past work, the photon indexes of RLQs ($\Gamma\approx1.5$--1.7) were often found to be 
significantly flatter than those of RQQs ($\Gamma\approx1.7$--2.3; \citealt{wilkes1987, reeves1997, page2005}),
though this notable spectral difference was not always confirmed (e.g. \citealt{lawson1992, sambruna1999, grandi2006}).
\citet{sambruna1999} found that the evidence is weak that broad-line radio galaxies (BLRGs) have
flatter \mbox{X-ray} spectra than those of radio-quiet Seyfert~1 galaxies.
Furthermore, previous studies that utilized RLQ samples containing SSRQs and treated them separately as a group
(i.e. \citealt{lawson1992, lawson1997, grandi2006}) generally noticed a similarity
between the photon indexes of SSRQs and those of RQQs.
However, these studies had only a small number of SSRQs in their samples,
which prevented solid conclusions from being drawn.

\citet{zhu2020} found that the nuclear X-ray emission of RLQs is generally dominated by the corona,
and the coronal X-ray luminosity depends on optical/UV luminosity and radio-loudness parameter,
which indicate a disk-corona interplay and corona-jet connection, respectively.
In Fig.~\ref{fig:gamma_relation}, we investigate the correlations of
the photon index of RLQs with other quasar properties. 
In particular, we are interested in testing if the corona-jet connection could affect,
in addition to the X-ray brightness of the corona, the shape of the coronal X-ray spectrum.
Specifically, correlations of the $\Gamma$ of coronal X-rays with $R$ and $L_\mathrm{x}/L_\mathrm{x,rqq}$
would provide insights into more details of the corona-jet connection.
Even though \citet{zhu2020} did not find a distinct jet \mbox{X-ray} component in SSRQs,
the X-ray emission of a small portion ($\lesssim10$\%) of FSRQs likely
contains a significant ($>30$\%) jet X-ray component.
We therefore treat FSRQs and SSRQs separately in Fig.~\ref{fig:gamma_relation},
which is enabled by our large sample size.
Furthermore, since statistical correlation analyses generally do not condsider the effects of
measurement errors, we apply a cut of SNR $>5$ to the spectra to reduce the scatter caused by measurement errors,
which results in 150 and 114 spectra of FSRQs and SSRQs, respectively, which are shown in Fig.~\ref{fig:gamma_relation}.

We show the correlations of $\Gamma$ with $R$, $L_\mathrm{x}/L_\mathrm{x,rqq}$, $L_\mathrm{radio}$,
and $z$ for FSRQs and SSRQs in Fig.~\ref{fig:gamma_relation}.
We perform Spearman rank-correlation tests and show the results in the upper-left corner of
each panel.
In addition, we group FSRQs and SSRQs into five and four bins of comparable size, respectively;
the median $\Gamma$ is calculated for each bin and shown as large open squares in Fig.~\ref{fig:gamma_relation},
the errors of which are estimated using bootstrapping.
Neither the $p$-values resulting from correlation tests nor the median photon indexes support
correlations (more significant than $p$-value $=10^{-3}$) between $\Gamma$
and $R$, $L_\mathrm{x}/L_\mathrm{x, rqq}$, $L_\mathrm{radio}$, and $z$ for SSRQs.
No strong correlation is found for FSRQs either, 
except for the anti-correlation between $\Gamma$ and $L_\mathrm{radio}$ and a less-significant anti-correlation between $\Gamma$ and $R$; 
the latter correlation is probably driven by the former.
There are 14 $\log L_\mathrm{radio}>34.3$ FSRQs in the lower-left panel of Fig.~\ref{fig:gamma_relation}, 
almost all of which have $\Gamma$ that is smaller than the median photon index (1.84) of all RLQs.
Instead, the photon indices of these most radio-luminous FSRQs are around 1.5--1.6.
If these 14 FSRQs are removed in the correlation tests, 
the tentative $\Gamma$-$L_\mathrm{radio}$ and $\Gamma$-$R$ correlations disappear
among the remaining 136 FSRQs.
Therefore, these FSRQs likely represent the rare population that has a strong jet X-ray component (\citealt{zhu2020}).

Finally, we matched the utilized RLQs and RQQs with the quasar-property catalog of \citet{shen2011}. 
There are 138 and 136 black-hole mass measurements (based on Mg~{\sc ii} or H$\beta$) for RLQs and RQQs, respectively.
The median Eddington ratio of the RLQs ($-0.93_{-0.06}^{+0.04}$) is about $0.2$ dex smaller than that of the RQQs ($-0.73_{-0.06}^{+0.03}$), 
which might explain the mild difference between the median $\Gamma$ values of RLQs and RQQs 
if both groups follow the previously known correlation between $\Gamma$ and Eddington ratio (e.g. \citealt{shemmer2008, brightman2013}).
However, a correlation between Eddington ratio and $\Gamma$ is not found for 
RLQs (see the top panel of Fig.~\ref{fig:Edd_gamma}), and
a correlation test using Spearman's $\rho$ results in $p$-value=0.76.
A similar correlation test for the RQQ sample results in a negative result as 
well (see the bottom panel of Fig.~\ref{fig:Edd_gamma}), 
suggesting that our data quality might play an important role in the negative result for RLQs.
Future RLQ samples with high-quality X-ray and infrared spectroscopy
can address better the correlations between $\Gamma$ and 
Eddington ratio and FWHM of H$\beta$ (e.g. \citealt{brandt1997, laor1997}).

\subsection{Iron-line emission and the Compton-reflection hump}
\label{sec:specStack}
The distribution of the fitting statistic in \S~\ref{sec:specResults} (i.e. the middle panel of Fig.~\ref{fig:nethist}) 
indicates that the power-law function is, globally, 
an acceptable description of the X-ray spectra of our individual RLQs;
other features that arise in a narrow rest-frame X-ray band (e.g. fluorescent iron line emission)
cannot be detected in individual spectra, given their limited numbers of counts.
We therefore utilize a stacking technique to constrain the average properties
of the fluorescent iron line and Compton reflection for our RLQs.
We only consider RLQs with a spectroscopic redshift (83\% of our sample) and an X-ray spectrum with SNR~$>5$.
We exclude 14 FSRQs with $\log L_\mathrm{radio}>34.3$ 
as these are potentially beamed objects (see \S~\ref{sec:specResults} and Fig.~\ref{fig:gamma_relation}) 
and 2 other RLQs with $>10^4$ net counts since we do not want individual objects 
to dominate the spectral stacking,\footnote{\label{fn:cnts} The $\Delta$Cash map utilized below is weighted by the number of counts in each spectrum.
After removing these two objects, 
the RLQ with the largest number of net counts contributes $\approx5$\% to the total net counts.
The data-to-model ratio is not weighted by the number of counts of single objects in stacking.}
which results in 216 RLQ spectra for the analyses below.
We first fit the individual \mbox{X-ray} spectra of RLQs using a power-law
model with fixed $\Gamma=1.84$ (cf. \S~\ref{sec:specResults}) excluding channels that correspond to the rest-frame $<2$~keV, \mbox{5--8~keV}, and $>10$~keV bands, 
where additional emission features might exist.\footnote{If $\Gamma$ is allowed to be freely varying, we obtain a median $\Gamma=1.87_{-0.04}^{+0.03}$, which is consistent with 1.84.  Therefore, the results in \S~\ref{sec:specResults} are not strongly biased given the strength of the iron-line emission and Compton reflection we found in this section.}
The single free parameter of the model, i.e., the normalization factor of the power-law function, 
is also fixed after fitting.
For each \mbox{X-ray} spectrum, the data-to-model ratio is calculated over a grid of 
pre-defined rest-frame energy bins.
Note that the background is subtracted from the total counts in each bin, 
and the model here refers to the Galactic absorbed power-law function.
Finally, the data-to-model ratios of different RLQs are averaged in each energy bin.

We show the stacked data-to-model ratio over the rest-frame \mbox{3--20}~keV band of 
58 RLQs at $z>2$ (with a total of $7.4\times10^3$ net counts) 
in the top panel of Fig.~\ref{fig:reflection},
where bins with a size of 1~keV are utilized. 
The error bars of each bin are estimated using bootstrapping.
The data points that deviate from unity 
(the grey dashed line in the top panel of Fig.~\ref{fig:reflection}) 
show the locations where excess spectral features are present in the stacked residual spectrum.
The vertical dotted line indicates the position (6.4~keV) of the fluorescent iron line
expected from X-rays illuminating neutral material.
The data between 6--7~keV signify that strong iron-line emission is present in the stacked spectrum.
At $\ge14$~keV, another strong feature is also apparent that might be associated with the Compton-reflection hump.
We thus utilize the pexrav model (\citealt{magdziarz1995}) of {\sc xspec} to assess
if the blue data points in Fig.~\ref{fig:reflection} (top) are consistent with
the prediction of Compton reflection for an X-ray source over a slab of neutral material.
We choose a set of typical parameters, $\Gamma=1.84, E_\mathrm{cut}=300$~keV, and $\cos i=0.5$, 
with other parameters being set at their defaults. 
The resulting prediction is shown as the orange curve in Fig.~\ref{fig:reflection} (top).
The blue data points at $E>14$~keV generally follow the Compton-reflection model.

The middle panel of Fig.~\ref{fig:reflection} shows the stacking results using all 216 RLQs over the rest-frame 2.95--8.05 keV band (with a total of $2.9\times10^4$ net counts)
with a bin size of 0.3 keV. We show the positions of rest-frame 6.4~keV, 6.7~keV, and 6.97~keV using vertical dotted lines.
From this plot, it is apparent that the iron-line emission is not localized in a single bin but is distributed from $<6$~keV up to $>7$~keV,
which cannot be explained by instrumental dispersion of a single narrow emission line. 
Possibly, both neutral and ionized material contributes to the iron-line production (e.g. \citealt{nandra1997}),
and Fig.~\ref{fig:reflection} (middle) shows the blending of several narrow lines.
Furthermore, since the excess emission also arises in the bin at $<6.4$ keV, 
a broad relativistic component is also likely present (e.g. \citealt{brenneman2009}).

To constrain the contribution from the narrow component of the neutral
iron line at 6.4 keV, we utilize a method that is able to obtain the optimal energy resolution.
We add a Gaussian function to the fixed power-law function to 
represent a narrow-line component for each quasar.
The FWHM of the Gaussian function is fixed to be 20 eV in the rest frame, which is much smaller than the instrumental dispersion.
The strength of the narrow emission line is parameterized using rest-frame EW.
We sequentially increase the location of the narrow emission line 
from rest-frame 5.5~keV to 7.5~keV, with a step size of 0.1~keV.
At each position, we iterate the EW of the line from 10~eV up 
to~200 eV, with a step size of 10 eV.
The corresponding Cash statistic is calculated over the $21\times20$ grid 
of line properties (i.e. location and strength).
The Cash statistic of the simple power-law model without the narrow emission line is 
then subtracted, resulting in a two-dimensional $\Delta$Cash map for each quasar.
We then summed the $\Delta$Cash maps of all quasars and 
show a contour plot of the results in Fig.~\ref{fig:reflection} (bottom).

The minimum value in the contour plot, 
which corresponds to the largest improvement to the power-law model, 
occurs when the contour is at $E=6.4$~keV and EW $=70$~eV.
The $\Delta$Cash value given these parameters is $\Delta$Cash $=-27.9$ and
indicates that the neutral iron line is detected at a 5.3$\sigma$ significance level.
The 90\% confidence uncertainty of the EW is ($-20,+30$)~eV,
which corresponds to a change in $\Delta$Cash of 2.71.
\citet{hu2019} measured a strength of EW $\approx50$~eV 
for a narrow emission line at 6.4~keV using the composite X-ray spectrum of 97 RL AGNs,
which is consistent with our measurement within uncertainties.
This method indicates that the strongest narrow component of the iron emission is 
located at $E=6.4$~keV; however, substantial emission remains in the stacked spectrum at 
rest-frame 6--7~keV.
Integrating the data-to-model ratio over 6--7~keV reveals a total iron-line emission EW of $189_{-44}^{+49}$~eV, 
leaving EW $\approx120$ eV to other iron species and the possible broad-line component.
We discuss the implications of the strong iron-line emission in \S~\ref{sec:discussion}.
Similar analyses for the matched RQQs result in EW $=100_{-30}^{+20}$~eV for
the narrow emission component at 6.4 keV.
Thus, the EWs for RLQs and RQQs appear statistically consistent.

\begin{figure*}
\centering
\includegraphics[width=0.90\textwidth, clip]{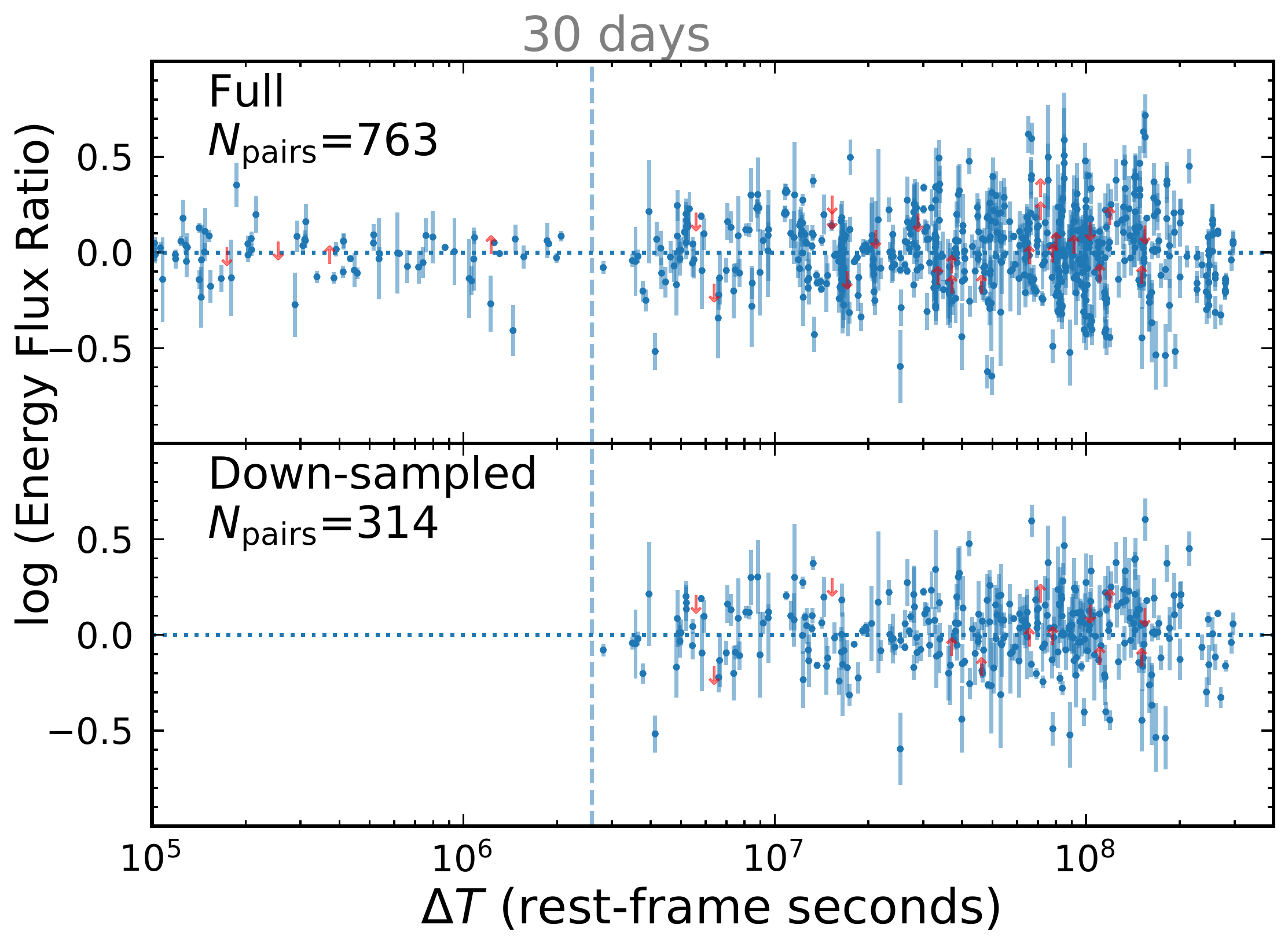}
\caption{Energy flux ratio as a function of the rest-frame timescale, where the red downward and upward arrows represent upper and lower limits, respectively.
The top panel shows all permutations of the full sample, while the bottom panel 
shows the permutations after removing repeated observations within 30 rest-frame days and
down-sampling the X-ray data so that the number of observations of any RLQ is less than or equal to seven.}
\label{fig:var_dT}
\end{figure*}

\begin{figure}
\centering
\includegraphics[width=0.48\textwidth, clip]{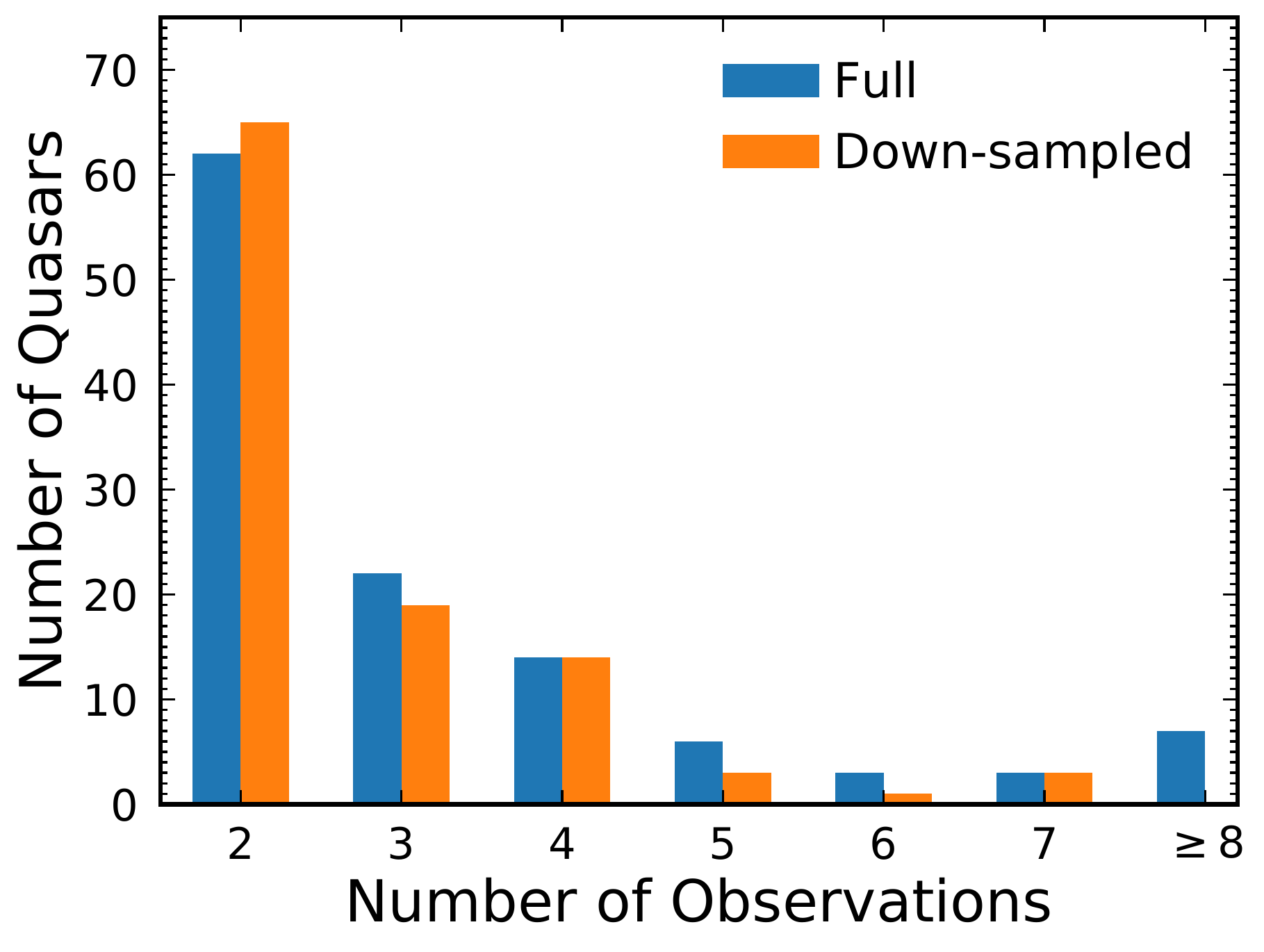}
\caption{The number of X-ray observations for each RLQ in our variability study.
The blue bars depict the 117 RLQs that have multiple {\it Chandra}/{\it XMM-Newton} observations with
sensitivity better than $f=3\times10^{-14}$ erg~cm$^{-2}$~s$^{-1}$.
The orange bars show the 105 RLQs remaining after removing repeated observations of the same RLQ
within 30 days (in the rest frame) and down-sampling the X-ray data 
    to reduce the maximum number of observations to seven per RLQ (see \S~\ref{sec:downSample}).}
\label{fig:Nobs}
\end{figure}

\section{Ensemble X-ray variability of RLQs}
\label{sec:var}

\subsection{Energy flux ratio and down-sampling method}
\label{sec:downSample}
There are 117 RLQs in our sample that have more than one \mbox{X-ray} observation, and thus we are able to investigate, with a large set of data, 
how the X-ray brightness of these RLQs varies over time (in the quasar rest frame).\footnote{The FSRQs with $\log L_\mathrm{radio}>34.3$ are excluded from consideration.}
For every pair of time-ordered data points in the X-ray light curve of a quasar in our sample, we calculate the ratio of the flux
of the later-epoch data over that of the earlier epoch. 
For each of these energy flux ratios, there is a corresponding timescale (denoted $\Delta T$),
which represents the time interval of the two observations corrected to the rest-frame of quasar.
The dependence of energy flux ratio on the timescale is shown in the top panel of Fig.~\ref{fig:var_dT}.

These 117 RLQs have 386 X-ray observations in total, 
where the number of observations of each quasar ranges from 2 to 20.
The distribution of the number of observations of each quasar is shown as the blue bar plot in Fig.~\ref{fig:Nobs}.
For a quasar with $n$ X-ray observations, $n(n-1)/2$ energy flux ratios can be computed using the method outlined above.
In Fig.~\ref{fig:Nobs}, more than half of the quasars (65/117) have only two X-ray observations, resulting in 65 energy flux ratios;
however, a single quasar with 20 X-ray observations produces 190 energy flux ratios.
Therefore, to prevent the distribution of energy flux ratio from being dominated by a small number of quasars with a relatively large number of observations, 
we elect to down-sample the X-ray light curves.
We first grouped any consecutive observations that are separated by <30 rest-frame
days,\footnote{The number of data points is too small
to constrain well variability at $\Delta T<30$ days in the top panel of Fig.~\ref{fig:var_dT}.}
and assigned each group an ID number.
Among the observations with the same group ID, 
we keep only the most-sensitive observation (smallest $f_\mathrm{det}$), 
which removes 84 of the 386 observations.
If the number of observations is still $>7$ after the grouping-and-selecting procedure above,
we randomly remove observations (excluding the first and last ones to keep the maximum light-curve baseline) until 
only 7 observations remain,\footnote{In this way, no single quasar can produce more than 10\% of the energy flux ratios.}
which removes 5 more observations.
Note that if we down-sample the data to a maximum number of observations of 3, 
the main results of \S~\ref{sec:var} are not affected.
The resulting distribution of the number of observations of each quasar is shown as the orange bars in Fig.~\ref{fig:Nobs},
and the down-sampled energy flux ratio vs. rest-frame timescale is shown in the bottom panel of Fig.~\ref{fig:var_dT}.
The down-sampled data are used in the following analyses.
After down-sampling the light-curves and restricting the timescale to $\Delta T\ge30$ rest-frame days,
297 observations of 105 RLQs remained, producing 314 energy flux ratios (see Table~\ref{tab:var}).

\subsection{X-ray variability of typical RLQs}
\begin{figure}
\centering
\includegraphics[width=0.48\textwidth, clip]{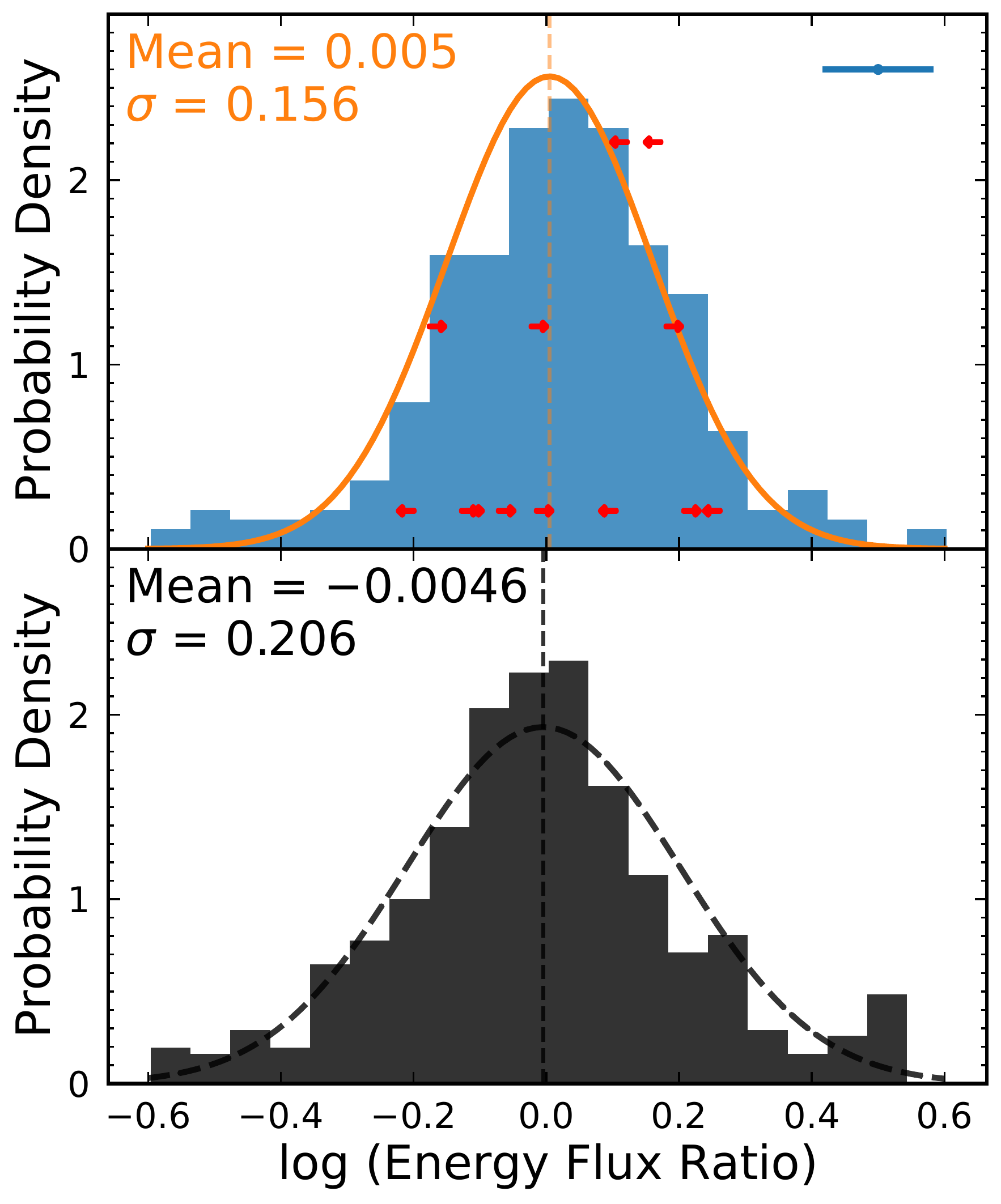}
\caption{The distribution of $\log$(energy flux ratio) after down-sampling for RLQs is shown in the top panel, 
where red arrows represent upper/lower limits.
The magnitude of the median measurement uncertainty is shown in the upper-right corner.
The orange curve depicts the deconvolved normal function that represents the intrinsic variability.
Note that the standard deviations of the distribution before and after deconvolution are 0.190 and 0.156, respectively.
Therefore, the intrinsic variability dominates the spread of $\log$~(energy flux ratio).
The mean and standard deviation of the intrinsic distribution are presented in the upper-left corner (orange).
In the bottom panel, the distribution in black represents the matched RQQ sample with its respective deconvolved variability distribution (black dashed curve).}
\label{fig:histMax7}
\end{figure}

We depict the distribution of the energy flux ratio in the top panel of Fig.~\ref{fig:histMax7} and 
the result of an AD normality test performed on the distribution of the energy fluxes
in the first row of Table~\ref{tab:statTest};
the distribution does not depart significantly from a normal distribution.
We calculated the kurtosis\footnote{The kurtosis largely indicates whether the tails of distribution contain excess data
(kurtosis > 0) compared to a normal distribution (kurtosis $=0$, e.g. \citealt{livesey2007, westfall2014, timlin2020}).} and estimated its uncertainties using bootstrapping.
The kurtosis of the energy flux distribution for the RLQs is consistent with that of a normal distribution as well.

To remove the broadening of the distribution of energy flux ratio that is
caused by measurement errors, 
and obtain the amplitude of the instrinsic X-ray variability,
we fit a normal function to the distribution in Fig.~\ref{fig:histMax7} (top) using 
a maximum-likelihood method that distinguishes the
measurement errors and intrinsic scatter (e.g. \citealt{macccaro1988}).
The likelihood function includes terms that account for the few upper and lower limits in the data (e.g. see Appendix~B of \citealt{zhu2019}).
The best-fit normal function that represents the intrinsic distribution of the energy flux ratio
is shown as the orange curve in Fig.~\ref{fig:histMax7} (top), where the model parameters are 
presented in the upper-left corner.

To assess the dependence of X-ray variability amplitude on quasar luminosity and timescale,
we divide the data at the median UV luminosity ($\log L_\mathrm{2500\angstrom}=30.45$) and median timescale ($\log \Delta T=7.79$).
We perform statistical tests on these sub-samples and report the results in Table~\ref{tab:statTest}.
The distributions of energy flux ratio are consistent with a normal distribution in both luminosity and timescale bins.
We also performed model fitting, as we did with the full sample, to the energy flux distributions of the luminosity and timescale bins
and list the results in Table~\ref{tab:list}.
There is strong evidence that the variability amplitude (i.e. $\sigma$) 
increases with rest-frame timescale (e.g. \citealt{timlin2020});
however, the dependence of variability amplitude on quasar luminosity is only marginal.

\begin{table*}
\centering
\caption{Statistical tests on distributions of energy flux ratio of RLQs.}
\label{tab:statTest}
\begin{threeparttable}[b]
\begin{tabularx}{\linewidth}{@{}Y@{}}
\begin{tabular}{lccc}
\hline
\hline
    Sample & $N_\mathrm{pairs}$ & Normality ($p$-value)\tnote{a} & Kurtosis\tnote{b} \\
\hline
    Down-sampled & 314 & 0.02 & $0.73_{-0.45}^{+0.38}$ \\
\hline
    High-luminosity ($\log L_\mathrm{2500\angstrom}\ge30.45$) & 184  & 0.08 & $0.66_{-0.43}^{+0.35}$ \\
    Low-luminosity ($\log L_\mathrm{2500\angstrom}< 30.45$) & 130 & 0.45 & $0.11_{-0.41}^{+0.31}$ \\
\hline
    Long-timescale ($\log \Delta T\ge 7.79$)  &  157 & 0.46 & $0.35_{-0.37}^{+0.29}$ \\
    Short-timescale ($\log \Delta T < 7.79$) & 157 & 0.41 & $0.27_{-0.47}^{+0.35}$\\
\hline
\end{tabular}
\end{tabularx}
\begin{tablenotes}
\item[a] The $p$-value resulting from the AD normality test.
\item[b] The kurtosis values do not strongly deviate from that of a normal distribution (kurtosis $=0$).
    The uncertainties of the kurtosis values are estimated using bootstrapping.
\end{tablenotes}
\end{threeparttable}
\end{table*}

\begin{figure}
\centering
\includegraphics[width=0.48\textwidth, clip]{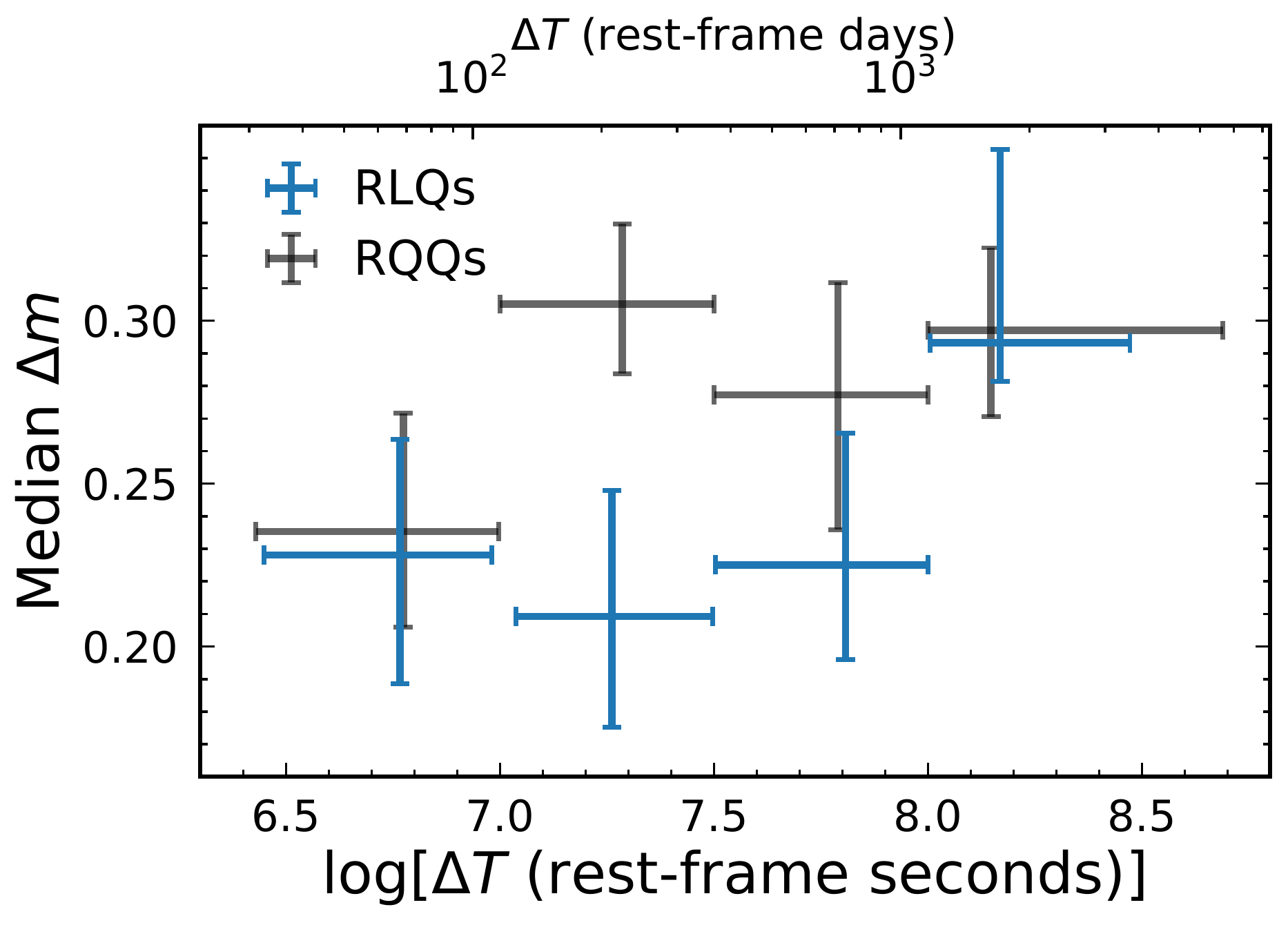}
\caption{The ensemble structure function for the X-ray variability of RLQs (blue) 
compared with that of matched RQQs (black).
The $x$-axis values are in units of both rest-frame seconds (bottom ticks) and days (top ticks).
The X-ray variability amplitude of RQQs seems to be insensitive to timescale above $\approx200$ rest-frame days.
However, the structure function of RLQs apparently increases with $\Delta T$ out to the last bin, 
where RLQs become as variable as RQQs.}
\label{fig:sf}
\end{figure}

\subsection{Comparing RLQs with matched RQQs}
\label{sec:matchedRQQ}
The energy flux ratios of RQQs are calculated and down-sampled following the method described in \S~\ref{sec:downSample}.
We then selected a matched sample of RQQs in the $z$, $\log L_\mathrm{2500\angstrom}$, and $\log \Delta T$ parameter spaces.
For each RLQs in the $z$-$\log L_\mathrm{2500\angstrom}$-$\log \Delta T$ space, 
we select the two nearest RQQs, without replacement.
We compare the distributions ($z$, $L_\mathrm{2500\angstrom}$, and $\Delta T$) of matched RQQs in Fig.~\ref{fig:mRQQs}. 
AD tests are performed for the distributions in the three panels of Fig.~\ref{fig:mRQQs}, 
which support that the RQQs and RLQs are suitably matched.

We depict the energy flux ratio distribution of matched RQQs in the bottom panel of Fig.~\ref{fig:histMax7}, where 
the black-dashed curve represents the normal function resulting from model fitting.
Model fitting is also performed for sub-samples of RQQs divided by luminosity and timescale, and the results are summarized in Table~\ref{tab:list}.
We find that, in our sample, the intrinsic X-ray variability amplitude of our 
RLQ sample is always smaller than that of the matched RQQ sample.

To further examine the dependence of X-ray flux variation on rest-frame timescale for the RLQ and RQQ samples,
we calculate the often-utilized ensemble structure function (SF).
We adopt the definition of SF from \citet{fiore1998},
\begin{equation}
    \Delta m_{ji}=\left|2.5\log [f(t_j)/f(t_i)]\right|,
\label{eq:sf}
\end{equation}
where $f(t_j)$ and $f(t_i)$ are X-ray fluxes of individual 
objects that are measured at epochs $t_j$ and $t_i$ ($t_j>t_i$), respectively.
Note that $\Delta m$ is equivalent to the 
absolute value of the previously computed energy flux ratio (Fig.~\ref{fig:var_dT}) scaled by a factor of 2.5.
We divide the data points into 4 timescale bins that are separated by $\Delta T=$ $10^7$, $10^{7.5}$, and $10^8$~s.
These dividing points are chosen such that each bin evenly spans 
$\approx0.5$ dex in timescale.
The four bins contain 39, 59, 130, and 86 data points, respectively.
The basic results are not affected if we merge the first 2 bins 
to make the distribution of data points in each bin more even.
In each bin, the median value and uncertainties (estimated using bootstrapping) are calculated, 
and the typical (i.e. median) scatter caused by flux-measurement error is subtracted in quadrature, 
which has little effect on the shape but slightly decreases the amplitude of the SF.
The SFs of RLQs and RQQs are shown in Fig.~\ref{fig:sf}.

The SF of RQQs increases with $\Delta T$ in the first two bins; however,
no significant increase is found at timescales above $\approx200$ days.
This result is consistent with \citet{shemmer2017}, who utilized a similar method
and created an ensemble SF that increases at relatively short timescales and shows no substantial
changes at relatively long timescales (see their Fig.~4).
On the other hand, although the SF of RLQs is generally smaller than that of matched RQQs,
it is consistent with increasing with $\Delta T$ out to the largest timescale bin,
where the two types of quasars have similar variability amplitudes.
The SF of RLQs is unconstrained at longer timescales
and requires X-ray data with an extended baseline to test if a flattening occurs at $>3000$ days in the rest frame.
(see \S~\ref{sec:futurework}).

\begin{table*}
\centering
\caption{Intrinsic variability distribution of RLQs compared with those of matched RQQs.}
\label{tab:list}
\begin{threeparttable}[b]
\begin{tabularx}{\linewidth}{@{}Y@{}}
\begin{tabular}{lccccc}
\hline
\hline
    Sample & \multicolumn{2}{c}{$m$\tnote{a}} & \multicolumn{2}{c}{$\sigma$\tnote{b}} \\
    & RLQs & RQQs & RLQs & RQQs \\
\hline
    Down-sampled & $0.005_{-0.010}^{+0.010}$ & $-0.004_{-0.009}^{+0.009}$& $0.156_{-0.007}^{+0.010}$ & $0.212_{-0.007}^{+0.009}$ \\
\hline
    High-luminosity ($\log L_\mathrm{2500\angstrom}\ge30.45$) & $0.016_{-0.014}^{+0.015}$ & $-0.020_{-0.016}^{+0.016}$ & $0.160_{-0.009}^{+0.015}$ & $0.190_{-0.011}^{+0.015}$ \\
    Low-luminosity ($\log L_\mathrm{2500\angstrom}< 30.45$) & $-0.007_{-0.014}^{+0.014}$ & $0.004_{-0.012}^{+0.012}$ & $0.150_{-0.010}^{+0.016}$ & $0.208_{-0.009}^{+0.011}$\\
\hline
    Long-timescale ($\log \Delta T\ge 7.79$)  & $-0.003_{-0.015}^{+0.016}$ & $0.002_{-0.015}^{+0.015}$ & $0.178_{-0.010}^{+0.016}$ & $0.211_{-0.010}^{+0.014}$ \\
    Short-timescale ($\log \Delta T < 7.79$) & $0.014_{-0.013}^{+0.014}$ & $-0.010_{-0.012}^{+0.013}$ & $0.131_{-0.008}^{+0.014}$ & $0.195_{-0.009}^{+0.012}$ \\
\hline
\end{tabular}
\end{tabularx}
\begin{tablenotes}
\item[a] The mean of the deconvolved energy flux ratio distribution.
\item[b] The intrinsic variability amplitude after removing the contribution of the measurement errors.
\end{tablenotes}
\end{threeparttable}
\end{table*}


\section{Discussion}
\label{sec:discussion}

\subsection{Comparison with previous sample-based X-ray spectral studies}
\label{sec:discuss_bias}
X-ray observations of RLQs starting in the {\it Einstein} era 
have repeatedly found in sample-based studies (see \S~\ref{sec:sampleSelection} and Table~\ref{tab:sample})
that their X-ray continua below a few tens of keV (rest-frame)
are significantly flatter than those of RQQs.
The spectral-fitting results of our investigation, however, show that
the median photon index of the X-ray spectra of 
our RLQ sample is $1.84_{-0.01}^{+0.01}$,
which is significantly steeper than that obtained in previous studies ($\Gamma\sim1.6$) and is
found to be closer to that of the similarly selected RQQ sample ($1.90_{-0.01}^{+0.02}$).
The median photon indexes of FSRQs and SSRQs are also consistent with each other.
One of the main differences between our current work and previous investigations is that 
our work utilizes RLQs selected in SDSS color space whereas 
previous investigations mainly utilize radio-selected quasars.
A bias was likely introduced when generalizing the high-energy properties of a small number of 
radio-selected RLQs to the full RLQ population in previous investigations.
Radio surveys tend to select the most radio-luminous quasars
in the Universe which may exhibit more extreme properties 
than their lower-luminosity counterparts which are included in optically selected RLQ samples
(e.g. \citealt{kellermann1994, gurkan2019}).
For example, the radio-selected FSRQs are generally considered to be beamed objects (e.g. \citealt{orr1982}), whereas 
most optically selected FSRQs seem to be an intermediate population between SSRQs and 
RQQs (e.g. \citealt{boroson1992, falcke1996, laor2000}), rather than highly beamed RLQs.
Indeed, we find in the most radio-luminous regime ($\log L_\mathrm{radio}>34.3$)
that FSRQs with flatter power-law slopes ($\Gamma~\sim1.55$) dominate the population (see \S~\ref{sec:specResults}).
Additionally, it is risky to draw conclusions from sample-based studies
that largely use targeted observations from archival X-ray data,
since targeted objects have a poorly defined selection function 
and tend to have unique properties that are atypical 
of those of their parent population.\footnote{The sample of \citet{zhou2020} in Table~\ref{tab:spec} is not affected by this selection effect since the 43 3CRR quasars have complete targeted X-ray coverage.}

\citet{siemiginowska2008} report the X-ray properties of a sample of 
RLQs that are also GPS and compact steep spectrum (CSS) radio sources.
They measured a median photon index of \mbox{$\Gamma=1.84$}, 
which is consistent with our results and indicates that the \mbox{X-ray}
spectral properties of GPS and CSS quasars 
are not necessarily different from those of typical RLQs.

\subsection{The X-ray reflection features of RLQs}
We detected strong iron K$\alpha$ emission in the rest-frame 6--7~keV band
and a feature at rest-frame $>14$~keV that is likely 
the Compton-reflection hump in \S~\ref{sec:specStack}.
Such features are expected if the X-ray emission of RLQs is dominated by their coronae
as for radio-quiet AGNs,
where the evidence for iron-line emission is ubiquitous (e.g. \citealt{nandra2007}).
Iron-line emission can be detected from
many radio-loud AGNs (e.g. \citealt{eracleous1996, lohfink2017, rani2018}) 
but is not ubiquitously observed (e.g. \citealt{eracleous1998}).
The general consensus is that the X-ray reflection features
of radio-loud AGNs generally seem to be weaker than those of
radio-quiet AGNs (e.g. \citealt{wozniak1998, eracleous2000}),
possibly due to dilution by a beamed X-ray continuum from jets (e.g. \citealt{ghisellini2019}),
an ionized accretion disk (e.g. \citealt{ballantyne2014}),
or an outflowing corona (e.g. \citealt{king2017}).

The iron-line emission from AGNs shows an X-ray Baldwin effect such that the EW of the
narrow component of the neutral iron line decreases with
increasing 2--10 keV luminosity (\citealt{iwasawa1993}).
This anti-correlation is significant for radio-quiet AGN samples (e.g. \citealt{bianchi2007, shu2012}),
but it remains unclear if radio-loud AGNs follow the same relation. 
Typically, radio-loud AGNs are removed from such correlation studies  
since their supposed jet-linked X-ray emission 
can increase continuum X-ray luminosity, reduce the EW of iron line emission, and cause
an artificial X-ray Baldwin effect (e.g. \citealt{jimnez2005, jiang2006}).
If the iron-line emission from our RLQs is produced by the primary X-ray continuum 
radiated from a \mbox{RQQ-like} corona
and is then diluted by a jet-linked continuum (with an intensity comparable to that of the coronal component), 
we would expect a typical strength of EW $\approx20$~eV (e.g. Eq.~1 in \citealt{bianchi2007}),
which is considerably weaker than the EW $=70_{-20}^{+30}$~eV that we found 
for our RLQ sample in \S~\ref{sec:specStack}.
If we instead consider the X-ray emission to be dominated by the corona,
the EW we measure is still somewhat stronger
than that predicted by the \mbox{X-ray} Baldwin effect (EW $\approx$ 40~eV).
Possibly, a broad component contributes some amount of the emission 
at 6.4~keV due to limited energy resolution (especially at high redshift).
Future X-ray missions with improved energy resolution and throughput
can probe the details of the profile of the iron lines (see \S~\ref{sec:futurework}).
It is also possible that RLQs follow a different \mbox{X-ray} Baldwin relation from that of RQQs
if the decrease of iron-line EW is related to but not
directly caused by the increase of $L_\mathrm{x}$.
For example, if the X-ray Baldwin effect is caused by a
receding torus (e.g. \citealt{lawrence1991, simpson2005})
and the inner radius (and thereby the covering factor) of the torus
is mainly determined by the UV luminosity,
RLQs of similar radio-loudness will have a distinct EW-$L_\mathrm{x}$ relation.
All in all, our investigation indicates that the iron-line emission of RLQs is 
not considerably diluted by a strong jet X-ray component.

The spectral shape and prominence of the iron line found in this work,
along with our other measurements of X-ray spectral properties,
put constraints on the model of the AGN corona.  For example,
a dynamic corona (e.g. \citealt{beloborodov1999, malzac2001}) is used to explain the 
X-ray properties of AGNs in some studies (e.g. \citealt{king2017}).
This model invokes a hot corona that moves away from the SMBH at a mildly relativistic velocity 
and is accelerated into the large-scale jet in radio-loud AGNs;
the hot corona in radio-quiet AGNs is static or moves at a lower velocity.
According to this model, the flatter \mbox{X-ray} continuum, lower level of X-ray reflection, 
and higher jet power of radio-loud AGNs are attributed to the higher outflow velocities of 
their coronae than those of radio-quiet AGNs.
However, our spectral-fitting and \mbox{X-ray} spectral-stacking results
do not support a substantial difference between the 
shapes of the \mbox{X-ray} continua of RLQs and RQQs or 
significantly weaker reflection in RLQs, 
and therefore rule out the idea that the outflow velocity of the corona is the 
key parameter controlling the X-ray and jet properties (e.g. \citealt{gupta2020}).

\subsection{The X-ray variability of RLQs}
\label{sec:discuss_var}
The typical X-ray variability amplitude of our optically selected RLQs
on timescales of months-to-years in the rest frame is $\approx40\%$,
and extreme variations are rare in our data (see Fig.~\ref{fig:var_dT}),
which is not consistent with jet-dominated objects that show 
frequent large-amplitude flares in their X-ray light curves 
(e.g. \citealt{ulrich1997, chatterjee2008, castignani2017, weaver2020}).
This qualitative similarity to the variability of RQQs (e.g. \citealt{timlin2020})
may indicate that the X-ray emission of our RLQs is coronal-dominated.
The overall variability amplitude is somewhat smaller than that of matched RQQs 
($\approx$ 60\%), and the X-ray variability amplitude increases with timescale,
consistent with a red-noise-type power spectrum 
(e.g. \citealt{uttley2005}).\footnote{\citet{soldi2014} compared the hard X-ray variability of radio galaxies to that of radio-quiet Seyfert galaxies selected from the BAT 58-month survey.
The variability amplitude of these radio-loud AGNs is consistent with that of radio-quiet AGNs in the 14--24 keV band,
while in the 35--100 keV band the former group varies more than the latter.
Considering the fact that these are hard X-ray-selected, low-luminosity objects,
a direct comparsion of their results with ours is probably unjustified.}
The fact that RLQs vary less than RQQs in the X-ray band at fixed timescale
might suggest that RLQs require a relatively longer timescale
to vary as much as RQQs, which is supported by the SFs depicted in Fig.~\ref{fig:sf}.

Assuming that the fluctuations of the accretion
flow propagate at similar speeds in both RQQs and RLQs,
the timescale of flux variability then reflects the 
physical scale of the X-ray emitting region;
the coronae of RLQs would be larger than those of matched RQQs.
It is possible that the size of the corona is larger in RLQs than in 
RQQs at fixed black-hole mass (e.g. \citealt{dogruel2020}).
An anti-correlation between the X-ray variability amplitude and black-hole mass among radio-quiet AGNs
is observed (e.g. \citealt{oneill2005, ponti2012}).
It is therefore also possible to consider a simpler scenario where the 
corona size scales with the black-hole mass similarly for both RLQs and RQQs,
provided that the SMBHs of our RLQs are generally
more massive than those of matched RQQs (e.g. \citealt{laor2000}).
Indeed, there are 47 and 94 black-hole mass measurements (based on either Mg~{\sc ii} or H$\beta$) in the catalog
of \citet{shen2011}
for the utilized RLQs and RQQs in our \mbox{X-ray} variability investigation,
and the median black-hole mass of the RLQs is $\approx0.2$~dex larger than that of the RQQs.
Furthermore, the median Eddington ratio of the former group is $\approx0.3$~dex smaller than that of the later,
which might play an independent role as well (e.g. \citealt{oneill2005}).
There is a characteristic timescale ($T_\mathrm{B}$) in the \mbox{X-ray} power spectrum of AGNs,
above which the increase of the variability power per decade timescale is not as significant as below $T_\mathrm{B}$.
This timescale depends on both the black-hole mass (e.g. \citealt{markowitz2003, papadakis2004}) 
and Eddington ratio such that $T_\mathrm{B}\propto M_\mathrm{BH}/\lambda_\mathrm{Edd}$ (e.g. \citealt{mchardy2006}).
Assuming that the X-ray power spectra of RLQs have a similar
shape to those of radio-quiet AGNs and the characteristic
timescale follows the same scaling relation above,
the most significant increase of the X-ray variability amplitude of RLQs will occur at timescales
$\approx0.5$~dex larger than those of RQQs, which is roughly consistent with Fig.~\ref{fig:sf}.
To further assess the relation between black-hole mass/Eddington ratio and X-ray variability,
we divide the 47 RLQs with measured black-hole masses 
into high-$M_\mathrm{BH}$/low-$\lambda_\mathrm{Edd}$ and 
low-$M_\mathrm{BH}$/high-$\lambda_\mathrm{Edd}$ groups at the median $M_\mathrm{BH}$\footnote{The two groups have nearly identical median $L_\mathrm{uv}$.  Therefore, $M_\mathrm{BH}$ and $\lambda_\mathrm{Edd}$ are related to each other, the role of which cannot be separated.}
and found that the variability amplitude of the former group ($\sigma=0.15_{-0.01}^{+0.02}$) is significantly smaller than that of the
later ($\sigma=0.22_{-0.02}^{+0.03}$), consistent with the assumption above.

Another possible explanation for the smaller variability amplitudes of RLQs is that
the fluctuations of the inner accretion flow are relatively weak in RLQs compared to RQQs at fixed timescale.
\citet{macleod2010} compared the optical/UV variability properties 
of quasars of different radio-loudness using SDSS Stripe 82 data. 
Except for the most radio-loud group that might be contaminated by 
jet emission in the optical/UV band,
the variability amplitudes of moderately radio-loud quasars are slightly (but significantly) 
smaller than those of radio-nondetected quasars (with controlled black-hole mass, luminosity, and Eddington ratio).
\citet{cai2019} also suggest that the UV-emitting inner 
accretion flows of RLQs are stabilized (relative to those of RQQs) by the jet-launching magnetic field. 

\subsection{The $L_\mathrm{x}$-$L_\mathrm{uv}$-$L_\mathrm{radio}$ relation of RLQs}
The X-ray spectral and variability properties obtained in this paper generally support the
result of \citet{zhu2020} that the \mbox{X-ray} emission of general RLQs is mainly radiated from their coronae.
In particular, the lack of a strong correlation of $\Gamma$ with $L_\mathrm{x}/L_\mathrm{x,rqq}$ further 
disputes an important role of a jet X-ray component.
\citet{zhu2020} suggested that only in a small portion of FSRQs is the jet-linked \mbox{X-ray} 
component apparent. The parameterized model fitting in the correlation analyses 
in \S~3.2 of \citet{zhu2020} supports that both 
the coronal and jet X-ray components of FSRQs increase with radio luminosity such that 
$L_\mathrm{x,corona}\propto L_\mathrm{radio}^{0.2}$ (following the corona-jet connection, 
$L_\mathrm{x,corona}\propto R^{0.2}$) and $L_\mathrm{x, jet}\propto L_\mathrm{radio}$, 
predicting that the jet-linked X-ray component most clearly emerges 
in the most radio-luminous objects.
Our spectral fitting results in \S~\ref{sec:specResults} (and the discussion above) of this paper
are consistent with such a picture; in the most radio-luminous (likely beamed) FSRQs the jet core 
dominates the X-ray emission, resulting in significantly flattened X-ray spectra.

\subsection{Connections with the hard X-ray emission of stellar-mass black holes}
Several lines of evidence support a unification of the properties of the disk-corona/jet system
surrounding accreting black holes spanning a wide range of mass
(e.g. \citealt{merloni2003, mchardy2006, kording2006, zhu2020}).
Black-hole X-ray binaries (BHXRBs) usually show 
a high-energy cut-off and Compton reflection in 
their hard \mbox{X-ray} spectra (e.g. \citealt{gierlinski1997}), similar to that of AGNs.
\citet{poutanen2003} pointed out that since the cut-off energy is always around several hundred keV, 
this feature is consistent with thermal Comptonization as well 
as a thermostat due to pair production but is not consistent with non-thermal synchrotron emission, 
the cut-off energy of which varies by large factors.\footnote{Fine tuning is 
probably also required to produce a nearly constant cut-off energy 
for the case of emission from non-thermal Comptonization.}
Moreover, the reflection features are not consistent with the idea that 
the X-ray source is beamed away from the disk (\citealt{poutanen2003}).
The lack of abrupt changes during state transitions of BHXRBs is not expected if a
radiatively inefficient jet dominates the X-ray emission (\citealt{maccacaro2005}).
Furthermore, the optical depth and temperature required by the jet-dominated model (e.g. \citealt{markoff2005})
are often physically unrealistic (\citealt{malzac2009}).
There is sometimes a non-thermal tail that emerges in
the MeV $\gamma$-ray spectra of BHXRBs (e.g. \citealt{mcconnell2002}), which might come from a jet.
However, this component is never energetically dominant in the \mbox{X-ray} band.
Therefore, a jet-dominated model for the hard X-ray emission is disfavored 
for both accreting stellar-mass black holes and SMBHs, supporting unification ideas.

\section{Summary and Future Prospects}
\label{sec:summary}

\subsection{Summary of main results}
We present X-ray spectral and long-term variability analyses of a large sample of
optically selected RLQs from \citet{zhu2020}. 
We utilize serendipitous X-ray data from the {\it Chandra} and {\it XMM-Newton} archives
that are more sensitive than $3\times10^{-14}$ erg cm$^{-2}$ s$^{-1}$ in the 0.5--7 keV energy band.
The spectral and temporal properties of these RLQs are compared with those of 
matched RQQs (\citealt{timlin2020}) using similarly selected X-ray data
(see \S~\ref{sec:sampleSelection} for details). The main results are summarized as follows:
\begin{enumerate}
\item The X-ray spectra of RLQs can be well described using a simple power-law model.
From our model fits, we find the median photon index of RLQs is $1.84_{-0.01}^{+0.01}$,
which is close to that of matched RQQs ($1.90_{-0.01}^{+0.02}$).
The median photon indices are similar between FSRQs and SSRQs.
We also investigated whether the $\Gamma$ values for FSRQs and SSRQs correlate with 
$R$, $L_\mathrm{x}/L_\mathrm{x,rqq}$, $L_\mathrm{radio}$, and $z$.
No strong correlations are found.
A small number of the most radio-luminous FSRQs ($\log L_\mathrm{radio}>34.3$), 
however, have flatter ($\Gamma\sim1.55$) X-ray spectra than those of other RLQs.
The stacked X-ray spectra of our RLQs show strong iron-line emission 
and a possible Compton-reflection hump. 
The narrow iron line at 6.4~keV is detected at a 5.3$\sigma$ significance level with EW = $70_{-20}^{+30}$~eV.
See \S~\ref{sec:spec}.

\item We quantify the X-ray variability amplitude of the RLQs
    with multiple serendipitous observations.
We required the interval between X-ray measurements to be larger
than 30 days in the quasar rest frame;
the median timescale is 1.7 year, and the longest timescale probed by our data is $\approx10$ yr.
Extreme large-amplitude variations are rare for our RLQs.
The intrinsic variability amplitude after removing the scatter caused by measurement errors
        is $0.156_{-0.007}^{+0.010}$ (43.2\%) for RLQs, 
which is significantly smaller than for the matched 
RQQs ($0.212_{-0.007}^{+0.009}$ or 62.9\%) on similar timescales.
The X-ray variability amplitude of RLQs increases with timescale,
and the most significant increase occurs at longer timescales than for RQQs.
See \S~\ref{sec:var}.

\item The X-ray spectral (including the power-law photon index of the 
primary continuum and the strength of superimposed reflection features) and variability analyses of our RLQs
support the results of \citet{zhu2020} indicating that the X-ray emission 
of typical RLQs is dominated by the disk/corona.
The previous claims that RLQs have substantially flatter X-ray spectra than those of RQQs
are likely affected by various selection effects.
Given the smaller variability amplitude of RLQs relative to RQQs matched in $L_\mathrm{uv}$, $z$, and timescale, 
the physical scale of the X-ray emitting region of the disk/corona system may be larger in RLQs, 
caused by their more massive SMBHs and/or lower Eddington ratio.
A predominantly coronal origin for the X-ray emission of RLQs is 
also consistent with the results for the hard X-ray emission from stellar-mass black holes.
See \S~\ref{sec:discussion}.
\end{enumerate}

\subsection{Future work}
\label{sec:futurework}
Despite the notable connection between the X-ray brightness of the disk/corona
and the jet efficiency of RLQs (\citealt{zhu2020}),
the basic spectral shape of the X-rays (i.e. $\Gamma$) radiated from the hot corona
does not seem to be strongly affected by such a connection,
demonstrated by the similarity between the median $\Gamma$ values of RLQs and RQQs and also 
by the fact that the $\Gamma$ values of individual RLQs do not seem to
correlate with either $R$ or $L_\mathrm{x}/L_\mathrm{x,rqq}$.
It therefore would be valuable to investigate 
how the physical properties of the corona (i.e. temperature, optical depth, and geometry)
change with $R$ and $L_\mathrm{x}/L_\mathrm{x,rqq}$,
which probably would require a powerful \mbox{X-ray} observatory with broadband response 
and excellent hard X-ray sensitivity (e.g. {\it HEX-P}; \citealt{harrison2017}).

The next-generation X-ray observatories (e.g. {\it Athena}; \citealt{athena2013})
with orders of magnitude increases in effective area and energy resolution relative to
current instruments will probe the iron emission lines and
the Compton-reflection humps of individual RLQs,
which can constrain the geometry and physical conditions
of the material spiraling onto the black hole.
Furthermore, it will be important to study
if RLQs follow a similar $\Gamma$-$\lambda_\mathrm{Edd}$ relation as for RQQs.


The advent of {\it eROSITA} (\citealt{erositaScienceBook}) will produce copious
well-sampled time-domain X-ray data that, when combined with existing archival data,
will allow for better constraints upon
the correlation between the variability amplitude of RLQs and timescale.
X-ray variability is one of the most-promising methods that can constrain the physical scale of the quasar corona,
and {\it eROSITA} measurements may allow testing of the variability scenarios described in \S~\ref{sec:discuss_var}.





\section*{Acknowledgements}
We thank the referee for a constructive review.
We have benefitted from discussions with Thomas Maccarone about the hard X-ray emission
of BHXRBs.
We thank Ari Laor for helpful comments that improved the manuscript.
SFZ, JDT, and WNB acknowledge support from
CXC grant AR1-22008X, NASA ADAP grant 80NSSC18K0878, and the Penn State ACIS
Instrument Team Contract SV4-74018 (issued by the {\it Chandra}
X-ray Center, which is operated by the Smithsonian Astrophysical
Observatory for and on behalf of NASA under contract
NAS8-03060).  The {\it Chandra} ACIS team Guaranteed Time Observations (GTO) utilized were
selected by the ACIS Instrument Principal Investigator, Gordon P.
Garmire, currently of the Huntingdon Institute for X-ray Astronomy, LLC,
which is under contract to the Smithsonian Astrophysical Observatory
via Contract SV2-82024.
This research has made use of data obtained from the {\it Chandra} Data Archive and
the {\it Chandra} Source Catalog.
Based on observations obtained with {\it XMM-Newton},
an ESA science mission with instruments and contributions directly funded by
ESA Member States and NASA.

\section*{Data availability}
The data used in this investigation are available in the
article and in its online supplementary material.

\bibliographystyle{mnras}
\bibliography{mn} 

\begin{thebibliography}{}
\makeatletter
\relax
\def\mn@urlcharsother{\let\do\@makeother \do\$\do\&\do\#\do\^\do\_\do\%\do\~}
\def\mn@doi{\begingroup\mn@urlcharsother \@ifnextchar [ {\mn@doi@}
  {\mn@doi@[]}}
\def\mn@doi@[#1]#2{\def\@tempa{#1}\ifx\@tempa\@empty \href
  {http://dx.doi.org/#2} {doi:#2}\else \href {http://dx.doi.org/#2} {#1}\fi
  \endgroup}
\def\mn@eprint#1#2{\mn@eprint@#1:#2::\@nil}
\def\mn@eprint@arXiv#1{\href {http://arxiv.org/abs/#1} {{\tt arXiv:#1}}}
\def\mn@eprint@dblp#1{\href {http://dblp.uni-trier.de/rec/bibtex/#1.xml}
  {dblp:#1}}
\def\mn@eprint@#1:#2:#3:#4\@nil{\def\@tempa {#1}\def\@tempb {#2}\def\@tempc
  {#3}\ifx \@tempc \@empty \let \@tempc \@tempb \let \@tempb \@tempa \fi \ifx
  \@tempb \@empty \def\@tempb {arXiv}\fi \@ifundefined
  {mn@eprint@\@tempb}{\@tempb:\@tempc}{\expandafter \expandafter \csname
  mn@eprint@\@tempb\endcsname \expandafter{\@tempc}}}

\bibitem[\protect\citeauthoryear{{Anderson} \& {Darling}}{{Anderson} \&
  {Darling}}{1952}]{ad1952}
{Anderson} T.~W.,  {Darling} D.~A.,  1952, \mn@doi [Ann. Math. Statist.]
  {10.1214/aoms/1177729437}, 23, 193

\bibitem[\protect\citeauthoryear{{Arcodia}, {Merloni}, {Nandra}  \&
  {Ponti}}{{Arcodia} et~al.}{2019}]{arcodia2019}
{Arcodia} R.,  {Merloni} A.,  {Nandra} K.,   {Ponti} G.,  2019, \mn@doi [\aap]
  {10.1051/0004-6361/201935874}, \href
  {https://ui.adsabs.harvard.edu/abs/2019A&A...628A.135A} {628, A135}

\bibitem[\protect\citeauthoryear{{Baker} \& {Cousins}}{{Baker} \&
  {Cousins}}{1984}]{baker1984}
{Baker} S.,  {Cousins} R.~D.,  1984, \mn@doi [Nuclear Instruments and Methods
  in Physics Research] {10.1016/0167-5087(84)90016-4}, \href
  {https://ui.adsabs.harvard.edu/abs/1984NIMPR.221..437B} {221, 437}

\bibitem[\protect\citeauthoryear{{Ballantyne} et~al.,}{{Ballantyne}
  et~al.}{2014}]{ballantyne2014}
{Ballantyne} D.~R.,  et~al., 2014, \mn@doi [\apj] {10.1088/0004-637X/794/1/62},
  \href {https://ui.adsabs.harvard.edu/abs/2014ApJ...794...62B} {794, 62}

\bibitem[\protect\citeauthoryear{{Becker}, {White}  \& {Helfand}}{{Becker}
  et~al.}{1995}]{becker1995}
{Becker} R.~H.,  {White} R.~L.,   {Helfand} D.~J.,  1995, \mn@doi [\apj]
  {10.1086/176166}, \href
  {https://ui.adsabs.harvard.edu/#abs/1995ApJ...450..559B} {450, 559}

\bibitem[\protect\citeauthoryear{{Beloborodov}}{{Beloborodov}}{1999}]{beloborodov1999}
{Beloborodov} A.~M.,  1999, \mn@doi [\apjl] {10.1086/311810}, \href
  {https://ui.adsabs.harvard.edu/abs/1999ApJ...510L.123B} {510, L123}

\bibitem[\protect\citeauthoryear{{Bianchi}, {Guainazzi}, {Matt}  \& {Fonseca
  Bonilla}}{{Bianchi} et~al.}{2007}]{bianchi2007}
{Bianchi} S.,  {Guainazzi} M.,  {Matt} G.,   {Fonseca Bonilla} N.,  2007,
  \mn@doi [\aap] {10.1051/0004-6361:20077331}, \href
  {https://ui.adsabs.harvard.edu/abs/2007A&A...467L..19B} {467, L19}

\bibitem[\protect\citeauthoryear{{Boroson} \& {Green}}{{Boroson} \&
  {Green}}{1992}]{boroson1992}
{Boroson} T.~A.,  {Green} R.~F.,  1992, \mn@doi [\apjs] {10.1086/191661}, \href
  {https://ui.adsabs.harvard.edu/abs/1992ApJS...80..109B} {80, 109}

\bibitem[\protect\citeauthoryear{{Brandt}, {Mathur}  \& {Elvis}}{{Brandt}
  et~al.}{1997}]{brandt1997}
{Brandt} W.~N.,  {Mathur} S.,   {Elvis} M.,  1997, \mn@doi [\mnras]
  {10.1093/mnras/285.3.L25}, \href
  {https://ui.adsabs.harvard.edu/abs/1997MNRAS.285L..25B} {285, L25}

\bibitem[\protect\citeauthoryear{{Brenneman} \& {Reynolds}}{{Brenneman} \&
  {Reynolds}}{2009}]{brenneman2009}
{Brenneman} L.~W.,  {Reynolds} C.~S.,  2009, \mn@doi [\apj]
  {10.1088/0004-637X/702/2/1367}, \href
  {https://ui.adsabs.harvard.edu/abs/2009ApJ...702.1367B} {702, 1367}

\bibitem[\protect\citeauthoryear{{Brightman} et~al.,}{{Brightman}
  et~al.}{2013}]{brightman2013}
{Brightman} M.,  et~al., 2013, \mn@doi [\mnras] {10.1093/mnras/stt920}, \href
  {https://ui.adsabs.harvard.edu/abs/2013MNRAS.433.2485B} {433, 2485}

\bibitem[\protect\citeauthoryear{{Broos}, {Feigelson}, {Townsley}, {Getman},
  {Wang}, {Garmire}, {Jiang}  \& {Tsuboi}}{{Broos} et~al.}{2007}]{broos2007}
{Broos} P.~S.,  {Feigelson} E.~D.,  {Townsley} L.~K.,  {Getman} K.~V.,  {Wang}
  J.,  {Garmire} G.~P.,  {Jiang} Z.,   {Tsuboi} Y.,  2007, \mn@doi [\apjs]
  {10.1086/512068}, \href
  {https://ui.adsabs.harvard.edu/abs/2007ApJS..169..353B} {169, 353}

\bibitem[\protect\citeauthoryear{Burke et~al.,}{Burke et~al.}{2020}]{db2020}
Burke D.,  et~al., 2020, sherpa/sherpa: Sherpa 4.12.1,
  \mn@doi{10.5281/zenodo.3944985}, \url
  {https://doi.org/10.5281/zenodo.3944985}

\bibitem[\protect\citeauthoryear{{Cai}, {Sun}, {Wang}, {Zhu}, {Gu}  \&
  {Yuan}}{{Cai} et~al.}{2019}]{cai2019}
{Cai} Z.,  {Sun} Y.,  {Wang} J.,  {Zhu} F.,  {Gu} W.,   {Yuan} F.,  2019,
  \mn@doi [Science China Physics, Mechanics, and Astronomy]
  {10.1007/s11433-018-9330-4}, \href
  {https://ui.adsabs.harvard.edu/abs/2019SCPMA..6269511C} {62, 69511}

\bibitem[\protect\citeauthoryear{{Cash}}{{Cash}}{1979}]{cash1979}
{Cash} W.,  1979, \mn@doi [\apj] {10.1086/156922}, \href
  {https://ui.adsabs.harvard.edu/abs/1979ApJ...228..939C} {228, 939}

\bibitem[\protect\citeauthoryear{{Castignani} et~al.,}{{Castignani}
  et~al.}{2017}]{castignani2017}
{Castignani} G.,  et~al., 2017, \mn@doi [\aap] {10.1051/0004-6361/201629775},
  \href {https://ui.adsabs.harvard.edu/abs/2017A&A...601A..30C} {601, A30}

\bibitem[\protect\citeauthoryear{{Chatterjee} et~al.,}{{Chatterjee}
  et~al.}{2008}]{chatterjee2008}
{Chatterjee} R.,  et~al., 2008, \mn@doi [\apj] {10.1086/592598}, \href
  {https://ui.adsabs.harvard.edu/abs/2008ApJ...689...79C} {689, 79}

\bibitem[\protect\citeauthoryear{{Chatterjee} et~al.,}{{Chatterjee}
  et~al.}{2011}]{chatterjee2011}
{Chatterjee} R.,  et~al., 2011, \mn@doi [\apj] {10.1088/0004-637X/734/1/43},
  \href {https://ui.adsabs.harvard.edu/abs/2011ApJ...734...43C} {734, 43}

\bibitem[\protect\citeauthoryear{{Condon}, {Cotton}, {Greisen}, {Yin},
  {Perley}, {Taylor}  \& {Broderick}}{{Condon} et~al.}{1998}]{condon1998}
{Condon} J.~J.,  {Cotton} W.~D.,  {Greisen} E.~W.,  {Yin} Q.~F.,  {Perley}
  R.~A.,  {Taylor} G.~B.,   {Broderick} J.~J.,  1998, \mn@doi [\aj]
  {10.1086/300337}, \href
  {https://ui.adsabs.harvard.edu/abs/1998AJ....115.1693C} {115, 1693}

\bibitem[\protect\citeauthoryear{{Dickey} \& {Lockman}}{{Dickey} \&
  {Lockman}}{1990}]{dickey1990}
{Dickey} J.~M.,  {Lockman} F.~J.,  1990, \mn@doi [Annual Review of Astronomy
  and Astrophysics] {10.1146/annurev.aa.28.090190.001243}, \href
  {https://ui.adsabs.harvard.edu/#abs/1990ARA&A..28..215D} {28, 215}

\bibitem[\protect\citeauthoryear{{Dogruel}, {Dai}, {Guerras}, {Cornachione}  \&
  {Morgan}}{{Dogruel} et~al.}{2020}]{dogruel2020}
{Dogruel} M.~B.,  {Dai} X.,  {Guerras} E.,  {Cornachione} M.,   {Morgan} C.~W.,
   2020, \mn@doi [\apj] {10.3847/1538-4357/ab879b}, \href
  {https://ui.adsabs.harvard.edu/abs/2020ApJ...894..153D} {894, 153}

\bibitem[\protect\citeauthoryear{{Elvis} et~al.,}{{Elvis}
  et~al.}{1994}]{elvis1994}
{Elvis} M.,  et~al., 1994, \mn@doi [\apjs] {10.1086/192093}, \href
  {https://ui.adsabs.harvard.edu/abs/1994ApJS...95....1E} {95, 1}

\bibitem[\protect\citeauthoryear{{Eracleous} \& {Halpern}}{{Eracleous} \&
  {Halpern}}{1998}]{eracleous1998}
{Eracleous} M.,  {Halpern} J.~P.,  1998, \mn@doi [\apj] {10.1086/306190}, \href
  {https://ui.adsabs.harvard.edu/abs/1998ApJ...505..577E} {505, 577}

\bibitem[\protect\citeauthoryear{{Eracleous}, {Halpern}  \&
  {Livio}}{{Eracleous} et~al.}{1996}]{eracleous1996}
{Eracleous} M.,  {Halpern} J.~P.,   {Livio} M.,  1996, \mn@doi [\apj]
  {10.1086/176871}, \href
  {https://ui.adsabs.harvard.edu/abs/1996ApJ...459...89E} {459, 89}

\bibitem[\protect\citeauthoryear{{Eracleous}, {Sambruna}  \&
  {Mushotzky}}{{Eracleous} et~al.}{2000}]{eracleous2000}
{Eracleous} M.,  {Sambruna} R.,   {Mushotzky} R.~F.,  2000, \mn@doi [\apj]
  {10.1086/309076}, \href
  {https://ui.adsabs.harvard.edu/abs/2000ApJ...537..654E} {537, 654}

\bibitem[\protect\citeauthoryear{{Evans} et~al.,}{{Evans}
  et~al.}{2019}]{evans2019}
{Evans} I.~N.,  et~al., 2019, in AAS/High Energy Astrophysics Division.
  AAS/High Energy Astrophysics Division.
p. 114.01

\bibitem[\protect\citeauthoryear{{Falcke}, {Patnaik}  \& {Sherwood}}{{Falcke}
  et~al.}{1996}]{falcke1996}
{Falcke} H.,  {Patnaik} A.~R.,   {Sherwood} W.,  1996, \mn@doi [\apjl]
  {10.1086/310386}, \href
  {https://ui.adsabs.harvard.edu/abs/1996ApJ...473L..13F} {473, L13}

\bibitem[\protect\citeauthoryear{{Fiore}, {Laor}, {Elvis}, {Nicastro}  \&
  {Giallongo}}{{Fiore} et~al.}{1998}]{fiore1998}
{Fiore} F.,  {Laor} A.,  {Elvis} M.,  {Nicastro} F.,   {Giallongo} E.,  1998,
  \mn@doi [\apj] {10.1086/306031}, \href
  {https://ui.adsabs.harvard.edu/abs/1998ApJ...503..607F} {503, 607}

\bibitem[\protect\citeauthoryear{{George} \& {Fabian}}{{George} \&
  {Fabian}}{1991}]{george1991}
{George} I.~M.,  {Fabian} A.~C.,  1991, \mn@doi [\mnras]
  {10.1093/mnras/249.2.352}, \href
  {https://ui.adsabs.harvard.edu/abs/1991MNRAS.249..352G} {249, 352}

\bibitem[\protect\citeauthoryear{{Ghisellini} et~al.,}{{Ghisellini}
  et~al.}{2019}]{ghisellini2019}
{Ghisellini} G.,  et~al., 2019, \mn@doi [\aap] {10.1051/0004-6361/201935750},
  \href {https://ui.adsabs.harvard.edu/abs/2019A&A...627A..72G} {627, A72}

\bibitem[\protect\citeauthoryear{{Gibson} \& {Brandt}}{{Gibson} \&
  {Brandt}}{2012}]{gibson2012}
{Gibson} R.~R.,  {Brandt} W.~N.,  2012, \mn@doi [\apj]
  {10.1088/0004-637X/746/1/54}, \href
  {https://ui.adsabs.harvard.edu/abs/2012ApJ...746...54G} {746, 54}

\bibitem[\protect\citeauthoryear{{Gierlinski}, {Zdziarski}, {Done}, {Johnson},
  {Ebisawa}, {Ueda}, {Haardt}  \& {Phlips}}{{Gierlinski}
  et~al.}{1997}]{gierlinski1997}
{Gierlinski} M.,  {Zdziarski} A.~A.,  {Done} C.,  {Johnson} W.~N.,  {Ebisawa}
  K.,  {Ueda} Y.,  {Haardt} F.,   {Phlips} B.~F.,  1997, \mn@doi [\mnras]
  {10.1093/mnras/288.4.958}, \href
  {https://ui.adsabs.harvard.edu/abs/1997MNRAS.288..958G} {288, 958}

\bibitem[\protect\citeauthoryear{{Grandi} \& {Palumbo}}{{Grandi} \&
  {Palumbo}}{2004}]{grandi2004}
{Grandi} P.,  {Palumbo} G. G.~C.,  2004, \mn@doi [Science]
  {10.1126/science.1101787}, \href
  {https://ui.adsabs.harvard.edu/abs/2004Sci...306..998G} {306, 998}

\bibitem[\protect\citeauthoryear{{Grandi}, {Malaguti}  \& {Fiocchi}}{{Grandi}
  et~al.}{2006}]{grandi2006}
{Grandi} P.,  {Malaguti} G.,   {Fiocchi} M.,  2006, \mn@doi [\apj]
  {10.1086/500100}, \href
  {https://ui.adsabs.harvard.edu/abs/2006ApJ...642..113G} {642, 113}

\bibitem[\protect\citeauthoryear{{Gupta}, {Sikora}  \& {Rusinek}}{{Gupta}
  et~al.}{2020}]{gupta2020}
{Gupta} M.,  {Sikora} M.,   {Rusinek} K.,  2020, \mn@doi [\mnras]
  {10.1093/mnras/stz3384}, \href
  {https://ui.adsabs.harvard.edu/abs/2020MNRAS.492..315G} {492, 315}

\bibitem[\protect\citeauthoryear{{G{\"u}rkan} et~al.,}{{G{\"u}rkan}
  et~al.}{2019}]{gurkan2019}
{G{\"u}rkan} G.,  et~al., 2019, \mn@doi [\aap] {10.1051/0004-6361/201833892},
  \href {https://ui.adsabs.harvard.edu/abs/2019A&A...622A..11G} {622, A11}

\bibitem[\protect\citeauthoryear{{Harrison} \& {HEX-P
  Collaboration}}{{Harrison} \& {HEX-P Collaboration}}{2017}]{harrison2017}
{Harrison} F.,  {HEX-P Collaboration} 2017, in APS April Meeting Abstracts. p.
  K9.002

\bibitem[\protect\citeauthoryear{{Hartman} et~al.,}{{Hartman}
  et~al.}{1992}]{hartman1992}
{Hartman} R.~C.,  et~al., 1992, \mn@doi [\apjl] {10.1086/186263}, \href
  {https://ui.adsabs.harvard.edu/abs/1992ApJ...385L...1H} {385, L1}

\bibitem[\protect\citeauthoryear{{Hayashida} et~al.,}{{Hayashida}
  et~al.}{2015}]{hayashida2015}
{Hayashida} M.,  et~al., 2015, \mn@doi [\apj] {10.1088/0004-637X/807/1/79},
  \href {https://ui.adsabs.harvard.edu/abs/2015ApJ...807...79H} {807, 79}

\bibitem[\protect\citeauthoryear{{Hu}, {Liu}, {Jin}  \& {Yuan}}{{Hu}
  et~al.}{2019}]{hu2019}
{Hu} J.,  {Liu} Z.,  {Jin} C.,   {Yuan} W.,  2019, \mn@doi [\mnras]
  {10.1093/mnras/stz2030}, \href
  {https://ui.adsabs.harvard.edu/abs/2019MNRAS.488.4378H} {488, 4378}

\bibitem[\protect\citeauthoryear{{Humphrey}, {Liu}  \& {Buote}}{{Humphrey}
  et~al.}{2009}]{humphrey2009}
{Humphrey} P.~J.,  {Liu} W.,   {Buote} D.~A.,  2009, \mn@doi [\apj]
  {10.1088/0004-637X/693/1/822}, \href
  {https://ui.adsabs.harvard.edu/abs/2009ApJ...693..822H} {693, 822}

\bibitem[\protect\citeauthoryear{{Ivezi{\'c}} et~al.,}{{Ivezi{\'c}}
  et~al.}{2002}]{ivezic2002}
{Ivezi{\'c}} {\v{Z}}.,  et~al., 2002, \mn@doi [\aj] {10.1086/344069}, \href
  {https://ui.adsabs.harvard.edu/\#abs/2002AJ....124.2364I} {124, 2364}

\bibitem[\protect\citeauthoryear{{Iwasawa} \& {Taniguchi}}{{Iwasawa} \&
  {Taniguchi}}{1993}]{iwasawa1993}
{Iwasawa} K.,  {Taniguchi} Y.,  1993, \mn@doi [\apjl] {10.1086/186948}, \href
  {https://ui.adsabs.harvard.edu/abs/1993ApJ...413L..15I} {413, L15}

\bibitem[\protect\citeauthoryear{{Jiang}, {Wang}  \& {Wang}}{{Jiang}
  et~al.}{2006}]{jiang2006}
{Jiang} P.,  {Wang} J.~X.,   {Wang} T.~G.,  2006, \mn@doi [\apj]
  {10.1086/503866}, \href
  {https://ui.adsabs.harvard.edu/abs/2006ApJ...644..725J} {644, 725}

\bibitem[\protect\citeauthoryear{{Jim{\'e}nez-Bail{\'o}n}, {Piconcelli},
  {Guainazzi}, {Schartel}, {Rodr{\'\i}guez-Pascual}  \&
  {Santos-Lle{\'o}}}{{Jim{\'e}nez-Bail{\'o}n} et~al.}{2005}]{jimnez2005}
{Jim{\'e}nez-Bail{\'o}n} E.,  {Piconcelli} E.,  {Guainazzi} M.,  {Schartel} N.,
   {Rodr{\'\i}guez-Pascual} P.~M.,   {Santos-Lle{\'o}} M.,  2005, \mn@doi
  [\aap] {10.1051/0004-6361:20041623}, \href
  {https://ui.adsabs.harvard.edu/abs/2005A&A...435..449J} {435, 449}

\bibitem[\protect\citeauthoryear{{Just}, {Brandt}, {Shemmer}, {Steffen},
  {Schneider}, {Chartas}  \& {Garmire}}{{Just} et~al.}{2007}]{just2007}
{Just} D.~W.,  {Brandt} W.~N.,  {Shemmer} O.,  {Steffen} A.~T.,  {Schneider}
  D.~P.,  {Chartas} G.,   {Garmire} G.~P.,  2007, \mn@doi [\apj]
  {10.1086/519990}, \href
  {https://ui.adsabs.harvard.edu/#abs/2007ApJ...665.1004J} {665, 1004}

\bibitem[\protect\citeauthoryear{{Kaastra}}{{Kaastra}}{2017}]{kaastra2017}
{Kaastra} J.~S.,  2017, \mn@doi [\aap] {10.1051/0004-6361/201629319}, \href
  {https://ui.adsabs.harvard.edu/abs/2017A&A...605A..51K} {605, A51}

\bibitem[\protect\citeauthoryear{{Kamraj}, {Harrison}, {Balokovi{\'c}},
  {Lohfink}  \& {Brightman}}{{Kamraj} et~al.}{2018}]{kamraj2018}
{Kamraj} N.,  {Harrison} F.~A.,  {Balokovi{\'c}} M.,  {Lohfink} A.,
  {Brightman} M.,  2018, \mn@doi [\apj] {10.3847/1538-4357/aadd0d}, \href
  {https://ui.adsabs.harvard.edu/abs/2018ApJ...866..124K} {866, 124}

\bibitem[\protect\citeauthoryear{{Kang}, {Wang}  \& {Kang}}{{Kang}
  et~al.}{2020}]{kang2020}
{Kang} J.,  {Wang} J.,   {Kang} W.,  2020, \mn@doi [\apj]
  {10.3847/1538-4357/abadf5}, \href
  {https://ui.adsabs.harvard.edu/abs/2020ApJ...901..111K} {901, 111}

\bibitem[\protect\citeauthoryear{{Kellermann}, {Sramek}, {Schmidt}, {Shaffer}
  \& {Green}}{{Kellermann} et~al.}{1989}]{kellermann1989}
{Kellermann} K.~I.,  {Sramek} R.,  {Schmidt} M.,  {Shaffer} D.~B.,   {Green}
  R.,  1989, \mn@doi [\aj] {10.1086/115207}, \href
  {https://ui.adsabs.harvard.edu/#abs/1989AJ.....98.1195K} {98, 1195}

\bibitem[\protect\citeauthoryear{{Kellermann}, {Sramek}, {Schmidt}, {Green}  \&
  {Shaffer}}{{Kellermann} et~al.}{1994}]{kellermann1994}
{Kellermann} K.~I.,  {Sramek} R.~A.,  {Schmidt} M.,  {Green} R.~F.,   {Shaffer}
  D.~B.,  1994, \mn@doi [\aj] {10.1086/117145}, \href
  {https://ui.adsabs.harvard.edu/abs/1994AJ....108.1163K} {108, 1163}

\bibitem[\protect\citeauthoryear{{King}, {Lohfink}  \& {Kara}}{{King}
  et~al.}{2017}]{king2017}
{King} A.~L.,  {Lohfink} A.,   {Kara} E.,  2017, \mn@doi [\apj]
  {10.3847/1538-4357/835/2/226}, \href
  {https://ui.adsabs.harvard.edu/abs/2017ApJ...835..226K} {835, 226}

\bibitem[\protect\citeauthoryear{{K{\"o}rding}, {Jester}  \&
  {Fender}}{{K{\"o}rding} et~al.}{2006}]{kording2006}
{K{\"o}rding} E.~G.,  {Jester} S.,   {Fender} R.,  2006, \mn@doi [\mnras]
  {10.1111/j.1365-2966.2006.10954.x}, \href
  {https://ui.adsabs.harvard.edu/abs/2006MNRAS.372.1366K} {372, 1366}

\bibitem[\protect\citeauthoryear{{Lacy} et~al.,}{{Lacy}
  et~al.}{2020}]{vlass2020}
{Lacy} M.,  et~al., 2020, \mn@doi [\pasp] {10.1088/1538-3873/ab63eb}, \href
  {https://ui.adsabs.harvard.edu/abs/2020PASP..132c5001L} {132, 035001}

\bibitem[\protect\citeauthoryear{{Laor}}{{Laor}}{2000}]{laor2000}
{Laor} A.,  2000, \mn@doi [\apjl] {10.1086/317280}, \href
  {https://ui.adsabs.harvard.edu/abs/2000ApJ...543L.111L} {543, L111}

\bibitem[\protect\citeauthoryear{{Laor}, {Fiore}, {Elvis}, {Wilkes}  \&
  {McDowell}}{{Laor} et~al.}{1997}]{laor1997}
{Laor} A.,  {Fiore} F.,  {Elvis} M.,  {Wilkes} B.~J.,   {McDowell} J.~C.,
  1997, \mn@doi [\apj] {10.1086/303696}, \href
  {https://ui.adsabs.harvard.edu/abs/1997ApJ...477...93L} {477, 93}

\bibitem[\protect\citeauthoryear{{Lawrence}}{{Lawrence}}{1991}]{lawrence1991}
{Lawrence} A.,  1991, \mn@doi [\mnras] {10.1093/mnras/252.4.586}, \href
  {https://ui.adsabs.harvard.edu/abs/1991MNRAS.252..586L} {252, 586}

\bibitem[\protect\citeauthoryear{{Lawson} \& {Turner}}{{Lawson} \&
  {Turner}}{1997}]{lawson1997}
{Lawson} A.~J.,  {Turner} M.~J.~L.,  1997, \mn@doi [\mnras]
  {10.1093/mnras/288.4.920}, \href
  {https://ui.adsabs.harvard.edu/abs/1997MNRAS.288..920L} {288, 920}

\bibitem[\protect\citeauthoryear{{Lawson}, {Turner}, {Williams}, {Stewart}  \&
  {Saxton}}{{Lawson} et~al.}{1992}]{lawson1992}
{Lawson} A.~J.,  {Turner} M.~J.~L.,  {Williams} O.~R.,  {Stewart} G.~C.,
  {Saxton} R.~D.,  1992, \mn@doi [\mnras] {10.1093/mnras/259.4.743}, \href
  {https://ui.adsabs.harvard.edu/abs/1992MNRAS.259..743L} {259, 743}

\bibitem[\protect\citeauthoryear{{Leighly}, {O'Brien}, {Edelson}, {George},
  {Malkan}, {Matsuoka}, {Mushotzky}  \& {Peterson}}{{Leighly}
  et~al.}{1997}]{leighly1997}
{Leighly} K.~M.,  {O'Brien} P.~T.,  {Edelson} R.,  {George} I.~M.,  {Malkan}
  M.~A.,  {Matsuoka} M.,  {Mushotzky} R.~F.,   {Peterson} B.~M.,  1997, \mn@doi
  [\apj] {10.1086/304276}, \href
  {https://ui.adsabs.harvard.edu/abs/1997ApJ...483..767L} {483, 767}

\bibitem[\protect\citeauthoryear{Livesey}{Livesey}{2007}]{livesey2007}
Livesey J.,  2007, \mn@doi [Clinical biochemistry]
  {10.1016/j.clinbiochem.2007.04.003}, 40, 1032

\bibitem[\protect\citeauthoryear{{Lohfink} et~al.,}{{Lohfink}
  et~al.}{2017}]{lohfink2017}
{Lohfink} A.~M.,  et~al., 2017, \mn@doi [\apj] {10.3847/1538-4357/aa6d07},
  \href {https://ui.adsabs.harvard.edu/abs/2017ApJ...841...80L} {841, 80}

\bibitem[\protect\citeauthoryear{{MacLeod} et~al.,}{{MacLeod}
  et~al.}{2010}]{macleod2010}
{MacLeod} C.~L.,  et~al., 2010, \mn@doi [\apj] {10.1088/0004-637X/721/2/1014},
  \href {https://ui.adsabs.harvard.edu/abs/2010ApJ...721.1014M} {721, 1014}

\bibitem[\protect\citeauthoryear{{Maccacaro}, {Gioia}, {Wolter}, {Zamorani}  \&
  {Stocke}}{{Maccacaro} et~al.}{1988}]{macccaro1988}
{Maccacaro} T.,  {Gioia} I.~M.,  {Wolter} A.,  {Zamorani} G.,   {Stocke} J.~T.,
   1988, \mn@doi [\apj] {10.1086/166127}, \href
  {https://ui.adsabs.harvard.edu/abs/1988ApJ...326..680M} {326, 680}

\bibitem[\protect\citeauthoryear{{Maccarone}}{{Maccarone}}{2005}]{maccacaro2005}
{Maccarone} T.~J.,  2005, \mn@doi [\mnras] {10.1111/j.1745-3933.2005.00047.x},
  \href {https://ui.adsabs.harvard.edu/abs/2005MNRAS.360L..68M} {360, L68}

\bibitem[\protect\citeauthoryear{{Magdziarz} \& {Zdziarski}}{{Magdziarz} \&
  {Zdziarski}}{1995}]{magdziarz1995}
{Magdziarz} P.,  {Zdziarski} A.~A.,  1995, \mn@doi [\mnras]
  {10.1093/mnras/273.3.837}, \href
  {https://ui.adsabs.harvard.edu/abs/1995MNRAS.273..837M} {273, 837}

\bibitem[\protect\citeauthoryear{{Malzac}, {Beloborodov}  \&
  {Poutanen}}{{Malzac} et~al.}{2001}]{malzac2001}
{Malzac} J.,  {Beloborodov} A.~M.,   {Poutanen} J.,  2001, \mn@doi [\mnras]
  {10.1046/j.1365-8711.2001.04450.x}, \href
  {https://ui.adsabs.harvard.edu/abs/2001MNRAS.326..417M} {326, 417}

\bibitem[\protect\citeauthoryear{{Malzac}, {Belmont}  \& {Fabian}}{{Malzac}
  et~al.}{2009}]{malzac2009}
{Malzac} J.,  {Belmont} R.,   {Fabian} A.~C.,  2009, \mn@doi [\mnras]
  {10.1111/j.1365-2966.2009.15553.x}, \href
  {https://ui.adsabs.harvard.edu/abs/2009MNRAS.400.1512M} {400, 1512}

\bibitem[\protect\citeauthoryear{{Markoff}, {Nowak}  \& {Wilms}}{{Markoff}
  et~al.}{2005}]{markoff2005}
{Markoff} S.,  {Nowak} M.~A.,   {Wilms} J.,  2005, \mn@doi [\apj]
  {10.1086/497628}, \href
  {https://ui.adsabs.harvard.edu/abs/2005ApJ...635.1203M} {635, 1203}

\bibitem[\protect\citeauthoryear{{Markowitz} et~al.,}{{Markowitz}
  et~al.}{2003}]{markowitz2003}
{Markowitz} A.,  et~al., 2003, \mn@doi [\apj] {10.1086/375330}, \href
  {https://ui.adsabs.harvard.edu/abs/2003ApJ...593...96M} {593, 96}

\bibitem[\protect\citeauthoryear{{Marscher}}{{Marscher}}{2006}]{marscher2006}
{Marscher} A.~P.,  2006, \mn@doi [Astronomische Nachrichten]
  {10.1002/asna.200510510}, \href
  {https://ui.adsabs.harvard.edu/abs/2006AN....327..217M} {327, 217}

\bibitem[\protect\citeauthoryear{{Marscher}, {Jorstad}, {G{\'o}mez}, {Aller},
  {Ter{\"a}sranta}, {Lister}  \& {Stirling}}{{Marscher}
  et~al.}{2002}]{marscher2002}
{Marscher} A.~P.,  {Jorstad} S.~G.,  {G{\'o}mez} J.-L.,  {Aller} M.~F.,
  {Ter{\"a}sranta} H.,  {Lister} M.~L.,   {Stirling} A.~M.,  2002, \mn@doi
  [\nat] {10.1038/nature00772}, \href
  {https://ui.adsabs.harvard.edu/abs/2002Natur.417..625M} {417, 625}

\bibitem[\protect\citeauthoryear{{Marscher} et~al.,}{{Marscher}
  et~al.}{2018}]{marscher2018}
{Marscher} A.~P.,  et~al., 2018, \mn@doi [\apj] {10.3847/1538-4357/aae4de},
  \href {https://ui.adsabs.harvard.edu/abs/2018ApJ...867..128M} {867, 128}

\bibitem[\protect\citeauthoryear{{Marshall} et~al.,}{{Marshall}
  et~al.}{2018}]{marshall2018}
{Marshall} H.~L.,  et~al., 2018, \mn@doi [\apj] {10.3847/1538-4357/aaaf66},
  \href {https://ui.adsabs.harvard.edu/abs/2018ApJ...856...66M} {856, 66}

\bibitem[\protect\citeauthoryear{{McConnell} et~al.,}{{McConnell}
  et~al.}{2002}]{mcconnell2002}
{McConnell} M.~L.,  et~al., 2002, \mn@doi [\apj] {10.1086/340436}, \href
  {https://ui.adsabs.harvard.edu/abs/2002ApJ...572..984M} {572, 984}

\bibitem[\protect\citeauthoryear{{McHardy}, {Lawson}, {Newsam}, {Marscher},
  {Robson}  \& {Stevens}}{{McHardy} et~al.}{1999}]{mchardy1999}
{McHardy} I.,  {Lawson} A.,  {Newsam} A.,  {Marscher} A.,  {Robson} I.,
  {Stevens} J.,  1999, \mn@doi [\mnras] {10.1046/j.1365-8711.1999.02959.x},
  \href {https://ui.adsabs.harvard.edu/abs/1999MNRAS.310..571M} {310, 571}

\bibitem[\protect\citeauthoryear{{McHardy}, {Koerding}, {Knigge}, {Uttley}  \&
  {Fender}}{{McHardy} et~al.}{2006}]{mchardy2006}
{McHardy} I.~M.,  {Koerding} E.,  {Knigge} C.,  {Uttley} P.,   {Fender} R.~P.,
  2006, \mn@doi [\nat] {10.1038/nature05389}, \href
  {https://ui.adsabs.harvard.edu/abs/2006Natur.444..730M} {444, 730}

\bibitem[\protect\citeauthoryear{{Merloni}, {Heinz}  \& {di Matteo}}{{Merloni}
  et~al.}{2003}]{merloni2003}
{Merloni} A.,  {Heinz} S.,   {di Matteo} T.,  2003, \mn@doi [\mnras]
  {10.1046/j.1365-2966.2003.07017.x}, \href
  {https://ui.adsabs.harvard.edu/abs/2003MNRAS.345.1057M} {345, 1057}

\bibitem[\protect\citeauthoryear{{Merloni} et~al.,}{{Merloni}
  et~al.}{2012}]{erositaScienceBook}
{Merloni} A.,  et~al., 2012, arXiv e-prints, \href
  {https://ui.adsabs.harvard.edu/abs/2012arXiv1209.3114M} {p. arXiv:1209.3114}

\bibitem[\protect\citeauthoryear{{Miller}, {Brandt}, {Schneider}, {Gibson},
  {Steffen}  \& {Wu}}{{Miller} et~al.}{2011}]{miller2011}
{Miller} B.~P.,  {Brandt} W.~N.,  {Schneider} D.~P.,  {Gibson} R.~R.,
  {Steffen} A.~T.,   {Wu} J.,  2011, \mn@doi [\apj]
  {10.1088/0004-637X/726/1/20}, \href
  {https://ui.adsabs.harvard.edu/#abs/2011ApJ...726...20M} {726, 20}

\bibitem[\protect\citeauthoryear{{Molina}, {Malizia}, {Bassani}, {Ursini},
  {Bazzano}  \& {Ubertini}}{{Molina} et~al.}{2019}]{molina2019}
{Molina} M.,  {Malizia} A.,  {Bassani} L.,  {Ursini} F.,  {Bazzano} A.,
  {Ubertini} P.,  2019, \mn@doi [\mnras] {10.1093/mnras/stz156}, \href
  {https://ui.adsabs.harvard.edu/abs/2019MNRAS.484.2735M} {484, 2735}

\bibitem[\protect\citeauthoryear{{Mushotzky}, {Done}  \& {Pounds}}{{Mushotzky}
  et~al.}{1993}]{mushotzky1993}
{Mushotzky} R.~F.,  {Done} C.,   {Pounds} K.~A.,  1993, \mn@doi [\araa]
  {10.1146/annurev.aa.31.090193.003441}, \href
  {https://ui.adsabs.harvard.edu/abs/1993ARA&A..31..717M} {31, 717}

\bibitem[\protect\citeauthoryear{{Nandra}, {George}, {Mushotzky}, {Turner}  \&
  {Yaqoob}}{{Nandra} et~al.}{1997}]{nandra1997}
{Nandra} K.,  {George} I.~M.,  {Mushotzky} R.~F.,  {Turner} T.~J.,   {Yaqoob}
  T.,  1997, \mn@doi [\apjl] {10.1086/310937}, \href
  {https://ui.adsabs.harvard.edu/abs/1997ApJ...488L..91N} {488, L91}

\bibitem[\protect\citeauthoryear{{Nandra}, {O'Neill}, {George}  \&
  {Reeves}}{{Nandra} et~al.}{2007}]{nandra2007}
{Nandra} K.,  {O'Neill} P.~M.,  {George} I.~M.,   {Reeves} J.~N.,  2007,
  \mn@doi [\mnras] {10.1111/j.1365-2966.2007.12331.x}, \href
  {https://ui.adsabs.harvard.edu/abs/2007MNRAS.382..194N} {382, 194}

\bibitem[\protect\citeauthoryear{{Nandra} et~al.,}{{Nandra}
  et~al.}{2013}]{athena2013}
{Nandra} K.,  et~al., 2013, arXiv e-prints, \href
  {https://ui.adsabs.harvard.edu/abs/2013arXiv1306.2307N} {p. arXiv:1306.2307}

\bibitem[\protect\citeauthoryear{{O'Neill}, {Nandra}, {Papadakis}  \&
  {Turner}}{{O'Neill} et~al.}{2005}]{oneill2005}
{O'Neill} P.~M.,  {Nandra} K.,  {Papadakis} I.~E.,   {Turner} T.~J.,  2005,
  \mn@doi [\mnras] {10.1111/j.1365-2966.2005.08860.x}, \href
  {https://ui.adsabs.harvard.edu/abs/2005MNRAS.358.1405O} {358, 1405}

\bibitem[\protect\citeauthoryear{{Orr} \& {Browne}}{{Orr} \&
  {Browne}}{1982}]{orr1982}
{Orr} M.~J.~L.,  {Browne} I.~W.~A.,  1982, \mn@doi [\mnras]
  {10.1093/mnras/200.4.1067}, \href
  {https://ui.adsabs.harvard.edu/abs/1982MNRAS.200.1067O} {200, 1067}

\bibitem[\protect\citeauthoryear{{Padovani} et~al.,}{{Padovani}
  et~al.}{2017}]{padovani2017}
{Padovani} P.,  et~al., 2017, \mn@doi [\aapr] {10.1007/s00159-017-0102-9},
  \href {https://ui.adsabs.harvard.edu/abs/2017A&ARv..25....2P} {25, 2}

\bibitem[\protect\citeauthoryear{{Page}, {Reeves}, {O'Brien}  \&
  {Turner}}{{Page} et~al.}{2005}]{page2005}
{Page} K.~L.,  {Reeves} J.~N.,  {O'Brien} P.~T.,   {Turner} M.~J.~L.,  2005,
  \mn@doi [\mnras] {10.1111/j.1365-2966.2005.09550.x}, \href
  {https://ui.adsabs.harvard.edu/abs/2005MNRAS.364..195P} {364, 195}

\bibitem[\protect\citeauthoryear{{Paliya}, {Ajello}, {Cao}, {Giroletti},
  {Kaur}, {Madejski}, {Lott}  \& {Hartmann}}{{Paliya}
  et~al.}{2020}]{paliya2020}
{Paliya} V.~S.,  {Ajello} M.,  {Cao} H.~M.,  {Giroletti} M.,  {Kaur} A.,
  {Madejski} G.,  {Lott} B.,   {Hartmann} D.,  2020, \mn@doi [\apj]
  {10.3847/1538-4357/ab9c1a}, \href
  {https://ui.adsabs.harvard.edu/abs/2020ApJ...897..177P} {897, 177}

\bibitem[\protect\citeauthoryear{{Papadakis}}{{Papadakis}}{2004}]{papadakis2004}
{Papadakis} I.~E.,  2004, \mn@doi [\mnras] {10.1111/j.1365-2966.2004.07351.x},
  \href {https://ui.adsabs.harvard.edu/abs/2004MNRAS.348..207P} {348, 207}

\bibitem[\protect\citeauthoryear{{P{\^a}ris} et~al.,}{{P{\^a}ris}
  et~al.}{2018}]{paris2018}
{P{\^a}ris} I.,  et~al., 2018, \mn@doi [\aap] {10.1051/0004-6361/201732445},
  \href {https://ui.adsabs.harvard.edu/#abs/2018A&A...613A..51P} {613, A51}

\bibitem[\protect\citeauthoryear{{Ponti}, {Papadakis}, {Bianchi}, {Guainazzi},
  {Matt}, {Uttley}  \& {Bonilla}}{{Ponti} et~al.}{2012}]{ponti2012}
{Ponti} G.,  {Papadakis} I.,  {Bianchi} S.,  {Guainazzi} M.,  {Matt} G.,
  {Uttley} P.,   {Bonilla} N.~F.,  2012, \mn@doi [\aap]
  {10.1051/0004-6361/201118326}, \href
  {https://ui.adsabs.harvard.edu/abs/2012A&A...542A..83P} {542, A83}

\bibitem[\protect\citeauthoryear{{Poutanen} \& {Zdziarski}}{{Poutanen} \&
  {Zdziarski}}{2003}]{poutanen2003}
{Poutanen} J.,  {Zdziarski} A.~A.,  2003, in {Durouchoux} P.,  {Fuchs} Y.,
  {Rodriguez} J.,  eds, New Views on Microquasars. p.~95 (\mn@eprint {arXiv}
  {astro-ph/0209186})

\bibitem[\protect\citeauthoryear{{Primini} \& {Kashyap}}{{Primini} \&
  {Kashyap}}{2014}]{primini2014}
{Primini} F.~A.,  {Kashyap} V.~L.,  2014, \mn@doi [\apj]
  {10.1088/0004-637X/796/1/24}, \href
  {https://ui.adsabs.harvard.edu/abs/2014ApJ...796...24P} {796, 24}

\bibitem[\protect\citeauthoryear{{Rani} \& {Stalin}}{{Rani} \&
  {Stalin}}{2018}]{rani2018}
{Rani} P.,  {Stalin} C.~S.,  2018, \mn@doi [\apj] {10.3847/1538-4357/aab356},
  \href {https://ui.adsabs.harvard.edu/abs/2018ApJ...856..120R} {856, 120}

\bibitem[\protect\citeauthoryear{{Reeves} \& {Turner}}{{Reeves} \&
  {Turner}}{2000}]{reeves2000}
{Reeves} J.~N.,  {Turner} M.~J.~L.,  2000, \mn@doi [\mnras]
  {10.1046/j.1365-8711.2000.03510.x}, \href
  {https://ui.adsabs.harvard.edu/abs/2000MNRAS.316..234R} {316, 234}

\bibitem[\protect\citeauthoryear{{Reeves}, {Turner}, {Ohashi}  \&
  {Kii}}{{Reeves} et~al.}{1997}]{reeves1997}
{Reeves} J.~N.,  {Turner} M.~J.~L.,  {Ohashi} T.,   {Kii} T.,  1997, \mn@doi
  [\mnras] {10.1093/mnras/292.3.468}, \href
  {https://ui.adsabs.harvard.edu/abs/1997MNRAS.292..468R} {292, 468}

\bibitem[\protect\citeauthoryear{{Risaliti} \& {Lusso}}{{Risaliti} \&
  {Lusso}}{2019}]{risaliti2019}
{Risaliti} G.,  {Lusso} E.,  2019, \mn@doi [Nature Astronomy]
  {10.1038/s41550-018-0657-z}, \href
  {https://ui.adsabs.harvard.edu/abs/2019NatAs...3..272R} {3, 272}

\bibitem[\protect\citeauthoryear{{Rosen} et~al.,}{{Rosen}
  et~al.}{2016}]{rosen2016}
{Rosen} S.~R.,  et~al., 2016, \mn@doi [\aap] {10.1051/0004-6361/201526416},
  \href {https://ui.adsabs.harvard.edu/#abs/2016A&A...590A...1R} {590, A1}

\bibitem[\protect\citeauthoryear{{Sambruna}}{{Sambruna}}{1997}]{sambruna1997}
{Sambruna} R.~M.,  1997, \mn@doi [\apj] {10.1086/304640}, \href
  {https://ui.adsabs.harvard.edu/abs/1997ApJ...487..536S} {487, 536}

\bibitem[\protect\citeauthoryear{{Sambruna}, {Eracleous}  \&
  {Mushotzky}}{{Sambruna} et~al.}{1999}]{sambruna1999}
{Sambruna} R.~M.,  {Eracleous} M.,   {Mushotzky} R.~F.,  1999, \mn@doi [\apj]
  {10.1086/307981}, \href
  {https://ui.adsabs.harvard.edu/abs/1999ApJ...526...60S} {526, 60}

\bibitem[\protect\citeauthoryear{{Schwartz} et~al.,}{{Schwartz}
  et~al.}{2020}]{schwartz2020}
{Schwartz} D.~A.,  et~al., 2020, \mn@doi [\apj] {10.3847/1538-4357/abbd99},
  \href {https://ui.adsabs.harvard.edu/abs/2020ApJ...904...57S} {904, 57}

\bibitem[\protect\citeauthoryear{{Shang} et~al.,}{{Shang}
  et~al.}{2011}]{shang2011}
{Shang} Z.,  et~al., 2011, \mn@doi [The Astrophysical Journal Supplement
  Series] {10.1088/0067-0049/196/1/2}, \href
  {https://ui.adsabs.harvard.edu/#abs/2011ApJS..196....2S} {196, 2}

\bibitem[\protect\citeauthoryear{{Shemmer}, {Brandt}, {Netzer}, {Maiolino}  \&
  {Kaspi}}{{Shemmer} et~al.}{2008}]{shemmer2008}
{Shemmer} O.,  {Brandt} W.~N.,  {Netzer} H.,  {Maiolino} R.,   {Kaspi} S.,
  2008, \mn@doi [\apj] {10.1086/588776}, \href
  {https://ui.adsabs.harvard.edu/abs/2008ApJ...682...81S} {682, 81}

\bibitem[\protect\citeauthoryear{{Shemmer}, {Brandt}, {Paolillo}, {Kaspi},
  {Vignali}, {Lira}  \& {Schneider}}{{Shemmer} et~al.}{2017}]{shemmer2017}
{Shemmer} O.,  {Brandt} W.~N.,  {Paolillo} M.,  {Kaspi} S.,  {Vignali} C.,
  {Lira} P.,   {Schneider} D.~P.,  2017, \mn@doi [\apj]
  {10.3847/1538-4357/aa8b78}, \href
  {https://ui.adsabs.harvard.edu/abs/2017ApJ...848...46S} {848, 46}

\bibitem[\protect\citeauthoryear{{Shen} et~al.,}{{Shen}
  et~al.}{2011}]{shen2011}
{Shen} Y.,  et~al., 2011, \mn@doi [The Astrophysical Journal Supplement Series]
  {10.1088/0067-0049/194/2/45}, \href
  {https://ui.adsabs.harvard.edu/#abs/2011ApJS..194...45S} {194, 45}

\bibitem[\protect\citeauthoryear{{Shu}, {Wang}, {Yaqoob}, {Jiang}  \&
  {Zhou}}{{Shu} et~al.}{2012}]{shu2012}
{Shu} X.~W.,  {Wang} J.~X.,  {Yaqoob} T.,  {Jiang} P.,   {Zhou} Y.~Y.,  2012,
  \mn@doi [\apjl] {10.1088/2041-8205/744/2/L21}, \href
  {https://ui.adsabs.harvard.edu/abs/2012ApJ...744L..21S} {744, L21}

\bibitem[\protect\citeauthoryear{{Siemiginowska}, {LaMassa}, {Aldcroft},
  {Bechtold}  \& {Elvis}}{{Siemiginowska} et~al.}{2008}]{siemiginowska2008}
{Siemiginowska} A.,  {LaMassa} S.,  {Aldcroft} T.~L.,  {Bechtold} J.,   {Elvis}
  M.,  2008, \mn@doi [\apj] {10.1086/589437}, \href
  {https://ui.adsabs.harvard.edu/abs/2008ApJ...684..811S} {684, 811}

\bibitem[\protect\citeauthoryear{{Simmonds}, {Buchner}, {Salvato}, {Hsu}  \&
  {Bauer}}{{Simmonds} et~al.}{2018}]{simmonds2018}
{Simmonds} C.,  {Buchner} J.,  {Salvato} M.,  {Hsu} L.~T.,   {Bauer} F.~E.,
  2018, \mn@doi [\aap] {10.1051/0004-6361/201833412}, \href
  {https://ui.adsabs.harvard.edu/abs/2018A&A...618A..66S} {618, A66}

\bibitem[\protect\citeauthoryear{{Simpson}}{{Simpson}}{2005}]{simpson2005}
{Simpson} C.,  2005, \mn@doi [\mnras] {10.1111/j.1365-2966.2005.09043.x}, \href
  {https://ui.adsabs.harvard.edu/abs/2005MNRAS.360..565S} {360, 565}

\bibitem[\protect\citeauthoryear{{Soldi} et~al.,}{{Soldi}
  et~al.}{2008}]{soldi2008}
{Soldi} S.,  et~al., 2008, \mn@doi [\aap] {10.1051/0004-6361:200809947}, \href
  {https://ui.adsabs.harvard.edu/abs/2008A&A...486..411S} {486, 411}

\bibitem[\protect\citeauthoryear{{Soldi} et~al.,}{{Soldi}
  et~al.}{2014}]{soldi2014}
{Soldi} S.,  et~al., 2014, \mn@doi [\aap] {10.1051/0004-6361/201322653}, \href
  {https://ui.adsabs.harvard.edu/abs/2014A&A...563A..57S} {563, A57}

\bibitem[\protect\citeauthoryear{{Tananbaum} et~al.,}{{Tananbaum}
  et~al.}{1979}]{tananbaum1979}
{Tananbaum} H.,  et~al., 1979, \mn@doi [\apj] {10.1086/183100}, \href
  {https://ui.adsabs.harvard.edu/#abs/1979ApJ...234L...9T} {234, L9}

\bibitem[\protect\citeauthoryear{{Timlin}, {Brandt}, {Zhu}, {Liu}, {Luo}  \&
  {Ni}}{{Timlin} et~al.}{2020}]{timlin2020}
{Timlin} John~D. I.,  {Brandt} W.~N.,  {Zhu} S.,  {Liu} H.,  {Luo} B.,   {Ni}
  Q.,  2020, \mn@doi [\mnras] {10.1093/mnras/staa2661}, \href
  {https://ui.adsabs.harvard.edu/abs/2020MNRAS.tmp.2079T} {}

\bibitem[\protect\citeauthoryear{Timlin, Zhu, Brandt  \& Laor}{Timlin
  et~al.}{2021}]{timlin2021}
Timlin J.,  Zhu S.,  Brandt W.~N.,   Laor A.,  2021, \mn@doi [Research Notes of
  the {AAS}] {10.3847/2515-5172/abfbe5}, 5, 101

\bibitem[\protect\citeauthoryear{{Ulrich}, {Maraschi}  \& {Urry}}{{Ulrich}
  et~al.}{1997}]{ulrich1997}
{Ulrich} M.-H.,  {Maraschi} L.,   {Urry} C.~M.,  1997, \mn@doi [\araa]
  {10.1146/annurev.astro.35.1.445}, \href
  {https://ui.adsabs.harvard.edu/abs/1997ARA&A..35..445U} {35, 445}

\bibitem[\protect\citeauthoryear{{Uttley} \& {McHardy}}{{Uttley} \&
  {McHardy}}{2005}]{uttley2005}
{Uttley} P.,  {McHardy} I.~M.,  2005, \mn@doi [\mnras]
  {10.1111/j.1365-2966.2005.09475.x}, \href
  {https://ui.adsabs.harvard.edu/abs/2005MNRAS.363..586U} {363, 586}

\bibitem[\protect\citeauthoryear{{Wachter}, {Leach}  \& {Kellogg}}{{Wachter}
  et~al.}{1979}]{wachter1979}
{Wachter} K.,  {Leach} R.,   {Kellogg} E.,  1979, \mn@doi [\apj]
  {10.1086/157084}, \href
  {https://ui.adsabs.harvard.edu/abs/1979ApJ...230..274W} {230, 274}

\bibitem[\protect\citeauthoryear{{Weaver} et~al.,}{{Weaver}
  et~al.}{2020}]{weaver2020}
{Weaver} Z.~R.,  et~al., 2020, \mn@doi [\apj] {10.3847/1538-4357/aba693}, \href
  {https://ui.adsabs.harvard.edu/abs/2020ApJ...900..137W} {900, 137}

\bibitem[\protect\citeauthoryear{{Weisskopf}, {Wu}, {Trimble}, {O'Dell},
  {Elsner}, {Zavlin}  \& {Kouveliotou}}{{Weisskopf}
  et~al.}{2007}]{weisskopf2007}
{Weisskopf} M.~C.,  {Wu} K.,  {Trimble} V.,  {O'Dell} S.~L.,  {Elsner} R.~F.,
  {Zavlin} V.~E.,   {Kouveliotou} C.,  2007, \mn@doi [\apj] {10.1086/510776},
  \href {https://ui.adsabs.harvard.edu/abs/2007ApJ...657.1026W} {657, 1026}

\bibitem[\protect\citeauthoryear{Westfall}{Westfall}{2014}]{westfall2014}
Westfall P.~H.,  2014, The American Statistician, 68, 191

\bibitem[\protect\citeauthoryear{{Wilkes} \& {Elvis}}{{Wilkes} \&
  {Elvis}}{1987}]{wilkes1987}
{Wilkes} B.~J.,  {Elvis} M.,  1987, \mn@doi [\apj] {10.1086/165822}, \href
  {https://ui.adsabs.harvard.edu/#abs/1987ApJ...323..243W} {323, 243}

\bibitem[\protect\citeauthoryear{{Willis} et~al.,}{{Willis}
  et~al.}{2005}]{willis2005}
{Willis} J.~P.,  et~al., 2005, \mn@doi [\mnras]
  {10.1111/j.1365-2966.2005.09473.x}, \href
  {https://ui.adsabs.harvard.edu/abs/2005MNRAS.363..675W} {363, 675}

\bibitem[\protect\citeauthoryear{{Worrall}, {Giommi}, {Tananbaum}  \&
  {Zamorani}}{{Worrall} et~al.}{1987}]{worrall1987}
{Worrall} D.~M.,  {Giommi} P.,  {Tananbaum} H.,   {Zamorani} G.,  1987, \mn@doi
  [\apj] {10.1086/164999}, \href
  {https://ui.adsabs.harvard.edu/#abs/1987ApJ...313..596W} {313, 596}

\bibitem[\protect\citeauthoryear{{Wozniak}, {Zdziarski}, {Smith}, {Madejski}
  \& {Johnson}}{{Wozniak} et~al.}{1998}]{wozniak1998}
{Wozniak} P.~R.,  {Zdziarski} A.~A.,  {Smith} D.,  {Madejski} G.~M.,
  {Johnson} W.~N.,  1998, \mn@doi [\mnras] {10.1046/j.1365-8711.1998.01831.x},
  \href {https://ui.adsabs.harvard.edu/abs/1998MNRAS.299..449W} {299, 449}

\bibitem[\protect\citeauthoryear{{Xue} et~al.,}{{Xue} et~al.}{2011}]{xue2011}
{Xue} Y.~Q.,  et~al., 2011, \mn@doi [\apjs] {10.1088/0067-0049/195/1/10}, \href
  {https://ui.adsabs.harvard.edu/abs/2011ApJS..195...10X} {195, 10}

\bibitem[\protect\citeauthoryear{{York} et~al.,}{{York}
  et~al.}{2000}]{york2000}
{York} D.~G.,  et~al., 2000, \mn@doi [\aj] {10.1086/301513}, \href
  {https://ui.adsabs.harvard.edu/#abs/2000AJ....120.1579Y} {120, 1579}

\bibitem[\protect\citeauthoryear{{Zamorani}, {Giommi}, {Maccacaro}  \&
  {Tananbaum}}{{Zamorani} et~al.}{1984}]{zamorani1984b}
{Zamorani} G.,  {Giommi} P.,  {Maccacaro} T.,   {Tananbaum} H.,  1984, \mn@doi
  [\apj] {10.1086/161764}, \href
  {https://ui.adsabs.harvard.edu/abs/1984ApJ...278...28Z} {278, 28}

\bibitem[\protect\citeauthoryear{{Zhou} \& {Gu}}{{Zhou} \&
  {Gu}}{2020}]{zhou2020}
{Zhou} M.,  {Gu} M.,  2020, \mn@doi [\apj] {10.3847/1538-4357/ab7dca}, \href
  {https://ui.adsabs.harvard.edu/abs/2020ApJ...893...39Z} {893, 39}

\bibitem[\protect\citeauthoryear{{Zhou} \& {Gu}}{{Zhou} \&
  {Gu}}{2021}]{zhou2021}
{Zhou} M.-H.,  {Gu} M.-F.,  2021, \mn@doi [Research in Astronomy and
  Astrophysics] {10.1088/1674-4527/21/1/4}, \href
  {https://ui.adsabs.harvard.edu/abs/2021RAA....21....4Z} {21, 004}

\bibitem[\protect\citeauthoryear{{Zhu}, {Brandt}, {Wu}, {Garmire}  \&
  {Miller}}{{Zhu} et~al.}{2019}]{zhu2019}
{Zhu} S.~F.,  {Brandt} W.~N.,  {Wu} J.,  {Garmire} G.~P.,   {Miller} B.~P.,
  2019, \mn@doi [\mnras] {10.1093/mnras/sty2832}, \href
  {https://ui.adsabs.harvard.edu/#abs/2019MNRAS.482.2016Z} {482, 2016}

\bibitem[\protect\citeauthoryear{{Zhu}, {Brandt}, {Luo}, {Wu}, {Xue}  \&
  {Yang}}{{Zhu} et~al.}{2020}]{zhu2020}
{Zhu} S.~F.,  {Brandt} W.~N.,  {Luo} B.,  {Wu} J.,  {Xue} Y.~Q.,   {Yang} G.,
  2020, \mn@doi [\mnras] {10.1093/mnras/staa1411}, \href
  {https://ui.adsabs.harvard.edu/abs/2020MNRAS.496..245Z} {496, 245}

\makeatother
\end{thebibliography}

\appendix
\section{RQQ sample}
\label{sec:appendix1}
The properties of the comparison RQQ sample (see the end of \S~\ref{sec:rqqSample}) in this work are presented in Table~\ref{tab:rqqsample}.
The serendipitous {\it Chandra} observations utilized for these RQQs,
along with resulting derived X-ray properties, are listed in Table~\ref{tab:rqqobs}.

\begin{table}
\centering
    \caption{The RQQ sample used in this paper.} 
\label{tab:rqqsample}
\begin{threeparttable}[b]
\begin{tabular}{cccc}
\hline
\hline
    Name &  $z$  & $m_i$& $\log L_\mathrm{2500\angstrom}$ \\
\hline
    000635.69$-$000721.04&0.891&20.65&29.47 \\
    002312.05$+$002545.99&2.657&20.64&30.48 \\
    002402.12$-$020101.13&2.614&19.90&30.77 \\
    002707.56$+$170617.88&0.947&20.25&29.69 \\
    002751.16$+$262437.66&1.173&20.40&29.77\\
\hline
\end{tabular}
\end{threeparttable}
\end{table}

\begin{table*}
\centering
\caption{The sample of RQQ X-ray observations.}
\label{tab:rqqobs}
\begin{threeparttable}[b]
\begin{tabularx}{\linewidth}{@{}Y@{}}
\begin{tabular}{ccccccccccc}
\hline
\hline
    Name & ObsID & MJD & Inst.& $\log f_\mathrm{det}$ & net & SNR & $\log f_\mathrm{x}$& xdet & $\Gamma$ & goodness-of-fit \\
\hline
    000635.69$-$000721.04&4096&52853.3&ACIS&	$-$14.02&3.1&1.5&$-$14.02&	0&	$-99$	&17.5/25.2/9.2\\
    000635.69$-$000721.04&5617&53579.5&ACIS&	$-$14.13&32.4&4.8&$-$13.55&	1&  $2.06_{-0.58}^{+0.69}$	&116.5/128.7/13.6\\
    002312.05$+$002545.99&2252&51915.4& ACIS& $-$14.45&105.2&8.8&$-$13.11&	1&	$1.14_{-0.29}^{+0.31}$	&121.4/117.4/12.1\\
    002402.12$-$020101.13&2099&52141.1& 	ACIS& $-$14.53&9.6&3.0&$-$13.96&	1&	$2.04_{-0.90}^{+1.0}$	&40.6/47.4/11.2\\
    002402.12$-$020101.13&8918&54768.1& ACIS& $-$14.71&7.7&2.5&$-$14.39&	1&	$2.05_{-1.09}^{+1.48}$	&44.9/50.7/12.2\\
\hline
\end{tabular}
\end{tabularx}
\vspace{1mm}
    {\it Notes:} The columns of this table are similar to those of Table~\ref{tab:obs}, where further explanations are provided.
\end{threeparttable}
\end{table*}

\bsp    
\label{lastpage}
\end{document}